\RequirePackage{fix-cm}
\documentclass{svjour3}    
\smartqed 

\usepackage{graphicx}
\usepackage{subcaption}
\usepackage{amsmath}
\usepackage{mathptmx}  
\usepackage[a4paper, total={6in, 10in}, textwidth = 450pt, rmargin= 1.0in]{geometry}
\usepackage[super]{nth}
\usepackage{caption}
\DeclareCaptionLabelSeparator{none}{ }
\usepackage{relsize}
\usepackage[export]{adjustbox}
\usepackage{float}
\floatstyle{plaintop}
\restylefloat{table}
\usepackage{xcolor}
\usepackage{ulem}
\usepackage{color,soul}
\usepackage[colorlinks=true,
            linkcolor=blue,
            urlcolor=blue,
            citecolor=blue]{hyperref}            
\usepackage{bm}
\usepackage{amsfonts}
\usepackage{mathtools}
\usepackage{amsmath}
\usepackage[T1]{fontenc}
\usepackage[cal = pxtx, scr = dutchcal]{mathalpha}
\DeclareMathAlphabet{\mathcal}{OMS}{cmsy}{m}{n}
\usepackage{titlesec}
\titleformat*{\subsubsection}{\bfseries}
\usepackage[thicklines]{cancel}

\usepackage{rotating}
\usepackage{amssymb}
\usepackage{booktabs}
\usepackage{tabularray}
\usepackage{array,multirow,multicol}

\newcommand{\sample}{\mathcal{S}}
\newcommand{\conv}{\mathcal{C}}
\renewcommand{\thefigure}{\arabic{figure}}
\makeatletter
\renewcommand{\p@subfigure}{\thefigure}      
 
\makeatother
\usepackage{enumerate}
\usepackage[misc]{ifsym}

\usepackage[authoryear]{natbib}
\bibpunct{(}{)}{}{a}{}{;}

\begin{document}

\title{The Bayesian Gaussian Process Latent Variable Model for Spatio-Temporal Stream Networks}

\author{Marno Basson         \and
        Tobias M. Louw        \and
        Theresa R. Smith}

\institute{Marno Basson \at
              Department of Chemical Engineering, Stellenbosch University, Stellenbosch, 7600, South Africa \\
              ORCID ID: 0000-0002-7871-7912 \\
              \email{marnob@sun.ac.za}   \\              
           \and
           Tobias M. Louw \at
              Department of Chemical Engineering, Stellenbosch University, Stellenbosch, 7600, South Africa \\
              ORCID ID: 0000-0002-4166-0799 \\
              \email{tmlouw@sun.ac.za}   \\              
           \and 
           Theresa R. Smith (\Letter) \at
              Department of Mathematical Sciences, University of Bath, BA2 7AY, Bath, UK \\
              ORCID ID: 0000-0002-7085-3864 \\
              \email{trs35@bath.ac.uk} \\}

\date{}

\maketitle

\vspace{-1in}

\begin{abstract}

A variational inference-based framework for training a multi-output Gaussian process latent variable model, specifically tailored to the tails-up spatio-temporal stream network, is developed. Training, given a censored observational data set subject to missing values, proceeds by maximising a secondary variational lower bound on the model log marginal likelihood using gradient-based optimisation. Consequently, the theoretical development for a new family of tails-up spatio-temporal stream network models is introduced which rely on the sparse Gaussian process inducing variable framework, the Bayesian Gaussian process latent variable model, and local variational methods. These spatio-temporal models use stream distance instead of Euclidean distance and capture spatial and temporal dependencies using auto/cross-correlation and process convolution, respectively, which allows for the development of valid separable spatio-temporal stream network-based covariance functions. Results from the simulation-based case studies indicate that the proposed framework performs well when considering benchmark comparisons and several performance metrics. 

          
\end{abstract}
\section{Introduction}\label{Introduction}
Stream and river networks are essential resources (\citealp{VerHoef2006},  \citealp{VerHoef2010}), providing water for use by residential, commercial, industrial, agricultural, and natural ecosystems \citep{Mihelcic2014}. Several studies have aimed at raising awareness about contaminants of emerging concern (CECs) which include excreted hormones, pharmaceutical products, pesticides, agricultural fertilisers, personal care products, etc., and their potentially detrimental effect on the environment and human health (\citealp{Kolpin2002}; \citealp{Kolpin2004}; \citealp{Glassmeyer2005}; \citealp{Bruce2010}; \citealp{Archer2017a}; \citealp{Archer2017b}; \citealp{Lorenzo2019}). In the past, a lack of observational data has prohibited the effective monitoring and management of ecological and environmental factors. However, the advent of river and stream network-based in-situ sensing has allowed practitioners to collect more ecological and environmental spatio-temporal data (\citealp{Santos2022}) making it possible to monitor attributes like water quality (\citealp{Stackpoole2017}) and nutrient dynamics (\citealp{Wollheim2017}). 

Several problems can arise during the collection of stream network-based ecological and environmental spatio-temporal data. Firstly, the observational data is likely to exhibit spatio-temporal dependencies which arise from the stream network structure, fluid flow direction, discharge magnitude, longitudinal connectivity, etc (\citealp{Peterson2007}; \citealp{Peterson2010}; \citealp{Santos2022}). Secondly, despite the advent of in-situ sensing, practitioners can still encounter data censoring. Data censoring occurs when practitioners have access to a partially observed spatio-temporal measurement. These partially observed measurements can arise when the measured value falls outside the measurement device/analysis procedure's sensitivity range (\citealp{Ertin2007}; \citealp{Groot2012}; \citealp{Gammelli2022}; \citealp{Basson2023}). Furthermore, several data entries can also be missing due to sensor or analysis equipment failure.
Thirdly, spatio-temporal response variables can be subject to physical constraints, e.g., positivity of concentrations (\citealp{Holcomb2018}). In traditional spatial, as well as spatio-temporal statistics, a general linear regression model is typically used for inference and prediction purposes. Associated with the general linear model is a symmetric positive definite covariance matrix $\boldsymbol{\Sigma}$ parameterised in terms of $\frac{N(N+1)}{2}$ unique parameters, where $ N $ corresponds to the total number of observations. Using, for example, Euclidean distance, the number of parameters can be reduced if a covariance function is used to construct the covariance matrix $\boldsymbol{\Sigma}$ (\citealp{Cressie1993}; \citealp{VerHoef2006}). The use of a distance metric leads to the last problem encountered during the spatio-temporal data collection procedure. Typically, the distance metric is either measured during the data collection procedure or estimated using a Geographical Information System (GIS) and, therefore, subject to measurement error (\citealp{VerHoef2006}). The measured or estimated distance values used to compute the covariance matrix $\boldsymbol{\Sigma}$ impose additional input uncertainty mapping through the $\boldsymbol{\Sigma}$ matrix (\citealp{Titsias2010}), which consequently affects the quality of the predictions. Traditional spatial, as well as spatio-temporal, general linear models only account for the measurement uncertainty in the observational data associated with the response variable (via the nugget effect) and ignore the additional input uncertainty associated with covariance matrix $\boldsymbol{\Sigma}$. 

As noted by \cite{Santos2022}, very few frameworks exist that describe the complex and unique set of problems associated with stream network-based spatio-temporal data. The existing stream network-based frameworks can broadly be categorised into $\{1\}$ times series-based models (see, for example, \citealp{Hague2014}; \citealp{Graf2018}; \citealp{Graf2021}), $\{2\}$ purely spatial-based models (see, for example, \citealp{Chantal2006}; \citealp{VerHoef2006}; \citealp{Cressie2006}; \citealp{VerHoef2010}; \citealp{Isaak2014}; \citealp{Neill2018}; \citealp{McManus2020}), and $\{3\}$ a handful of spatio-temporal models (see, for example, \citealp{Money2009a}; \citealp{Money2009b}; \citealp{Donnell2014}; \citealp{Holcomb2018}; \citealp{Jackson2018}; \citealp{Tang2020}; \citealp{Santos2022}). 

Of particular interest to this study is the work of \cite{VerHoef2006} which outlines an autocorrelation-based methodology for constructing valid Gaussian process covariance functions that use hydrologic distance and flow data from the stream network to make predictions about the response variable of interest, for example, a CEC, at unsampled spatial locations. The stream network-based covariance functions are developed by integrating a moving-average function, otherwise referred to as a smoothing kernel (\citealp{Alvarez2010}), against a Gaussian white noise process. By running the smoothing kernel upstream from a location on the stream network, the authors develop covariance functions that explicitly incorporate the stream network flow via a weighting matrix. The autocorrelation-based procedure results in covariance functions that, by construction, produce valid covariance matrices that depend on the hydrologic distance between locations and only allow for covariance between stream locations that are flow-connected. This model was named the tails-up model based on the methodology underpinning the covariance construction procedure (\citealp{Peterson2010}). In this paper, the tails-up model will be referred to as the tails-up spatial stream network (SSN) model to distinguish it from the spatio-temporal counterpart developed in subsequent sections. The flow-connected behaviour encoded by the tails-up SSN approach makes the framework particularly useful for modelling the passive downstream movement of materials like CECs or other waterborne chemicals in a stream network. However, note that the tails-up SSN model also treats the estimated/measured values for the hydrologic distances and weighting parameters as deterministic, effectively ignoring the additional input uncertainty that arises from using these estimated/measured values for the hydrologic distances and weighting parameters.

This paper aims to develop a new theoretical framework for inferring CEC concentration latent function profiles from spatio-temporal stream network-based observational data. This will be achieved, within the context of extending the tails-up SSN model of \cite{VerHoef2006}, by developing an alternative Gaussian process-based latent variable model representation that accounts for \{1\} spatio-temporal dependencies underpinning stream network-based data in the multi-output (i.e., Co-Kriging) setting using separable kernel functions, \{2\} additional input uncertainty which arises from the measured/estimated tails-up model hydrologic distances and weighting parameters, \{3\} positivity constraints on the CEC concentration latent function profiles inferred from observational data, and \{4\} limitations associated with censoring in spatio-temporal data sets. Points \{1\} and \{2\} form the foundation of this study whereas points \{3\} and \{4\} are optional attributes that are introduced and explored. Despite imposing the latent function positivity constraint and developing the framework to account for data censoring, the theoretical methodology presented in this paper is completely general and can be applied to any latent function of interest for which an uncensored fully observed, or partially observed (i.e., missing), data set is available.To the best of the author's current knowledge, a spatial, as well as a spatio-temporal, multivariate (i.e., Co-Kriging) extension of the work of \cite{VerHoef2006} does not yet exist in the current Geostatistics literature.

Moreover, this paper aims to extend the work of \cite{VerHoef2006} even further by providing a mathematical tool that allows practitioners to derive a closed-form variational lower bound on the log marginal likelihood of the developed probabilistic latent variable model. This can be achieved by applying the variational sparse Gaussian process regression (\citealp{Titsias2008, Titsias2009}) and Bayesian Gaussian process latent variable (\citealp{Titsias2010}) models in conjunction with local variational methods (\citealp{Jordan1999}; \citealp{Nickisch2008}; \citealp{Bishop2009}). Although the variational sparse Gaussian process (GP) regression framework of \cite{Titsias2008, Titsias2009} was originally developed to facilitate computational speedups, in this paper, the authors leverage the framework as a mathematical tool to induce a lower bound on the log marginal likelihood of the developed probabilistic model (\citealp{Titsias2010}; \citealp{Gredilla2011}; \citealp{Damianou2011}; \citealp{Titsias2013}; \citealp{Damianou2013}; \citealp{Damianou2016}; \citealp{Zhao2016}). This allows the authors to variationally integrate over the hydrological distances and weighting parameters of the tails-up SSN framework resulting in a variational lower bound that can be used to perform Bayesian model training and inference. 

Although computational efficiency is not used as a primary motivating factor in this study, an interesting consequence of using the sparse GP-based methodology (\citealp{Titsias2008, Titsias2009}) stems from the fact that the lower bound developed in this paper does not require inverting the matrix $ \boldsymbol{\Sigma} $, which can be a numerical bottleneck during gradient-based optimisation and prediction (see Sect. \ref{Section2.1} and \citealp{Santos2022}). Instead, the lower bound requires inverting a smaller $ M \times M $ matrix $ \boldsymbol{K}_{MM} $ which facilitates computational speedups while maintaining prediction accuracy (see \cite{Titsias2008, Titsias2009} for more details). The developed framework is demonstrated with simulation-based case studies and results indicate that the proposed methodology performs well when considering benchmark comparisons and several performance metrics.

The remainder of the article is structured as follows. Section \ref{Section2} reformulates the tails-up SSN methodology of \cite{VerHoef2006} into an equivalent Gaussian process regression representation. 
Section \ref{SectionM} proceeds by introducing the multi-output (i.e., Co-Kriging) spatio-temporal prior density and likelihood function model specification procedure, as well as a mechanism for addressing censored observational data. 
Section \ref{Section3} introduces plausible approaches to solving the additional latent variable input uncertainty propagation problem and why these approaches fail. Building on Sect. \ref{Section3}, the solution to the input uncertainty propagation problem, as introduced in Sect. \ref{Section4}, comes in the form of the Bayesian Gaussian Process latent variable model (BGP-LVM) for stream networks. 
Section \ref{Section9} demonstrates the ability of the proposed BGP-LVM framework for stream networks to learn latent function representations from the different simulated observational data set scenarios, whereas Sect. \ref{Section10} ends with a discussion and some concluding remarks about the limitations of the proposed BGP-LVM framework for stream networks. Further details regarding model derivation, the latent function predictive equations, and other theoretical minutiae can be found in the accompanying Supplementary Information. 

\vspace{-0.45 cm}
\section{Developing An Equivalent Spatial Stream Network Representation}\label{Section2}

In this section, the authors outline the Gaussian process-based interpretation of the tails-up SSN model developed by \cite{VerHoef2006} The GP-based reformulation leads to several important insights about the tails-up SSN model structure, assumptions, and practical limitations.

\vspace{-0.45 cm}

\subsection{The Spatial Stream Network Model}\label{Section2.1}

The novel stream network-based framework proposed by \cite{VerHoef2006} can be formulated as a general spatial linear model according to Eq. \eqref{Eq1}.

\vspace{-0.35 cm}

\begin{equation}\label{Eq1}
    \boldsymbol{y_\kappa = Z\beta + \epsilon_\kappa} 
\end{equation}

The $N \times 1$ vector $\boldsymbol{y}_\kappa$ denotes the noise-corrupted observational data. The subscript $ \kappa $ indicates the original data set, which may exist on some subset $\mathcal{Y}_\kappa $ of the real line, i.e. $ y_\kappa \in \mathcal{Y}_\kappa \subseteq \mathbb{R} $. In subsequent sections, the authors will employ a transformation $ \mathcal{Y}_\kappa \to \mathbb{R} $ and perform inference on the transformed data set for which the symbol $ \boldsymbol{y} \in \mathbb{R}^n$ is reserved (see Section \ref{SectionM}). For example, strictly positive observations in the set $ \mathcal{Y}_\kappa = \{y_\kappa \in \mathbb{R} | y_\kappa > 0 \}$ may be transformed to the unconstrained space using a log-transform, i.e., $ y = \ln(y_\kappa ) $. 

\begin{figure*}[!ht]
    \centering
    \hspace{-0.80cm}
    \includegraphics[scale = 1]{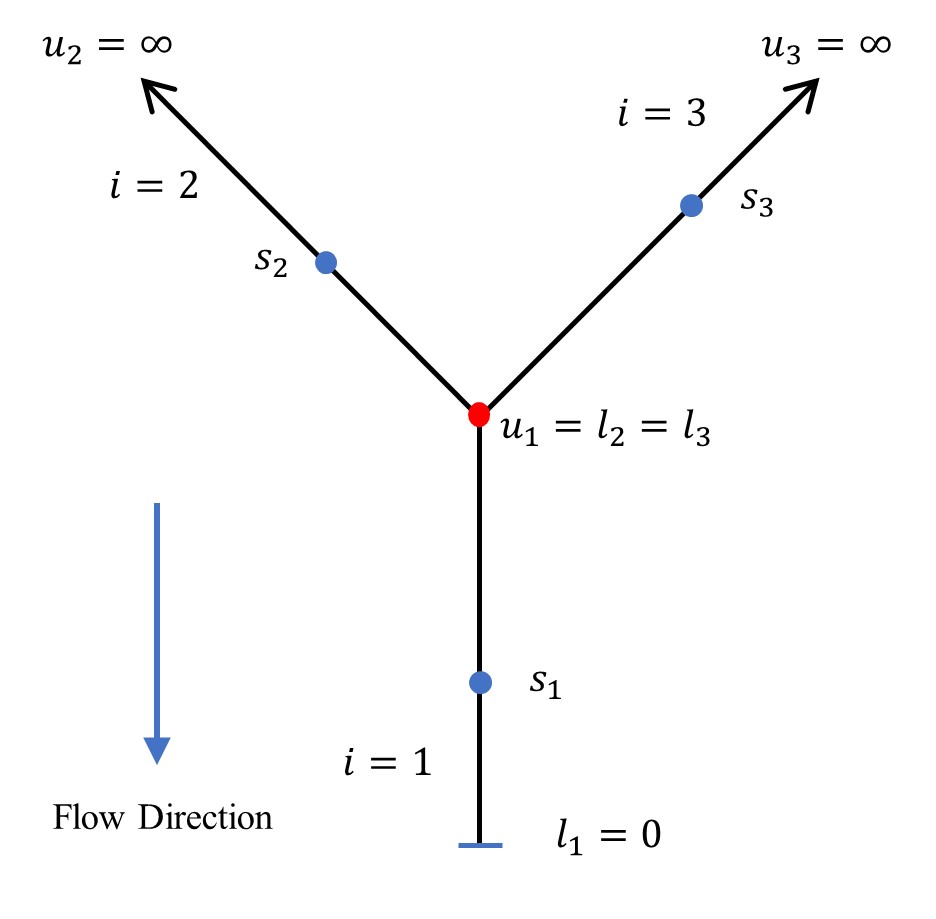}
    \caption{Hypothetical stream network consisting of three stream segments labelled $ i = 1, i = 2 $, and $ i = 3 $, respectively, with three sampled spatial locations denoted by $s_1, s_2 $, and $ s_3 $, respectively. Figure \ref{fig:Figure1} has been reproduced and adjusted from the work of \cite{VerHoef2006}. \vspace{-0.35 cm}}
    \label{fig:Figure1}
\end{figure*}

An $ M\times 1$ covariate vector $ \boldsymbol{z}  = [z_0,z_1,\cdots, z_{M-1}]^T $ models the relationship between the response variable and the known covariates at each sampled spatial location. The set of covariates at all sampled spatial locations are summarized in the design matrix $\boldsymbol{Z}$. The $ M\times 1$ parameter vector  $ \boldsymbol{\beta} = [\beta_0,\beta_1, \cdots,\beta_{M-1}]^T $ captures the relationship between the response and explanatory variables. In traditional spatial statistics, $\boldsymbol{Z}$ and $\boldsymbol{\beta}$ are typically treated as deterministic, while $\boldsymbol{\epsilon}_\kappa$ is assumed to be normally distributed with zero-mean and covariance $\boldsymbol{\Sigma}$. However, $\boldsymbol{\Sigma}$ contains too many parameters to estimate without imposing additional structure. Practitioners bypass this problem using assumptions such as ergodicity and stationarity  allowing them to formulate a distance dependent covariance using an appropriate kernel function (\citealp{Cressie1993}). \cite{VerHoef2006} extended the spatial linear model to stream networks and developed a new class of kernel functions which incorporates the stream network structure and relies on the hydrological distance (i.e., the distance travelled along a stream between two spatial locations). A generalised representation of a stream network is shown in Fig. \ref{fig:Figure1}, which will form the foundation for the probabilistic model developed in this paper. 

The generalised stream network in Fig. \ref{fig:Figure1} has a single most downstream spatial location $ l_1 $ set to a reference location of zero. A stream network will consist of finitely many stream segments, and many spatial locations on separate segments may have the same upstream distance relative to $ l_1 = 0$. In order to uniquely identify locations on a stream network, each stream segment is arbitrarily indexed by $ i = 1,2, \cdots, \mathcal{B} $ and the $ n^\text{th} $ spatial location on the $ i^\text{th} $ stream segment is identified by $ s_{i,n} $. A numerical value is assigned to each $ s_{i,n} \in \mathbb{R}$ equal to the length of the continuous line from the single most downstream location $ l_1 $ to the location identified by $ s_{i,n} $. If two locations are flow-connected (as defined below), then $ s_{i,n} > s_{d,m} $ implies that $ s_{i,n} $ is upstream of $s_{d,m} $. Note that in Fig. \ref{fig:Figure1} the second spatial index $ n $ has been dropped since there is a single spatial location per $ i^\text{th} $ stream segment and no ambiguity exists. 

To complete the description, the most downstream location on the $ i^{\text{th}} $ stream segment is denoted by $ l_i $ and the most upstream location by $ u_i $. The most upstream location on one stream segment will necessarily be equivalent to the most downstream location on the upstream segments of the confluence, e.g., $ u_1=l_2=l_3$ in Fig. \ref{fig:Figure1}. If there are no more upstream segments (e.g., $ i = 2 $ or $ i = 3 $ in Fig. \ref{fig:Figure1}) then $ u_i \to \infty $ (\citealp{VerHoef2006}). 

The set $ \mathcal{D}_{s_{i,n}} \subseteq \{ 1, 2, ... \mathcal{B} \}$ is defined as the index set of all stream segments downstream of spatial location $ s_{i,n} $ into which fluid passing through $ s_{i,n} $ flows, including $ i $. Two spatial locations $ s_{i,n} > s_{d,m} $ are said to be flow-connected if $ \mathcal{D}_{s_{i,n}} \cap \mathcal{D}_{s_{d,m}} =  \mathcal{D}_{s_{d,m}} $. In Fig. \ref{fig:Figure1}, $\mathcal{D}_{{s}_1} = \{1\} $ while $ \mathcal{D}_{{s}_2} = \{1,2\}$ and $ \mathcal{D}_{{s}_3} = \{1,3\}$. Spatial locations $ s_1 $ and $ s_2 $ ($s_1$ and $s_3$) are flow-connected since $ \mathcal{D}_{{s}_1} \cap \mathcal{D}_{{s}_2} = \mathcal{D}_{{s}_1} $ ($ \mathcal{D}_{{s}_1} \cap \mathcal{D}_{{s}_3} = \mathcal{D}_{{s}_1} $),  but $ s_2 $ and $ s_3 $ are not flow-connected since $ \mathcal{D}_{{s}_2} \cap \mathcal{D}_{{s}_3} \neq \mathcal{D}_{{s}_2} $ or $ \mathcal{D}_{{s}_3} $. With the definition of flow-connected sites, the hydrological distance between two spatial locations $ s_{i,n} $ and $ s_{d,m}$ is defined as follows (\citealp{VerHoef2006})

\vspace{-0.25 cm}

\begin{equation}\label{Eq24}
     d(s_{i,n},s_{d,m}) =
        \begin{cases} 
         \vert s_{i,n} - s_{d,m} \vert   
         &\text{ if sites $ s_{i,n} $ and $ s_{d,m} $ are flow-connected }\\
         (s_{i,n} - u) + (s_{d,m} - u) 
         &\text{ otherwise, with $ u = $ max $ \{u_k: k \in \mathcal{D}_{s_{i,n}} \cap \mathcal{D}_{s_{d,m}} \} $}
        \end{cases}
\end{equation}

The symbol $ u $ represents the nearest downstream junction point common to both sampled spatial locations. Using the proposed moving-average-based construction approach, \cite{VerHoef2006} showed that it is possible to derive the following valid kernel function for stream networks

\vspace{-0.35 cm}

\begin{equation}\label{Eq3}
     C \bigl(s_{i,n},s_{d,m}\vert \boldsymbol{\theta}_K, \theta_0 \bigr) = 
        \begin{cases} 
         0  
         &\text{ if sites are not flow-connected } \\
         \conv(0  \vert \boldsymbol{\theta}_K) + \theta_0 
         &\text{ if sites are flow-connected and $ d(s_{i,n},s_{d,m}) = 0$ }\\[1.5ex]
        \Biggl[ \mathlarger{\mathlarger{\prod}}_{\text{ } k \in B_{s_{i,n}, s_{d,m}} }  \sqrt{w_k} \Biggr]\conv \bigl(d(s_{i,n},s_{d,m}) \vert \boldsymbol{\theta}_K \bigr) &\text{ if sites are flow-connected and $ d(s_{i,n},s_{d,m}) > 0$} \\[1.5ex] 
        \end{cases}
\end{equation}

\vspace{-0.15 cm}

\begin{equation}\label{Eq4}
    \conv \bigl(d(s_{i,n},s_{d,m})\vert \boldsymbol{\theta}_K \bigr) = \int\limits_{s_{i,n}}^{\infty} g(x - s_{i,n}\vert \boldsymbol{\theta}_K)g(x - s_{d,m}\vert \boldsymbol{\theta}_K)dx = \int\limits_{0}^{\infty} g(x + d(s_{i,n},s_{d,m})\vert \boldsymbol{\theta}_K)g(x\vert \boldsymbol{\theta}_K)dx
\end{equation}

This specific variant of the SSN model is known as the tails-up model (\citealp{VerHoef2006}, \citealp{Peterson2010}) and relies on the weighting parameters $ w_k $  to maintain stationarity of the model variances. The set $B_{s_{i,n}, s_{d,m}} = \mathcal{D}_{s_{i,n}} \setminus \mathcal{D}_{s_{d,m}} $ represents the stream segments between two spatial locations $ s_{i,n} > s_{d,m}$, including $i$ but excluding $d$. The moving-average function, also referred to as the smoothing kernel (\citealp{Alvarez2010}), is represented by $g(\cdot \vert \boldsymbol{\theta}_K)$. For the tails-up model, a truncated smoothing kernel $ g(x|\boldsymbol{\theta}_K) = 0 $ when $ x < 0 $ is typically chosen. Equations \eqref{Eq1} to \eqref{Eq4} induce a marginal density over observations $ \boldsymbol{y}_{\kappa} $ (see Sect. \ref{Section2.2}), as given by Eq. \eqref{Eq5} below.

\vspace{-0.25 cm}

\begin{equation}\label{Eq5}
    p(\boldsymbol{y}_\kappa \vert \boldsymbol{\beta},\theta_0,\boldsymbol{\theta}_K) =  \mathcal{N}(\boldsymbol{y}_\kappa \vert \boldsymbol{Z\beta},\boldsymbol{\Sigma}) 
\end{equation}

The symbol $ \mathcal{N}(\cdot) $ denotes the Gaussian density function, and the elements of the covariance matrix $ \boldsymbol{\Sigma}$ are computed by evaluating Eq. \eqref{Eq3}. The marginal density in Eq. \eqref{Eq5} depends on the parameters $\boldsymbol{\theta}_K $, $ \theta_0$, and $ \boldsymbol{\beta} $. Note that the covariance matrix can be decomposed as $ \boldsymbol{\Sigma} = \boldsymbol{\Sigma}_K + \theta_0 \boldsymbol{I}_{NN} $, where $ \boldsymbol{\Sigma}_K $ is constructed using the stream network-based kernel function (Eq. \eqref{Eq3}) neglecting the nugget effect.  This key insight will become important in Sect. \ref{Section2.2} when establishing the connection between the tails-up SSN model and an alternative Bayesian probabilistic interpretation known as Gaussian process regression. 

Point estimates for the parameters $ \boldsymbol{\theta}_K $, $\theta_0$, and $ \boldsymbol{\beta} $ can be obtained by maximising the log marginal likelihood $\ln{p(\boldsymbol{y}_\kappa \vert \boldsymbol{\beta},\theta_0,\boldsymbol{\theta}_K)}$ (\citealp{Rasmussen2006}; \citealp{Bishop2009}; \citealp{Garreta2010}). The point estimates for $ \boldsymbol{\theta}_K $, $\theta_0$, and $ \boldsymbol{\beta} $  can then be used in conjunction with the Kriging equations to make stream network-based spatial latent function predictions as well as to quantify the uncertainty in the spatial latent function predictions. The application of the general spatial linear model requires inverting the $ N \times N $ covariance matrix $ \boldsymbol{\Sigma}$ during gradient-based optimisation and prediction. The (numerical) inversion process can become prohibitively slow since the time complexity scales as $ \mathcal{O}(N^3) $ where $ N $ corresponds to the number of stream network-based spatial observations (\citealp{Bishop2009}; \citealp{Santos2022}; \citealp{Basson2023}).

\vspace{-0.45 cm}

\subsection{The Gaussian Process Regression Model}\label{Section2.2}

The tails-up SSN model of \cite{VerHoef2006}, as discussed in Sect. \ref{Section2.1}, can equivalently be expressed as a Bayesian probabilistic model known as Gaussian process regression (GPR). Given an observational data set consisting of pairs $ \{(x_i,y_{\kappa,i})\}_{i = 1}^{N} $, the standard GPR model assumes that each observation $ y_{\kappa,i} $ is a noisy, independent realisation of some unknown latent function $ f_{\kappa,i} = f_\kappa(x_i), $ at the training input location $ x_i $, with additive Gaussian distributed noise with zero-mean and unknown variance $ \sigma_{y_\kappa}^2 $ (\citealp{Rasmussen2006}; \citealp{Bishop2009}). In other words,

\vspace{-0.25 cm}

\begin{equation}\label{Eq8}
    y_{\kappa,i} = f_{\kappa,i} + \epsilon_{\kappa,i}; \quad \epsilon_{\kappa,i} \sim \mathcal{N}(\epsilon_{\kappa,i} \vert 0,\theta_0)
\end{equation}

For a total of $ N $ observations, Eq. \eqref{Eq8} induces a multivariate Gaussian conditional density of the form $ p(\boldsymbol{y}_{\kappa} \vert \boldsymbol{f}_{\kappa},\theta_0) = \mathcal{N}(\boldsymbol{y}_{\kappa} \vert \boldsymbol{f}_{\kappa},\theta_0\boldsymbol{I}_{N N}) $ which can be interpreted as a joint Gaussian likelihood function. Next, a Gaussian Process (GP) prior $ f_\kappa \sim \mathcal{GP}( \boldsymbol{\beta}^T\boldsymbol{z}(x_i),k(x_i,x_j)) $ is specified 
with mean function $ \boldsymbol{\beta}^T\boldsymbol{z}(x_i) $ and covariance $ k(x_i,x_j) $. For the finite set of training input locations $ \boldsymbol{x} $ associated with the latent function vector $ \boldsymbol{f}_\kappa $, the GP prior follows a multivariate Gaussian density with mean vector $ \boldsymbol{Z\beta} $ (see Sect. \ref{Section2.1}) and covariance matrix $ \boldsymbol{\Sigma}_{K} $, constructed using the user-specified kernel function $ k(x_i,x_j) $ on the training input locations $ \boldsymbol{x} $. With the mean vector $ \boldsymbol{Z\beta} $ and covariance matrix $ \boldsymbol{\Sigma}_{K} $, the GP prior takes the following form

\vspace{-0.45 cm}

\begin{equation}\label{Eq12}
    p(\boldsymbol{f}_\kappa \vert \boldsymbol{\beta},{\boldsymbol{\theta}_K}) = \mathcal{N}(\boldsymbol{f}_\kappa \vert \boldsymbol{Z\beta},\boldsymbol{\Sigma}_{K})
\end{equation}

Point estimates for the kernel function parameters $ {\boldsymbol{\theta}_K} $, the noise variance parameter $ \theta_0 $, and the covariate parameter vector $ \boldsymbol{\beta} $ (collectively denoted by $ \boldsymbol{\theta} $) can be obtained by maximising the model log marginal likelihood $\ln p(\boldsymbol{y}_\kappa \vert \boldsymbol{\theta})$, see Eq. \eqref{Eq13} below, using gradient-based optimisation. The point estimate for $ \boldsymbol{\theta} $ can then be used in conjunction with the GPR prediction equations to make latent function predictions (\citealp{Rasmussen2006}; \citealp{Bishop2009}).

\vspace{-0.35 cm}

\begin{equation}\label{Eq13}
   \ln p(\boldsymbol{y}_\kappa \vert \boldsymbol{\theta}) = \ln \int\limits_{\boldsymbol{f}_\kappa} p(\boldsymbol{y}_\kappa,\boldsymbol{f}_\kappa\vert \boldsymbol{\theta})d\boldsymbol{f}_\kappa = \ln \int\limits_{\boldsymbol{f}_\kappa} p(\boldsymbol{y}_\kappa \vert \boldsymbol{f}_\kappa,\theta_0)p(\boldsymbol{f}_\kappa|\boldsymbol{\beta},\boldsymbol{\theta}_K)d\boldsymbol{f}_\kappa = \ln \mathcal{N}(\boldsymbol{y}_\kappa \vert \boldsymbol{Z\beta},\underbrace{\boldsymbol{\Sigma}_{K} + \theta_0\boldsymbol{I}_{NN}}_{\boldsymbol{\Sigma}})
\end{equation}

From Eqs. \eqref{Eq8} to \eqref{Eq13}, the GPR probabilistic perspective provides an alternative interpretation for the work of \cite{VerHoef2006}. More specifically, note how the prior density over latent function values (i.e., Eq. \eqref{Eq12}) depends on the covariance matrix $ \boldsymbol{\Sigma}_K $ which can constructed using the stream network-based kernel function (see Eq. \eqref{Eq3}) without accounting for the nugget effect $ \theta_0 $. Consequently, the moving-average-based construction proposed in \cite{VerHoef2006} 
can be interpreted as a framework for constructing a valid GP prior covariance matrix over spatial stream network-based latent functions. Equivalently, the tails-up SSN model marginal likelihood in Eqs. \eqref{Eq5} and \eqref{Eq13} can be interpreted as averaging over all the spatial stream network-based latent functions as supported under the prior density (Eq. \eqref{Eq12}). Consequently, the tails-up SSN model can be interpreted as a particular GPR model tailored to stream networks (similar arguments extend to the tails-down, the variance component, and Euclidean-based models (\citealp{VerHoef2010};  \citealp{Peterson2010})). Therefore, to extend the tails-up SSN model of \cite{VerHoef2006} to a spatio-temporal multivariate (i.e. Co-Kriging) setting, it is sufficient to specify both the likelihood function (i.e., the data-generating mechanism) and the GP prior density. Without loss of generality, the remainder of this paper assumes that the GP prior is centred on a zero-mean vector such that Eq. \eqref{Eq12} simplifies to $ p(\boldsymbol{f}_\kappa \vert {\boldsymbol{\theta}}_K) = \mathcal{N}(\boldsymbol{f}_\kappa \vert \boldsymbol{0},\boldsymbol{\Sigma}_{K}) $.  

\vspace{-0.35 cm}

\section{Co-Kriging Model Extension and Specification}\label{SectionM}

\begin{figure*}[!ht]
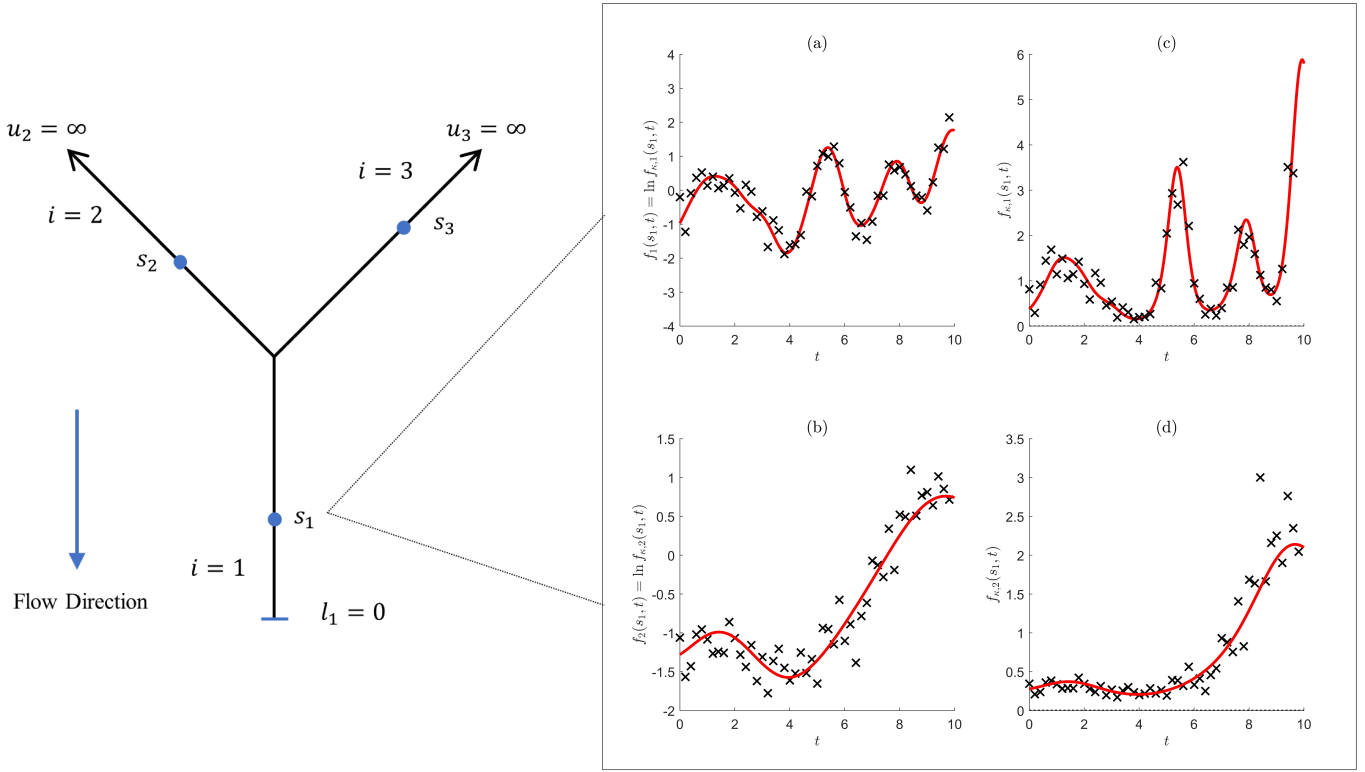

    \hspace{-1.4in}  
    \sbox0{\includegraphics{Fig2.jpeg}}
    \includegraphics[scale = 0.050]{Fig2.jpeg}
    \caption{The hypothetical stream network depicted in Fig. \ref{fig:Figure1} together with two positive underlying latent functions $ f_{\kappa,1}(s_1,t) $ and $ f_{\kappa,2}(s_1,t) $ (labelled (c) and (d), respectively), as well as the associated log-transformed latent functions $ f_1(s_1,t) = \ln f_{\kappa,1}(s_1,t) $ and $ f_2(s_1,t) = \ln f_{\kappa,2}(s_1,t) $ (labelled (a) and (b) respectively), viewed as a function of time $ t $ at sampled spatial location $ s_1 $. In (c) and (d) the black crosses denote the fully observed spatio-temporal observational data set at sampled spatial location $ s_1 $ viewed as a function of time $ t $ per latent function, respectively, whereas in (a) and (b) the black crosses denote the corresponding log-transformed observational data. Figure \ref{fig:Figure2} has been reproduced and adjusted from the work of \cite{VerHoef2006}. \vspace{-0.25 cm}}
    \label{fig:Figure2}
\end{figure*}

Next, the authors introduce the proposed spatio-temporal multi-output (i.e., Co-Kriging) extension for the tails-up SSN model discussed in Sect. \ref{Section2.1} and originally introduced in \cite{VerHoef2006}. As will become evident in Sect. \ref{Section6.1}, the authors will assume separability in both the space and time components of the derived covariance/cross-covariance functions as a means to develop the proposed spatio-temporal Co-Kriging extension. For readers who are already familiar with the spatial-based covariance development of \cite{VerHoef2006}, continue reading from below Eq. \eqref{Eq163} where the authors present the spatial cross-covariance function results for the proposed Co-Kriging model. Figure \ref{fig:Figure2} depicts the hypothetical spatio-temporal stream network-based scenario with the three sampled spatial locations, as introduced in Fig. \ref{fig:Figure1}. The red curves on the rightmost side of Fig. \ref{fig:Figure2} depict two underlying latent functions, denoted by $ f_{\kappa,1}(\cdot) $ and $ f_{\kappa,2}(\cdot) $, respectively, at sampled spatial location $ s_1 $, viewed as a function of time. The latent functions are positive and, for example, correspond to noise-free CEC concentration profiles. The corresponding black crosses denote the fully observed spatio-temporal observational data. The general GPR model and tails-up SSN model do not preserve the positivity of either the data or latent functions because the Gaussian density has support over the entire real line. Instead, the spatio-temporal data can be log-transformed (\citealp{Holcomb2018}) to give $ f_1(s_1,t) = \ln f_{\kappa,1}(s_1,t) $ and $ f_2(s_1,t) = \ln f_{\kappa,2}(s_1,t) $, as depicted by the left set of red curves in Fig. \ref{fig:Figure2}. 

Let $\boldsymbol{y}_{1} \in \mathbb{R}^{N_1 \times 1 }$ and $ \boldsymbol{y}_{2} \in \mathbb{R}^{N_2 \times 1 }$ denote the two (potentially transformed or pre-processed) observational data sets that are associated with the two underlying latent functions $\boldsymbol{f}_{1} \in \mathbb{R}^{N_1 \times 1 }$ and $ \boldsymbol{f}_{2} \in \mathbb{R}^{N_2 \times 1 }$, respectively.  Based on the standard GPR probabilistic model
, the joint density can be written as 

\begin{equation}\label{Eq2001}
   p \left (\begin{bmatrix} \boldsymbol{y}_{1} \\ \boldsymbol{y}_{2} \end{bmatrix},      
   \begin{bmatrix} \boldsymbol{f}_{1} \\ \boldsymbol{f}_{2} \end{bmatrix} \right ) = p \left ( \begin{bmatrix} \boldsymbol{y}_{1} \\ \boldsymbol{y}_{2} \end{bmatrix} \biggr \vert \begin{bmatrix} \boldsymbol{f}_{1} \\ \boldsymbol{f}_{2} \end{bmatrix} \right )p \left ( \begin{bmatrix} \boldsymbol{f}_{1} \\ \boldsymbol{f}_{2} \end{bmatrix}\right ) = p(\boldsymbol{y},\boldsymbol{f}) = p(\boldsymbol{y} \vert \boldsymbol{f})p(\boldsymbol{f})
\end{equation}

Note that in Eq. \eqref{Eq2001}, the dependence on the remaining model parameters has been omitted. Furthermore, observe from Eq. \eqref{Eq2001} that the observations and underlying latent functions were stacked into $ (N_1 + N_2) \times 1 $ vectors $\boldsymbol{y} = [\boldsymbol{y}_{1}^T,\boldsymbol{y}_{2}^T]^T$ and $ \boldsymbol{f} = [\boldsymbol{f}_{1}^T,\boldsymbol{f}_{2}^T]^T $, respectively.
The conditional density $ p(\boldsymbol{y} \vert \boldsymbol{f}) $, when viewed as a function of $\boldsymbol{f} $, is called the joint Co-Kriging likelihood function whereas $ p(\boldsymbol{f}) $ denotes the joint Co-Kriging prior density.  The rest of this section outlines the proposed spatio-temporal, multi-output extension for the tails-up SSN prior,  $ p(\boldsymbol{f}) $, followed by the Co-Kriging likelihood function $p(\boldsymbol{y} \vert \boldsymbol{f})$.

\vspace{-0.45 cm}

\subsection{Specifying The Joint Co-Kriging Prior Density} \label{Section6.1}

Following the framework of the zero-mean GPR prior with two latent function outputs, the prior density for $\boldsymbol{f}$ can be written as

\vspace{-0.35 cm}

\begin{equation}\label{Eq143}
    p(\boldsymbol{f} \vert \boldsymbol{h,w}) = \mathcal{N} \Biggl( \begin{bmatrix} \boldsymbol{f}_{1} \\ \boldsymbol{f}_{2} \end{bmatrix} \Biggl\vert \begin{bmatrix} \boldsymbol{0} \\ \boldsymbol{0} \end{bmatrix}; \begin{bmatrix} \boldsymbol{K}_{N_1N_1} & \boldsymbol{K}_{N_1N_2} \\ \boldsymbol{K}_{N_2N_1} & \boldsymbol{K}_{N_2N_2} \end{bmatrix} \Biggr)
\end{equation}

Note that the vectors $\boldsymbol{h}$ and $\boldsymbol{w}$ represent the uncertain hydrological distance and flow weighting parameter inputs, respectively. For the tails-up SSN model, as applied in the current Geostatistics literature, the uncertain hydrological distance and flow weighting parameter inputs are fixed to the measured/estimated values. Consequently, the uncertain inputs $\boldsymbol{h}$ and $\boldsymbol{w}$ can, theoretically, be dropped from the conditioning set, i.e., $ p(\boldsymbol{f} \vert \boldsymbol{h,w}) $ reverts to $ p(\boldsymbol{f}) $, as outlined in Eq. \eqref{Eq2001}. However, in Eq. \eqref{Eq143}, to authors retain the conditioning on the inputs $\boldsymbol{h}$ and $\boldsymbol{w}$ to emphasise that these variables should be treated as uncertain inputs. Incorporation of the input uncertainty is deferred to Section \ref{Section3}. Next, observe that specifying the prior density corresponds to constructing the covariance and cross-covariance sub-matrices associated with the underlying latent functions in Eq. \eqref{Eq143}. Specifically, in this paper, separability in the space and time components of the covariance matrices is assumed. The mean and covariance functions associated with Eq. \eqref{Eq143} are constructed via integration of moving-average functions (also referred to as smoothing kernels) against white noise Gaussian processes. For a more general overview of this type of construction, see \cite{VerHoef1998} and \cite{Gelfand2021}. A related derivation for the spatial component can be found in the Appendix of \cite{Money2009b}.
 Spatial correlation of a random variable associated with a latent function (e.g., $ \boldsymbol{f}_{1} $) depends on whether the spatial locations are flow-connected. Two locations are flow-connected if one location is downstream of the other, in which case it is assumed that a correlation exists as any event at the upstream location, such as the introduction of a waterborne compound, may affect the downstream location. Referring to Fig. \ref{fig:Figure2} as an example, sites $ s_1 $ and $ s_2 $ are flow connected but $ s_2 $ and $ s_3 $ are not. 

\begin{figure}[!ht]
  \centering
  \hspace{-4.2 cm}
  \begin{subfigure}[t]{0.48\textwidth}
    \centering    \captionsetup{justification=centering,singlelinecheck=false,margin={0.50cm,0cm}}
    \includegraphics[scale = 0.49]{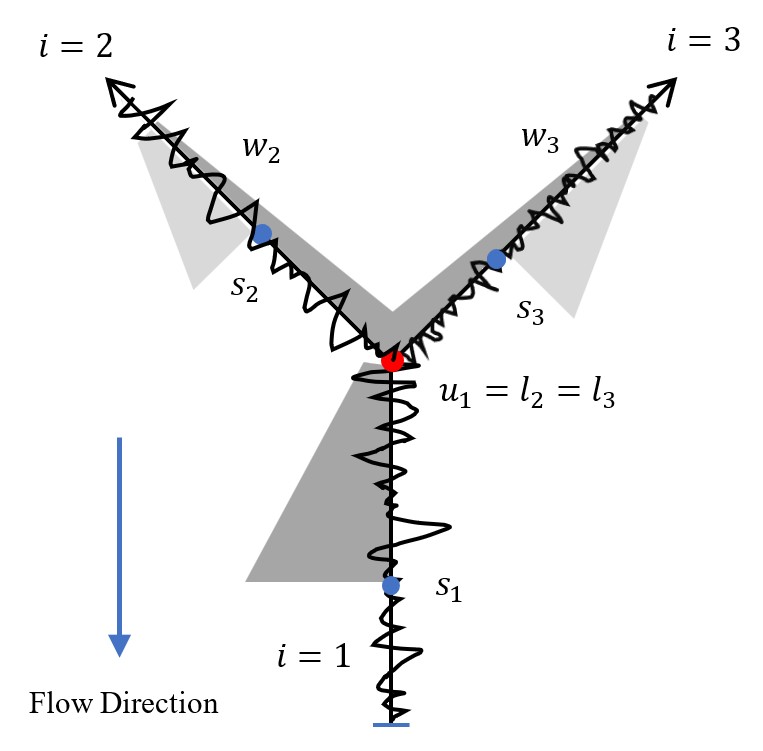}
    \caption{}
    \label{fig:Figure20-a}
  \end{subfigure}
  \hspace{-1.10 cm}
  \begin{subfigure}[t]{0.48\textwidth}
  \captionsetup{justification=centering,singlelinecheck=false,margin={5cm,0cm}}
    \centering
    \includegraphics[scale = 0.51]{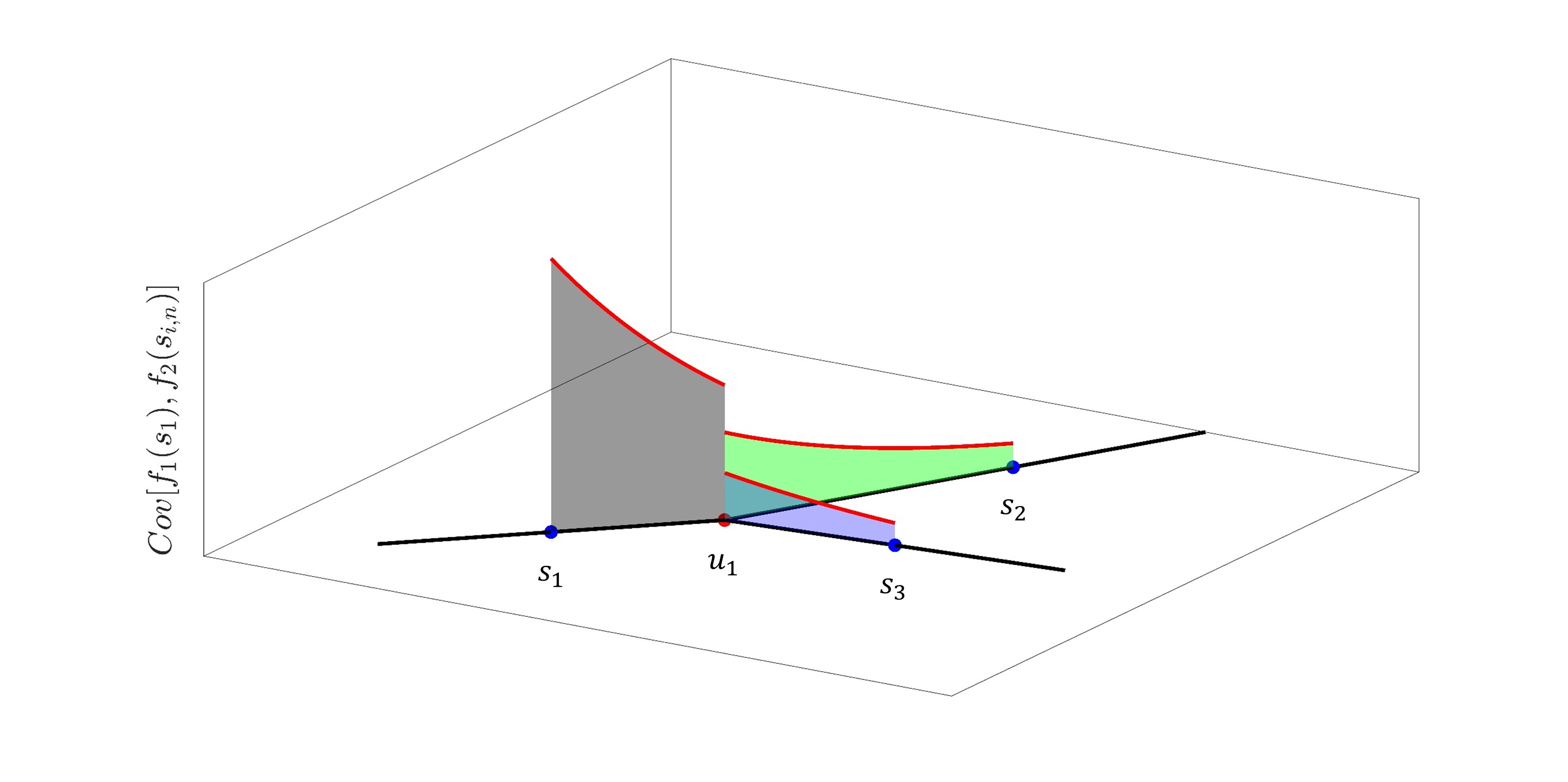}
    \caption{}
    \label{fig:Figure20-b}
  \end{subfigure}
  \caption{Panel (a) visually illustrates the methodology underpinning the proposed latent function random variable construction procedure. The depicted hypothetical network consists of three stream segments/branches labelled $ i = 1,2, $ and $ 3 $, respectively, together with three sampled spatial locations denoted by $ s_1, s_2 $, and $ s_3 $. The shaded grey areas denote the user-specified spatial smoothing kernel  $ g_1(\cdot) $ whereas the wavy lines represent the spatial white noise Gaussian process $ W(\cdot) $. The stream network junction point is denoted by $ u_1 $.  Panel (a) has been reproduced and adjusted from the work of \cite{VerHoef2006}. Panel (b) provides a perspective view of the hypothetical stream network. The solid red line depicts the cross-covariance $ \mathbb{C}\text{ov}[f_{1}(s_1),f_{2}(s_{i,n})] $ between underlying latent function $ f_{1}(s_1) $ at spatial location $ s_1 $ and underlying latent function $ f_{2}(s_{i,n}) $ at various arbitrary upstream locations $ s_{i,n} $. Shading in grey, blue, and green is purely illustrative to guide the reader's eye and should not be taken as conveying additional information. To construct the solid red line, all moving-average function parametric values were set to unity. \vspace{-0.05 cm}}
  \label{fig:Figure20}
\end{figure}

The grey shaded areas depicted in Fig. \ref{fig:Figure20-a} correspond to the moving-average function, also referred to as a smoothing kernel (\citealp{Alvarez2010}), that will play an important role in constructing random variables with this behaviour. In general, the smoothing kernel can go in both directions, in other words, upstream and downstream, relative to the fluid flow direction. However, restricting the smoothing kernel to move upstream is particularly useful as it allows for the development of a statistical framework that can be used to model the physical phenomena of flow-connected behaviour between sites. More specifically, by restricting the smoothing kernel such that $ g(x|\boldsymbol{\theta}_K) = 0 $ when $ x < 0 $, it is possible to develop a statistical model that correlates spatial locations $ s_1 $ and $ s_2 $ while, simultaneously, imposing no correlation between locations $ s_2 $ and $ s_3 $. This particular model, which explicitly relies on the flow-connected behaviour between spatial locations, is known as the tails-up model (see \citealp{VerHoef2006}; \citealp{VerHoef2010}, and Sect \ref{Section2.1} for more details) and is particularly useful for modelling the passive downstream movement of waterborne compounds in a stream network. In the following derivations, a truncated smoothing kernel $ g(x) = 0 \ \forall \ x<0$ is assumed in order to enforce the tails-up model constraints.

The standard procedure of constructing random variables as the integration of a smoothing kernel over a white noise Gaussian process (\citealp{Yaglom1987}) is typically defined for a line on $ \mathbb{R}^1 $. However, for stream networks, such as the network depicted in Figs. \ref{fig:Figure2} and \ref{fig:Figure20-a}, the line must split at the junction point $ u_1 $ to form two stream segments and the standard procedure does not apply. The problem is resolved by splitting the smoothing kernel into two contributing parts at the junction point $ u_1 $, as shown by the dark grey shaded areas in Fig. \ref{fig:Figure20-a}.  The latent function can then be constructed by the piece-wise summation over all contributing segments that contain the smoothing kernel. \cite{VerHoef2006} proposed constructing the spatial process at sampled spatial location $ s_1 $ as the integration of a smoothing kernel against a white noise Gaussian process such that

\vspace{-0.15 cm}

\begin{equation}\label{Eq144}
\begin{gathered}
    f_{1}(s_1) = \int\limits_{s_1}^{u_1} g_1(x_1 - s_1)W(x_1)dx_1 + \sqrt{w_2} \int\limits_{u_1}^{\infty} g_1(x_2 - s_1)W(x_2)dx_2 + \text{ } \sqrt{w_3} \int\limits_{u_1}^{\infty} g_1(x_3 - s_1)W(x_3)dx_3
    \end{gathered}
\end{equation}

The symbol $ g_1(x_i) $, where the dependence on any model parameters has been omitted for notational convenience, denotes the user-specified spatial smoothing kernel associated with latent function 1, whereas $ W(x_i) $ denotes the spatial white noise process depicted by the wavy lines around each of the stream segments in Fig. \ref{fig:Figure20-a}. 

If stationarity is desired, care must be taken when splitting the smoothing kernel $ g_1(\cdot) $ at the junction point $ u_1 $. 
\cite{VerHoef2006} showed that this problem can be circumvented by incorporating weighting parameters corresponding to $ \sqrt{w_k} $ for each stream segment $k$ after a split in the smoothing kernel occurs. Incorporating the appropriate weighting parameters then gives rise to the latent function random variable $ f_{1}(s_1) $, at spatial location $ s_1 $, construction procedure outlined in Eq. \eqref{Eq144}.   To maintain stationarity of the variances due to the necessary weighting scheme, the condition $ w_2 + w_3 = 1 $ must be satisfied. In other words, when a split occurs at a junction point, the weighting parameters at that junction point must sum to 1. As another example, consider constructing the latent function random variable $ f_{1}(s_2) $ at spatial location $ s_2 $. Observe from Fig. \ref{fig:Figure20-a} that there are no junction points upstream of the stream segment on which $ s_2 $ is located, in other words, no weighting parameters are required since the smoothing kernel does not split further upstream. 
More generally, \cite{VerHoef2006} showed that the latent function $ f_{1} $, at $ s_{i,n} $, can be constructed as follows

\vspace{-0.6 cm}

\begin{equation}\label{Eq146}
\begin{split}
    f_{1}(s_{i,n}) &= \int\limits_{s_{i,n}}^{u_i} g_1(x_i - s_{i,n})W(x_i)dx_i \text{ } + \mathlarger{\mathlarger{\sum}}_{j \in U_{s_{i,n}}\setminus i} \mathlarger{\Biggl(} \Biggl[ \mathlarger{\mathlarger{\prod}}_{\text{ } k \in B_{s_{i,n},[j]}}  \sqrt{w_k} \Biggr] \int\limits_{l_j}^{u_j} g_1(x_j - s_{i,n})W(x_j)dx_j \mathlarger{\Biggr)} \\ 
    &= \mathlarger{\mathlarger{\sum}}_{j \in U_{s_{i,n}}} \mathlarger{\Biggl(} \Biggl[ \mathlarger{\mathlarger{\prod}}_{\text{ } k \in B_{s_{i,n},[j]}}  \sqrt{w_k} \Biggr] \int\limits_{l_j}^{u_j} g_1(x_j - s_{i,n})W(x_j)dx_j \mathlarger{\Biggr)}
\end{split}
\end{equation}

Building on the notation introduced in Sect. \ref{Section2.1}, $ U_{s_{i,n}} $ is the set of stream segments upstream of spatial location $ s_{i,n} $, {including} stream segment $ i $. The set $ B_{s_{i,n},[j]} $ includes stream segments between spatial location $ s_{i,n} $ and upstream segment $ j $, which includes $ j $ but {excludes} $ i $. For example, $ U_{s_1} = \{1,2,3\}$, $ B_{s_1,[2]}  = \{2\} $ and $ B_{s_1,[3]}  = \{3\}$. The first integral can be gathered into the summation by using the convention that the product over an empty set is equal to one, and noting that the lower limit of integration may be extended from $s_{i,n}$ to $l_i$ given that a truncated kernel $ g(x) = 0 \ \forall \ x < 0$ is used.
Since each $ f_{1}(s_{i,n}) $ is constructed via the integration of a smoothing kernel over a white noise Gaussian process, $ f_{1} $ must also be a Gaussian process (see \citealp{Raissi2017} and \citealp{Agrell2019}) and is completely defined by its mean and covariance function. A zero-mean white noise Gaussian process, with $ \mathbb{E}_{p(W)}[W(\cdot)] = 0 $, is typically used. If $f_1(s_{d,m})$ at a downstream location $s_{d,m}$ is also constructed using the approach given by Eq. \eqref{Eq146}, then the covariance $\mathbb{C}$ov$[f_{1}(s_{i,n}),f_{1}(s_{d,m})]$ between $ f_{1} $ at $s_{i,n}$ and $s_{d,m}$ can be computed as

\vspace{-0.15 cm}

\begin{equation}\label{Eq194}
\begin{gathered}
   \mathlarger{\mathlarger{\sum}}_{j \in U_{s_{i,n}}} \text{ } \mathlarger{\mathlarger{\sum}}_{j^{\prime} \in U_{s_{d,m}}} \mathlarger{\Biggl(} A_{j,j^{\prime}} \int\limits_{l_j}^{u_j} g_1(x_j - s_{i,n}) \Biggl[ \text{ } \int\limits_{l_{j^{\prime}}}^{u_{j^{\prime}}} g_1(x^{\prime}_{j^{\prime}} - s_{d,m}) \mathbb{E}_{p(W)}[W(x_j)W(x^{\prime}_{j^{\prime}})]dx^{\prime}_{j^{\prime}} \Biggr]dx_j \mathlarger{\Biggr)}
    \end{gathered}
\end{equation}
\begin{equation}\label{Eq157}
\begin{gathered}
    A_{j,j^{\prime}} = \Biggl[ \mathlarger{\mathlarger{\prod}}_{\text{ } k \in B_{s_{i,n},[j]}}  \sqrt{w_k} \Biggr] \Biggl[ \mathlarger{\mathlarger{\prod}}_{\text{ } k^{\prime} \in B_{s_{d,m},[j^{\prime}]}}  \sqrt{w_{k^{\prime}}} \Biggr]
    \end{gathered}
\end{equation}

The covariance of the zero-mean white noise Gaussian process $ W(\cdot) $ corresponds to 

\vspace{-0.15 cm}

\begin{equation}\label{Eq158}
\begin{gathered}
    \mathbb{C}\text{ov}[ W(x_j),W(x^{\prime}_{j^{\prime}})] = \mathbb{E}_{p(W)}[W(x_j)W(x^{\prime}_{j^{\prime}})] = \sigma_{p}^{2}\delta_{j,j^{\prime}}\delta(x^{\prime}_{j^{\prime}} - x_j)
\end{gathered}
\end{equation}

\noindent The symbols $ \delta_{j,j^{\prime}} $ and $ \delta(\cdot) $ denote the Kronecker delta and the Dirac delta, respectively. The former ensures zero covariance when $j$ and $j'$ refer to separate stream segments. Setting (without loss of generality) $ \sigma_{p}^{2} = 1$, using Eq. \eqref{Eq158}, and noting that when $ s_{d,m} $ is located downstream of $ s_{i,n} $ and the locations are flow-connected, the set $ U_{s_{i,n}} $ must be a subset of $ U_{s_{d,m}} $ and $ U_{s_{i,n}} \cap \text{ } U_{s_{d,m}} = U_{s_{i,n}} $, such that Eq. \eqref{Eq194} can be rewritten for flow-connected sites to obtain

\vspace{-0.35 cm}

\begin{equation}\label{Eq173}
\begin{gathered}
    \mathbb{C}\text{ov}[f_{1}(s_{i,n}),f_{1}(s_{d,m})] = 
    \mathlarger{\mathlarger{\sum}}_{j \in U_{s_{i,n}} } \mathlarger{\Biggl(} \sqrt{\Biggl[ \mathlarger{\mathlarger{\prod}}_{\text{ } k \in B_{s_{i,n},[j]}} w_k \Biggr] \Biggl[ \mathlarger{\mathlarger{\prod}}_{\text{ } k^{\prime} \in B_{s_{d,m},[j]}}  w_{k^{\prime}} \Biggr]} \text{ } \int\limits_{l_j}^{u_j} g_1(x_j - s_{i,n})g_1(x_j - s_{d,m})  dx_j \mathlarger{\Biggr)} 
    \end{gathered}
\end{equation}

When spatial locations $ s_{d,m} $ and $ s_{i,n} $ are not flow-connected, $ U_{s_{i,n}} \cap \text{ } U_{s_{d,m}} = \{\emptyset\} $ and Eq. \eqref{Eq194} reduces to 0. For flow-connected sites, the weighting parameters must sum to $ 1 $ at junction points (\citealp{VerHoef2006}; \citealp{Money2009b}) simplifying Eq. \eqref{Eq173} to

\vspace{-0.25 cm}

\begin{equation}\label{Eq163}
\begin{gathered}
    \mathbb{C}\text{ov}[f_{1}(s_{i,n}),f_{1}(s_{d,m})] = \Biggl[ \mathlarger{\mathlarger{\prod}}_{\text{ } k \in B_{s_{i,n},s_{d,m}}}  \sqrt{w_k} \Biggr] \int\limits_{s_{i,n}}^{\infty} g_1(x_i - s_{i,n})g_1(x_i - s_{d,m})  dx_i
    \end{gathered}
\end{equation}

Consequently, Eq. \eqref{Eq163} forms the foundation for a statistically independent application of the tails-up SSN model of \cite{VerHoef2006} and can used to construct the spatial component of the covariance sub-matrices $ \boldsymbol{K}_{N_{1}N_{1}} $ and $ \boldsymbol{K}_{N_{2}N_{2}} $ where $ f_1(\cdot) $ and $ g_1(\cdot) $ can be substituted with $ f_2(\cdot) $ and $ g_2(\cdot) $, respectively. A dependence structure can naturally be introduced for the multi-output case using the same moving-average-based construction procedure proposed by \cite{VerHoef2006}. However, care should be taken when considering which spatial location belongs to which underlying latent function. For example,  for $ s_{d,m} $ located downstream of $ s_{i,n} $, the covariance $ \mathbb{C}\text{ov}[f_{1}(s_{i,n}),f_{2}(s_{d,m})] $ can be computed

\vspace{-0.35 cm}

\begin{equation}\label{Eq168}
\begin{gathered}
    \mathbb{C}\text{ov}[f_{1}(s_{i,n}),f_{2}(s_{d,m})] = \Biggl[ \mathlarger{\mathlarger{\prod}}_{\text{ } k \in B_{s_{i,n},s_{d,m}}}  \sqrt{w_k} \Biggr] \int\limits_{s_{i,n}}^{\infty} g_1(x - s_{i,n})g_2(x - s_{d,m})  dx
    \end{gathered}
\end{equation}

\noindent When the spatial locations swap across latent functions, in other words, $ \mathbb{C}\text{ov}[f_{1}(s_{d,m}),f_{2}(s_{i,n})] $, then

\vspace{-0.25 cm}

\begin{equation}\label{Eq169}
\begin{gathered}
    \mathbb{C}\text{ov}[f_{1}(s_{d,m}),f_{2}(s_{i,n})] = \Biggl[ \mathlarger{\mathlarger{\prod}}_{\text{ } k \in B_{s_{i,n},s_{d,m}}}  \sqrt{w_k} \Biggr] \int\limits_{s_{i,n}}^{\infty} g_1(x - s_{d,m})g_2(x - s_{i,n})  dx
    \end{gathered}
\end{equation}

The results in Eqs. \eqref{Eq168} and \eqref{Eq169} provide a mechanism for constructing a valid cross-covariance between the two underlying latent functions $ f_{1} $ and $ f_{2} $ at the arbitrary spatial locations $ s_{d,m} $ and $ s_{i,n} $ provided that $ s_{d,m} $ is located downstream of spatial location $ s_{i,n} $ and that the sites are flow-connected. Consequently, Eqs. \eqref{Eq168} and \eqref{Eq169} can be used to construct the spatial component of the cross-covariance sub-matrix $ \boldsymbol{K}_{N_1N_2} $, and by symmetry also $ \boldsymbol{K}_{N_2N_1} = \boldsymbol{K}_{N_1N_2}^T $. For illustration purposes, Fig. \ref{fig:Figure20-b} visually depicts the cross-covariance (red solid line) between underlying latent function $ f_{1}(s_1) $, at sampled spatial location $ s_1 $, and underlying latent function $ f_{2}(s_{i,n}) $, at various arbitrary upstream locations $ s_{i,n} $, with all the moving-average function parametric values set to unity - see Eq. SI-70 in the Supplementary Information for more details. Note that at the junction point $ u_1 $ the covariance function $ \mathbb{C}\text{ov}[f_{1}(s_1),f_{2}(s_{i,n})]  $ is weighted proportional to the weighting parameter $ \sqrt{w_k} $ to account for the relative flow contribution from each river branch segment.

\noindent In general, given a stream network with $ K_f $ underlying latent functions, the spatial component of the cross-covariance between latent function $a$ at downstream location $s_{d,m}$ and latent function $b$ at upstream location $s_{i,n}$, can be compactly written to obtain

\vspace{-0.6 cm}

\begin{equation}\label{Eq177}
\begin{gathered}
    C_{{f_{a},f_{{b}}}}^{(s)}(s_{d,m},s_{i,n} \vert \boldsymbol{\theta}_a^{(s)}, \boldsymbol{\theta}_b^{(s)} ) =  
    \begin{cases} 
    0 
    &\text{ if sites are not flow-connected } \\[1.5ex] 
    \Biggl[ \mathlarger{\mathlarger{\prod}}_{\text{ } k \in B_{s_{i,n},s_{d,m}}}  \sqrt{w_k} \Biggr] \conv_{{f_{a},f_{{b}}}}^{(s)}(d(s_{d,m},s_{i,n}) \vert \boldsymbol{\theta}_a^{(s)}, \boldsymbol{\theta}_b^{(s)})
    &\text{ if sites are flow-connected  }
    \end{cases}
    \end{gathered}
\end{equation}

\begin{equation}\label{Eq178}
\begin{split}
    \conv_{{f_{a},f_{{b}}}}^{(s)}(d(s_{d,m},s_{i,n}) \vert \boldsymbol{\theta}_a^{(s)}, \boldsymbol{\theta}_b^{(s)}) = \int\limits_{0}^{\infty} g_a(x + d(s_{d,m},s_{i,n})  \vert \boldsymbol{\theta}_a^{(s)})g_b(x \vert \boldsymbol{\theta}_b^{(s)})  dx \text{ } \text{ } \text{ } \forall \text{ } \text{ } a,b = 1,\cdots,K_f
    \end{split}
\end{equation}

The notation $ C_{f_a,f_b}^{(s)} $ was introduced to denote the spatial component of the covariance, indicated by superscript $(s)$, in anticipation of the inclusion of a temporal component to the covariance. The most straightforward approach to introduce temporal attributes into the Co-Kriging prior would be to assume that the spatial $(s)$ and temporal $(t)$ components of the covariance are separable (\citealp{Tang2020}, \citealp{Christakos2017}). Consider a latent function $f_a$ at downstream location $s_{d,m}$ and time $t_p$, and another latent function $f_b$ at upstream location $s_{i,n}$ and time $t_q$. Given the assumed separable spatio-temporal covariance structure, the covariance between the two latent functions (at the given spatial locations and time points) is given by

\vspace{-0.45 cm}

\begin{equation}\label{Eq184}
\begin{split}
   \mathbb{C} \text{ov}[ f_a(s_{d,m}, t_p), f_b(s_{i,n}, t_q)  \vert 
   \boldsymbol{\theta}_a, \boldsymbol{\theta}_b ] = 
   C_{{f_{a},f_{{b}}}}^{(s)}(s_{d,m},s_{i,n} \vert \boldsymbol{\theta}_a^{(s)}, \boldsymbol{\theta}_b^{(s)} ) \times 
   C_{f_a,f_b}^{(t)}(t_p, t_q \vert\boldsymbol{\theta}_a^{(t)}, \boldsymbol{\theta}_b^{(t)} )
    \end{split}
\end{equation}

\noindent Superscript $(t)$ denotes the temporal covariance component which depends on the temporal instances $ t_p $ and $ t_q $ as well as the temporal covariance parameters $ \boldsymbol{\theta}_a^{(t)} $ and $ \boldsymbol{\theta}_b^{(t)} $. Parameter vector $ \boldsymbol{\theta}_a = \{ \boldsymbol{\theta}_a^{(s)}, \boldsymbol{\theta}_a^{(t)} \} $ is used to collectively denote the spatial and temporal covariance parameters for latent function $f_a$. A similar argument holds for $ \boldsymbol{\theta}_b $.

The practical benefit of assuming a separable spatio-temporal covariance stems from the fact that the spatial and temporal components can be specified independently. The work of \cite{Alvarez2010}, \cite{Alvarez2011} and \cite{Alvarez2012} is followed to construct the temporal component contribution. More specifically, similar to \cite{VerHoef2006}, the latent function random variable can be constructed as the integral of a smoothing kernel over a base latent process, however, now using the following convolution integral
\vspace{-0.40 cm}

\begin{equation}\label{Eq186}
\begin{split}
    {f}_{a}(t_p) = \int\limits_{-\infty}^{\infty} G_{a}(t_p - z \vert \boldsymbol{\theta}_a^{(t)})W_{t}(z)dz
    \end{split}
\end{equation}

Similar to the spatial case, it is also assumed that the base process is a zero-mean temporal white noise Gaussian process with covariance $ \delta(z^{\prime} - z) $. Following a similar derivation procedure to that outlined for the spatial covariance component contribution, the temporal covariance between two arbitrary latent functions $ {f}_{a}(t_p) $ and $ {f}_b(t_q) $ at temporal instances $ t_p $ and $ t_q $ can be computed as
\vspace{-0.50 cm}

\begin{equation}\label{Eq189}
\begin{split}
    C_{f_a,f_b}^{(t)}(t_p, t_q \vert\boldsymbol{\theta}_a^{(t)}, \boldsymbol{\theta}_b^{(t)} ) 
    &= \int\limits_{-\infty}^{\infty} G_{a}(z + (t_p - t_q) \vert \boldsymbol{\theta}_a^{(t)})G_{b}(z\vert \boldsymbol{\theta}_b^{(t)}) dz
    \end{split}
\end{equation}

Hence, the final separable spatio-temporal covariance for the Co-Kriging prior corresponds to

\vspace{-0.45 cm}

\begin{equation}\label{Eq192}
\begin{gathered}
    \mathbb{C} \text{ov}[ f_a(s_{d,m}, t_p), f_b(s_{i,n}, t_q)  \vert 
   \boldsymbol{\theta}_a, \boldsymbol{\theta}_b ] = 
   \begin{cases} 
    0 
    &\text{ if sites are not flow-connected } \\[1.5ex] 
    \Biggl[ \mathlarger{\mathlarger{\prod}}_{\text{ } k \in B_{s_{i,n},s_{d,m}}}  
    \sqrt w_k \Biggr] \conv_{f_a, f_b} ^{(s\times t)} (d(s_{d,m},s_{i,n}), t_p - t_q \mathlarger{\mathlarger{\vert}} \boldsymbol{\theta}_a, \boldsymbol{\theta}_b ) 
    &\text{ if sites are flow-connected  }
    \end{cases}
    \end{gathered}
\end{equation}

\vspace{-0.65 cm}

\begin{equation}\label{Eq193}
\begin{split}
    \conv_{f_a, f_b}^{(s\times t)} (d(s_{d,m},s_{i,n}) ,t_p - t_q \mathlarger{\mathlarger{\vert}} \boldsymbol{\theta}_a, \boldsymbol{\theta}_b )   
    &= \Biggl[ \text{ } \int\limits_{0}^{\infty} g_a(x + d(s_{d,m},s_{i,n})  \vert \boldsymbol{\theta}_a^{(s)})g_b(x \vert \boldsymbol{\theta}_b^{(s)})  dx \Biggr] \\ &\times 
    \Biggl[ \text{ } \int\limits_{-\infty}^{\infty} G_{a}(z + (t_p - t_q) \vert \boldsymbol{\theta}_a^{(t)})G_{b}(z\vert \boldsymbol{\theta}_b^{(t)}) dz  \Biggr]  \\\text{ } \text{ } \text{ }
    \end{split}
\end{equation}

\vspace{-0.75 cm}

Observe from Eqs. \eqref{Eq192} and \eqref{Eq193} that another practical benefit that implicitly arises from assuming a separable spatio-temporal covariance (see Eq. \eqref{Eq184}) includes the fact that the assumed kernel function structure preserves the flow-connected behaviour between spatial locations, as originally introduced in \cite{VerHoef2006}, over time. For example, from Eq. \eqref{Eq192}, if spatial locations $ s_{i,n} $ and $ s_{d,m} $ are not flow-connected across space, the flow unconnected behaviour is also preserved across temporal instances. If spatial locations  $ s_{i,n} $ and $ s_{d,m} $ are flow-connected across space, the locations are also correlated across time. 

\vspace{-0.45 cm}

 \subsection{Specifying The Joint Co-Kriging Likelihood Function}\label{Specifying The Joint Co-Kriging Likelihood}

Next, the likelihood function for the observed spatio-temporal data, conditioned on the latent functions, can be constructed. 
Inference is performed using the log-transformed observational data, but predictions can be computed on the original scale by reversing the bijective log transformation to draw conclusions about the positive underlying latent functions. Following the GPR generative structure outlined in Sect. \ref{Section2.2}, each observation is a noisy, independent realisation of some unknown latent function with additive Gaussian distributed noise with zero-mean and unknown variance. For example, observation $ y_{a}(s_{i,n},t_q) = \ln y_{\kappa,a}(s_{i,n},t_q) $, at time $ t_q $ and spatial location $ s_{i,n} $, can be described by
\begin{equation}\label{Eq84}
    \ln y_{\kappa,a}(s_{i,n},t_q) = \ln f_{\kappa,a}(s_{i,n},t_q) + \epsilon_{\kappa,a}(s_{i,n},t_q); \quad \epsilon_{\kappa,a}(s_{i,n},t_q) \sim \mathcal{N}(\epsilon_{\kappa,a}(s_{i,n},t_q) \vert 0,\sigma_{\ln{y_{\kappa,a}}}^2)
\end{equation} 
Note that $ f_{a}(s_{i,n},t_q) = \ln f_{\kappa,a}(s_{i,n},t_q) $ is the $ a^{\text{th}} $ (transformed) latent function at the spatio-temporal training input location $ (s_{i,n},t_q) $, and $ \sigma_{\ln{y_{\kappa,a}}}^2 $ is the unknown variance. The conditional density for a (transformed) observation at sampled spatial location $s_{i,n}$ and temporal instance $t_q$, therefore, corresponds to

\vspace{-0.25 cm}

\begin{equation}\label{Eq88}
    p \bigl(y_a(s_{i,n},t_q) \vert f_a(s_{i,n},t_q) \bigr) =  \mathcal{N} 
    \bigl(y_a(s_{i,n},t_q) \vert f_a(s_{i,n},t_q),\sigma_{a}^2 \bigr)
\end{equation}

\noindent Note that $ \sigma_a = \sigma_{\ln{y_{\kappa,a}}} $. 
Making the usual assumption of independence across the sampled spatial and temporal locations conditioned on the underlying latent function, the likelihood function associated with the $ a^\text{th} $ underlying latent function is given by Eq. \eqref{Eq91}.

\vspace{-0.25 cm}

\begin{equation}\label{Eq91}
    p \bigl(\boldsymbol{y}_a \vert \boldsymbol{f}_a \bigr) 
    = \mathlarger{\prod_{s \in \sample_a^{(s)}}} 
    \Bigl[ \mathlarger{\prod_{t \in \sample_{a,s}^{(t)}}} \text{ } 
    p \bigl(y_a(s, t) \vert f_a(s, t) \bigr) \Bigr] 
    = \mathlarger{\prod_{s \in \sample_a^{(s)}}} 
    \Bigl[ \mathlarger{\prod_{t \in \sample_{a,s}^{(t)}}} \text{ } 
    \mathcal{N} \bigl(y_a(s, t) \vert f_a(s,t),\sigma_a^2 \bigr) \Bigr]
\end{equation}

 \noindent In Eq. \eqref{Eq91}, $s \in \sample_a^{(s)} $ is the set of sampled spatial locations associated with the observational data $\boldsymbol{y}_a$. 
Similarly, $ t \in \sample_{a,s}^{(t)} $ denotes the set of sampled temporal instances associated with $ s $ and $\boldsymbol{y}_a$. The joint Co-Kriging likelihood function, assuming independence between latent function observations, for a fully observed spatio-temporal data set with $K_f$ underlying latent functions, then becomes

\vspace{-0.25 cm}

\begin{equation}\label{Eq93}
    p(\boldsymbol{y} \vert \boldsymbol{f}) = \underbrace{ \mathlarger{\prod_{a = 1}^{K_f}} \text{ } \Biggl[ \underbrace{ \mathlarger{\prod_{s\in \sample_a^{(s)}}} \Bigl[ \underbrace{\mathlarger{\prod_{t \in \sample_{a,s}^{(t)}}}  \mathcal{N} \bigl(y_a(s,t) \vert f_a(s,t),\sigma_a^2 \bigr) }_{\text{Temporal Independence}} \Bigr] }_{\text{Spatial Independence}} \Biggr] }_{\text{Independence Between Latent Function Observations}}
\end{equation} 

\vspace{-0.50 cm}

\subsection{Censored Observational Data}\label{Section5.2}

\vspace{-0.15 cm}

Recall that despite the advent of in-situ sensing, a practical problem that practitioners can encounter is data censoring. Data censoring occurs when practitioners have access to a partially observed measurement. For example,  suppose a surface water sample is analysed to determine the concentration of a CEC. If the CEC concentration value falls outside the sensitivity range of the measurement device/analysis procedure, either the upper or lower sensitivity range value is reported/returned resulting in a censored spatio-temporal observation. 
In this work, the authors are primarily concerned with the limits of detection and quantification, as associated with an analysis procedure/measurement device. The limit of detection, denoted by $ l_{d_a} $, can intuitively be thought of as the lowest true concentration value that produces a signal the practitioner can reliably distinguish from background noise using an analysis procedure/measurement device. However, the numeric concentration value is too uncertain for routine reporting. The below quantification (but still detectable) limit, denoted by $ l_{q_a} $, relates to the smallest concentration value that can be measured and reported with acceptable accuracy and precision using an analysis procedure/measurement device. A practical solution to the censoring problem would be to adjust the likelihood function to account for censored observational data explicitly. Tobit-based models are a particularly popular class of censored regression models and can be traced back to the work of \cite{Tobin1958}. 
To construct a Tobit-based likelihood function that accounts for data censoring, the qualitative arguments outlined in the work of \cite{Gammelli2022} and \cite{Basson2023} are followed, and
 the following adjusted mixed-likelihood, denoted by the symbol $ p_m(\cdot \vert \cdot) $, for the $ a^\text{th} $ latent function, which explicitly accounts for data censoring, can be defined

\begin{equation}\label{Eq106}
p_m \bigl(y_a(s_{i,n},t_q) \vert f_a(s_{i,n},t_q) \bigr) =
    \begin{cases}
    \mathcal{N} \bigl(y_a(s_{i,n},t_q) \vert f_a(s_{i,n},t_q),\sigma_a^2 \bigr) 
    &\text{ if } \ y_a(s_{i,n},t_q) \ > \ l_{q_a} \\[1.5ex]
    {\color{white}\begin{rcases}
    \color{black}{\Phi \bigl(l_{q_a} \vert f_a(s_{i,n},t_q),\sigma_a^2 + {\color{black}\sigma_{q_ad_a}^2} \bigr) -} \\[0.5ex]
    \color{black}{\Phi \bigl(l_{d_a} \vert f_a(s_{i,n},t_q),\sigma_a^2 + {\color{black}\sigma_{q_ad_a}^2}  \bigr)} 
    \end{rcases}}
    &\text{ if } y_a(s_{i,n},t_q)  =  l_{q_a}  \\[3.5ex]
    \Phi \bigl(l_{d_a} \vert f_a(s_{i,n},t_q),\sigma_a^2 + {\color{black}\sigma_{d_a}^2}  \bigr) 
    &\text{ if } \text{ } y_a(s_{i,n},t_q) \text{ } = \text{ } l_{d_a}
    \end{cases}
\end{equation}

In Eq. \eqref{Eq106}, the symbol $ \Phi(\cdot \vert \cdot) $ denotes the general Gaussian cumulative distribution function (cdf). For notational convenience, the symbol $ \Phi (l_{d_a} \vert f_a(s_{i,n},t_q),\sigma_a^2 + {\color{black}\sigma_{q_ad_a}^2} ) $ is used to denote $ \Phi \left(\frac{l_{d_a} - f_a(s_{i,n},t_q)}{\sigma_a^2 + {\color{black}\sigma_{q_ad_a}^2}} \right) $.  Note that, in latent function regions where a spatio-temporal data point is observed, the basic GPR data generating structure is preserved. Otherwise, the expression for data points between the limits of quantification and detection corresponds to $\mathbb{P}(l_{d_a} < Y_a(s_{i,n},t_q) \leq l_{q_a})$ and, for data points below the limit of detection, $\mathbb{P}(Y_a(s_{i,n},t_q) \leq l_{d_a})$. Here $ \mathbb{P}(\cdot) $ denotes the probability value whereas $ \mathcal{N}(\cdot) $ denotes the Gaussian probability density function, both associated with the data random variable $ Y_a(\cdot) $. 

Note that for each spatio-temporal training input location $ (s_{i,n},t_q) $ associated with the below detection threshold $ l_{d_a} $, a constant (with respect to the temporal and spatial domain) heteroskedastic model is assumed with a total variance contribution that is the sum of a shared likelihood variance parameter and a tuning/regulating variance parameter denoted by $ \sigma_{d_a}^2 $ (see \citealp{Basson2023} for more details). A similar argument holds for the spatio-temporal training input locations associated with the below quantification (but still detectable) threshold $ l_{q_a} $. Given a censored spatio-temporal data set, a mixed Co-Kriging likelihood function can be constructed from Eqs. \eqref{Eq93} and \eqref{Eq106} that explicitly accounts for data censoring as follows

\vspace{-0.25 cm}

\begin{equation}\label{Eq107}
\begin{gathered}    \
    p(\boldsymbol{y} \vert \boldsymbol{f}) = \mathlarger{\prod_{a = 1}^{K_f}} \text{ } \Biggl[ \mathlarger{\prod_{\text{ }s \in \sample_a^{(s)}}} {\color{black}\Biggl [} \text{ } 
    \mathlarger{\prod_{t \in \{\sample_{a,s}^{(t)} | y_a(s,t) > l_{q_a}\}}} \text{ } \mathcal{N} \bigl(y_a(s,t) \vert f_a(s,t),\sigma_a^2 \bigr) \times \\[1.5ex]
    \mathlarger{\prod_{t \in \{\sample_{a,s}^{(t)} | y_a(s,t) = l_{q_a} \}}} \text{ } \Bigl[ \Phi \bigl( f_a(s,t) \vert l_{d_a},\sigma_a^2 + {\color{black}\sigma_{q_ad_a}^2} \bigr) - \Phi \bigl(f_a(s,t) \vert l_{q_a},\sigma_a^2 + {\color{black}\sigma_{q_ad_a}^2}  \bigr) \Bigr] \times \\ 
    \mathlarger{\prod_{t \in \{\sample_{a,s}^{(t)} | y_a(s,t) = l_{d_a} \}}} \text{ } \Bigl[1 - \Phi \bigl( f_a(s,t) \vert l_{d_a},\sigma_a^2 + {\color{black}\sigma_{d_a}^2}  \bigr) \Bigr]  {\color{black}\Biggr]} \Biggl]
\end{gathered}    
\end{equation} 

To arrive at Eq. \eqref{Eq107}, note that the Gaussian cdf property $ \Phi(y \vert x, \sigma^2) = 1 - \Phi(x \vert y, \sigma^2) $ was used (\citealp{Pishro2014}) and that the adjusted mixed Co-Kriging likelihood function depends on the additional tuning/regulating variance parameters that have been dropped from the conditioning set for notational convenience. 
Unfortunately, due to the presence of the cdf terms in Eq. \eqref{Eq107}, the adjusted mixed Co-Kriging likelihood function structure cannot be rewritten into the compact multivariate Gaussian functional form. With the specification of the Co-Kriging likelihood function and the GP prior density complete, which serves as the spatial, as well as the spatio-temporal, multivariate (i.e., Co-Kriging) extension of the work of \cite{VerHoef2006}, the authors now turn their attention to mechanisms for propagating the additional sources of input uncertainty that is associated with the hydrological distances and the weighting parameters.
\vspace{-0.45 cm}

\section{Approaches To Solving The Uncertainty Propagation Problem}\label{Section3}

\vspace{-0.15 cm}

Recall that the covariance matrix $ \boldsymbol{\Sigma}_K $ that is computed from Eq. \eqref{Eq3}, and by extension also the Co-Kriging covariance matrix associated with Eq. \eqref{Eq143}, depends on the measured hydrological distances $ d(s_{i,n},s_{d,m}) $ and weighting parameters $ w_k $ which results in additional unaccounted for input uncertainty that is introduced into the GPR model structure. Consequently, the tails-up SSN model as currently applied in the literature can be interpreted as a GPR model with uncertain inputs (\citealp{Girard2003}; \citealp{Mchutchon2011}; \citealp{Damianou2016}). The performance and predictive quality of the tails-up SSN model not only depend on the choice of covariance function but may also be sensitive to the hydrological distance and weighting parameter input uncertainty. Therefore, just like modelling the measurement uncertainty associated with the response variable (via the nugget effect), it may also be necessary to model and propagate the additional input uncertainty associated with the measured/estimated hydrological distances and weighting parameters. 

The hydrological distance $ d(s_{i,n},s_{d,m}) $ can be decomposed into the sum of distances $ h_k $ between intermediate points, which includes all the sampled spatial locations and junction points in the flow path between $ s_{i,n}$ and $ s_{d,m}$. For example, with reference to Figs. \ref{fig:Figure9} and SI.2, $ d(s_1,s_2) = h_1 + h_2 $, $ d(s_1,s_3) = h_1 + h_3 $, and, if a second sampled location on stream segment $ i = 2 $ were to be included a distance $ h_4$ above $s_2$, then $ d(s_1,s_{2,2}) = h_1 + h_2 + h_4 $.
The distances $ h_k $ and weighting parameters $ w_k $ can be collected in vectors $\boldsymbol{h} = [h_1, h_2, h_3]^T $ and $ \boldsymbol{w} = [w_1, w_2]^T $ (see Eq. \eqref{Eq143}). Note that the hydrological distances $ \boldsymbol{h} $ are either measured during the data collection procedure or estimated using a GIS. The weighting parameters $ \boldsymbol{w} $ should ideally be calculated from the measured stream segment flow rates but is in practice often estimated using proxies such as stream order, hydrologic basin area, watershed area, etc. (\citealp{VerHoef2006}; \citealp{VerHoef2010}; \citealp{Santos2022}). Note that these measurement or estimation methods invariably contain a measure of uncertainty.

 A key contribution of this paper will be incorporating the uncertainty in the hydrological distance $\boldsymbol{h}$ and weighting parameter $\boldsymbol{w}$ inputs, which maps through the GP prior covariance $ \boldsymbol{\Sigma}_{K}(\boldsymbol{h},\boldsymbol{w})  $ (see Eq. \eqref{Eq12}), and by extension also the Co-Kriging covariance matrix associated with Eq. \eqref{Eq143}. As will become evident in Sect. \ref{Section3.1}, a naive extension of the standard GPR model to account for additional input uncertainty is analytically intractable, and hence some form of approximate inference is required. Typically, either MCMC sampling or variational inference is considered, with the latter pursued in this work. Mean-field variational inference is typically used, but as will be discussed in Sect. \ref{Section3.2}, even standard mean-field variational inference still fails to produce an analytically tractable approximation to the standard GPR model with uncertain inputs. The solution comes in the form of the Bayesian Gaussian Process Latent Variable Model (BGP-LVM), first introduced by \cite{Titsias2010}. The BGP-LVM will allow the additional sources of input uncertainty to be propagated through the standard GPR model whilst simultaneously allowing for closed-form approximate solutions that can be used for Bayesian model training and inference. The authors now consider each of the above-mentioned approaches in turn.

\vspace{-0.55 cm}

\subsection{Standard Bayesian Inference}\label{Section3.1}

\vspace{-0.25 cm}
Starting, for example, from Eqs. \eqref{Eq8} to \eqref{Eq13}, the joint density associated with the tails-up SSN model can be augmented with the (assumed independent) random variables $ \boldsymbol{h} $ and $ \boldsymbol{w} $ to give Eq. \eqref{Eq29}, where the dependence on any model parameters ${\boldsymbol{\theta}}$ has been dropped for notational convenience, such that 

\vspace{-0.25 cm}

\begin{equation}\label{Eq29}
   p(\boldsymbol{y},\boldsymbol{f},\boldsymbol{h},\boldsymbol{w}) = p(\boldsymbol{y} \vert \boldsymbol{f})p(\boldsymbol{f} \vert \boldsymbol{h},\boldsymbol{w})p(\boldsymbol{h})p(\boldsymbol{w})
\end{equation}

The model parameters $ \boldsymbol{\theta} $ would typically be obtained via a point estimation procedure (e.g., maximising the log marginal likelihood) whereas the vectors $ \boldsymbol{h} $ and $ \boldsymbol{w} $ are modelled with a probability density to account for the additional sources of input uncertainty. Conditioning on the observation vector $ \boldsymbol{y} $ (i.e., deriving the posterior density) results in

\begin{equation}\label{Eq30}
   p(\boldsymbol{f},\boldsymbol{h},\boldsymbol{w} \vert \boldsymbol{y}) = \frac{p(\boldsymbol{y} \vert \boldsymbol{f})p(\boldsymbol{f} \vert \boldsymbol{h},\boldsymbol{w})p(\boldsymbol{h})p(\boldsymbol{w})}{p(\boldsymbol{y})} 
\end{equation}

The model marginal likelihood $ p(\boldsymbol{y}) $, see Eq. \eqref{Eq31}, is obtained by marginalising over the latent function values $ \boldsymbol{f} $, as in the standard GPR model (see Sect. \ref{Section2.2}), and the latent variables $ \boldsymbol{h} $ and $ \boldsymbol{w} $, respectively.

\vspace{-0.25 cm}

\begin{equation}\label{Eq31}
  p(\boldsymbol{y}) = \int\limits_{\boldsymbol{f}} \int\limits_{\boldsymbol{h}} \int\limits_{\boldsymbol{w}} p(\boldsymbol{y} \vert \boldsymbol{f})p(\boldsymbol{f} \vert \boldsymbol{h},\boldsymbol{w})p(\boldsymbol{h})p(\boldsymbol{w}) d\boldsymbol{w} d\boldsymbol{h} d\boldsymbol{f} = \int\limits_{\boldsymbol{f}} p(\boldsymbol{y} \vert \boldsymbol{f}) \textcolor{black}{\left[ \int\limits_{\boldsymbol{h}} \int\limits_{\boldsymbol{w}} p(\boldsymbol{f} \vert \boldsymbol{h},\boldsymbol{w})p(\boldsymbol{h})p(\boldsymbol{w})  d\boldsymbol{w} d\boldsymbol{h} \right]} d\boldsymbol{f}
\end{equation}

Computing the integral in Eq. \eqref{Eq31} with respect to $\boldsymbol{h}$ and $\boldsymbol{w}$ is challenging and, in the case of the tails-up SSN model of \cite{VerHoef2006}, analytically intractable, rendering exact Bayesian inference infeasible (\citealp{Damianou2015}). This analytical intractability also persists for the Co-Kriging extension developed in Sect. \ref{SectionM}. The problem can be circumvented by Markov Chain Monte Carlo (MCMC) sampling, but these methods scale poorly to large data and parameter dimensionality settings. Variational inference provides a computationally tractable alternative and aims to approximate the true underlying posterior density by lower bounding the model log marginal likelihood $ \ln p(\boldsymbol{y}) $ (\citealp{Jordan1999}; \citealp{Wainwright2008}). Variational inference tends to scale better to large data and parameter dimensionality settings, when compared to MCMC, and often provides significant improvements in computational efficiency (\citealp{Bishop2009}; \citealp{Blei2017}). 

\vspace{-0.45 cm}

\subsection{Mean-Field Variational Inference}\label{Section3.2}

\vspace{-0.15 cm}

Standard variational inference aims to minimise the Kullback-Leibler ($ \mathcal{K}\mathcal{L} $) divergence between the true underlying posterior density $p(\boldsymbol{f},\boldsymbol{h},\boldsymbol{w} \vert \boldsymbol{y})$ over the latent variables of interest and an approximate posterior density $ q(\boldsymbol{f},\boldsymbol{h},\boldsymbol{w}) \in Q $ from a constrained set of density functions $ Q $. In other words, variational inference recasts the inference problem into an optimisation problem, the complexity of which is controlled by the set $ Q $ (see \cite{Blei2017} for more details). In this work, the application of variational inference is investigated as a means to propagate the additional sources of input uncertainty associated with $ p(\boldsymbol{h}) $ and $ p(\boldsymbol{w}) $. Introducing the Evidence Lower BOund (ELBO, \citealp{Blei2017}) $ \mathcal{F}[q(\boldsymbol{f},\boldsymbol{h},\boldsymbol{w})] $ (Eq. \eqref{Eq33}), also called the variational free energy (\citealp{Mackay2004}), the log marginal likelihood $ \ln p(\boldsymbol{y}) $ can be decomposed to yield Eq. \eqref{Eq32} (\citealp{Bishop2009}) below.

\vspace{-0.15 cm}

\begin{equation}\label{Eq33}
  \mathcal{F}\bigl[ q(\boldsymbol{f},\boldsymbol{h},\boldsymbol{w}) \bigr] =  \int\limits_{\boldsymbol{f}} \int\limits_{\boldsymbol{h}} \int\limits_{\boldsymbol{w}} q(\boldsymbol{f},\boldsymbol{h},\boldsymbol{w}) \ln \frac{p(\boldsymbol{y} \vert \boldsymbol{f})p(\boldsymbol{f} \vert \boldsymbol{h},\boldsymbol{w})p(\boldsymbol{h})p(\boldsymbol{w})}{q(\boldsymbol{f},\boldsymbol{h},\boldsymbol{w})} d\boldsymbol{w} d\boldsymbol{h} d\boldsymbol{f}
\end{equation}

\begin{equation}\label{Eq32}
  \ln p(\boldsymbol{y}) = \mathcal{F} \bigl[ q(\boldsymbol{f},\boldsymbol{h},\boldsymbol{w}) \bigr] \text{} + \text{} \mathcal{K}\mathcal{L} \bigl[ q(\boldsymbol{f},\boldsymbol{h},\boldsymbol{w}) \vert\vert p(\boldsymbol{f},\boldsymbol{h},\boldsymbol{w} \vert \boldsymbol{y}) \bigr]
\end{equation}

Since the Kullback-Leibler divergence satisfies Gibbs' inequality, i.e., $ \mathcal{K}\mathcal{L}[q(\boldsymbol{f},\boldsymbol{h},\boldsymbol{w}) \vert\vert p(\boldsymbol{f},\boldsymbol{h},\boldsymbol{w} \vert \boldsymbol{y})] \geq 0 $, 
the variational free energy lower bounds the model log marginal likelihood $ \ln p(\boldsymbol{y}) $ (\citealp{Mackay2004}) such that

\vspace{-0.25 cm}

\begin{equation}\label{Eq34}
  \ln p(\boldsymbol{y}) \geq \mathcal{F} \bigl[ q(\boldsymbol{f},\boldsymbol{h},\boldsymbol{w}) \bigr]
\end{equation}

A common choice for the set $ Q $ of approximate posterior densities is the mean-field variational family, where the model latent variables are assumed to be mutually independent factors (\citealp{Parisi1988}; \citealp{Bishop2009}; \citealp{Opper2009}; \citealp{Blei2017}) such that the joint variational posterior density factorises according to Eq. \eqref{Eq35}. Using Eq. \eqref{Eq35} to expand Eq. \eqref{Eq33} yields the variational lower bound given by Eq. \eqref{Eq36}.

\vspace{-0.25 cm}

\begin{equation}\label{Eq35}
  q(\boldsymbol{f},\boldsymbol{h},\boldsymbol{w}) = q(\boldsymbol{f})q(\boldsymbol{h})q(\boldsymbol{w})
\end{equation}

\vspace{-0.50 cm}

\begin{equation}\label{Eq36}
  \begin{split}
  \ln p(\boldsymbol{y}) \geq \int\limits_{\boldsymbol{f}} q(\boldsymbol{f}) \textcolor{black}{\left [\int\limits_{\boldsymbol{h}} \int\limits_{\boldsymbol{w}} q(\boldsymbol{h})q(\boldsymbol{w}) \ln p(\boldsymbol{f} \vert \boldsymbol{h},\boldsymbol{w}) d\boldsymbol{w} d\boldsymbol{h} \right]} d\boldsymbol{f}  + \int\limits_{\boldsymbol{f}} \int\limits_{\boldsymbol{h}} \int\limits_{\boldsymbol{w}} q(\boldsymbol{f})q(\boldsymbol{h})q(\boldsymbol{w}) \ln \frac{p(\boldsymbol{y} \vert \boldsymbol{f})p(\boldsymbol{h})p(\boldsymbol{w})}{q(\boldsymbol{f})q(\boldsymbol{h})q(\boldsymbol{w})} d\boldsymbol{w} d\boldsymbol{h} d\boldsymbol{f} 
  \end{split}
\end{equation}

The variational lower bound requires computing the expected value of $ \ln p(\boldsymbol{f} \vert \boldsymbol{h},\boldsymbol{w}) $ under the variational posterior densities $ q(\boldsymbol{h}) $ and $ q(\boldsymbol{w}) $, in other words, the integral in square brackets in Eq. \eqref{Eq36}. From Eq. \eqref{Eq3}, when ignoring the nugget effect and only considering the prior density over spatial stream network-based latent function (see Sects. \ref{Section2.1} and \ref{Section2.2}), note that the integral in square brackets remains analytically intractable as the latent variables $ \boldsymbol{h} $ and $ \boldsymbol{w} $ appear non-linearly in $ \boldsymbol{\Sigma}_K^{-1} $ and $ \ln \vert \boldsymbol{\Sigma}_K \vert $ for the standard tails-up SSN model of \cite{VerHoef2006}. This analytical intractability also persists for the Co-Kriging extension developed in Sect. \ref{SectionM}. Consequently, there is no closed-form solution available for the variational lower bound in Eq. \eqref{Eq36}. Therefore, to apply variational inference, an alternative strategy is required to propagate the additional sources of input uncertainty. 

\vspace{-0.45 cm}
\section{The BGP-LVM For Stream Networks}\label{Section4}

\vspace{-0.05 cm}

To this end, the Variational - also referred to as the Bayesian - Gaussian Process Latent Variable Model (BGP-LVM) provides the theoretical foundation for propagating the additional sources of input uncertainty in an analytically tractable manner (see \citealp{Lawrence2005}, \citealp{Titsias2010}, \citealp{Damianou2011}, \citealp{Damianou2016} and \citealp{Zhao2016} for more details).
More specifically, the standard variational Gaussian Process latent variable model variant, as applied to GPs with uncertain inputs, for independent and identically distributed (i.i.d.) data \cite[see Sects. 3.1.1 and 6.1]{Damianou2016}. However, to use the proposed BGP-LVM model, it will be necessary for the authors to introduce the notion of spatio-temporal inducing variables (see Sect. \ref{Section6.2} and \cite{Titsias2008, Titsias2009}), which, intuitively, can be thought of as a summary statistic that allows for closed-form and computationally efficient approximate inference. Here, the inducing variable summary statistic assumption is underpinned by the idea that the posterior distribution can be constructed at a smaller number of optimised temporal locations, while maintaining the full predictive resolution over the spatial domain, leading to the idea of temporal variational compression, which allows for a computationally efficient and analytically tractable implementation of the BGP-LVM. This will all be achieved within an optimisation-based framework where the goal will be to minimise the $ \mathcal{K}\mathcal{L} $-divergence between the true underlying posterior distribution and a set of assumed approximate posterior densities $ Q $. Here, the authors will take $ Q $ to be the set of densities from the mean-field variational family, which assumes that certain model latent variables are mutually independent, and will allow the authors to induce closed-form solutions that can be used for Bayesian model training and inference. Next, the authors turn their attention to specifying the prior densities that are required to implement the BGP-LVM for stream networks.

\vspace{-0.5 cm}

\subsection{Specifying The Uncertain Input Prior Densities}\label{Specifying The Uncertain Input Prior Densities}

\vspace{-0.15 cm}

Recall that hydrological distances are positive quantities and the weighting parameters are restricted to the domain between 0 and 1. These physical properties limit the types of densities that can be used to model the quantities $ \boldsymbol{h} $ and $ \boldsymbol{w} $ while simultaneously resulting in an analytically tractable variational lower bound. To circumvent these problems, this paper opts to use an alternative latent variable parameterisation for the hydrological distances and weighting parameters that preserve the physical constraints associated with these latent variables while also allowing for a closed-form solution to the variational lower bound. For example, from Fig. SI.2 in Sect. SI.5 of the Supplementary Information, the hydrological distance $ d(s_1,s_2) = h_1 + h_2$ between the sampled spatial locations $ s_1 $ and $ s_2 $ can alternatively be expressed as $ d(s_1,s_2) = \tau_1^2 + \tau_2^2$.

The quantity $ \tau $ represents an alternative latent variable parameterisation for the hydrological distance $ h = \tau^2$, which preserves the positivity constraint associated with $ h $. This will become important in the Supplementary Information (see Sect. SI.1) when propagating the additional sources of input uncertainty associated with the hydrological distances. In a similar manner, $ d(s_1,s_3) = \tau_1^2 + \tau_3^2 $ and $ d(s_2,s_3) = \tau_2^2 + \tau_3^2 $. 
 The weighting parameters $ w_2 $ and $ w_3 $ in Fig. SI.2, which are required to maintain stationarity of the variances, can also be defined in terms of an alternative latent variable parameterization. More specifically, in this paper, the $ k^\text{th} $ weighting parameter $ w_k $ is parameterised in terms of $ \sqrt{w_k} = \Phi(\gamma_k) $ such that $ w_k = {\Phi^2(\gamma_k)} $ where $ \Phi(\cdot) $ denotes the standard Gaussian cdf which preserves the property that $ 0 < {\Phi^2(\gamma_i)} < 1 $. In other words, for the stream network depicted in Fig. SI.2, $ w_2 $ and $ w_3 $ can be parameterised as $ w_2 = \Phi^2(\gamma_2) $ and $ w_3 = \Phi^2(\gamma_3) $, respectively. This alternative parameterisation will also become important in the Supplementary Information (see Sect. SI.1) when propagating the additional input uncertainty associated with the weighting parameters. Note that the constraint $ \Phi^2(\gamma_2) + \Phi^2(\gamma_3) = 1 $ will eventually be replaced with a variational analogue (see Eq. \eqref{Eq53} and Sect. SI. 3.2) after computing the required expected values, however, despite the aforementioned constraint being inherent to the developed probabilistic model, it must be imposed through an external mechanism. In this work, the authors rely on constrained-based numerical optimisation to enforce this model constraint.

 Using the alternative parameterisations, the spatial cross-covariance $ C_{f_a, f_b}^{(s)} $ given by Eqs. \eqref{Eq177} to \eqref{Eq178} can be rewritten as

 \vspace{-0.55 cm}

\begin{equation}\label{Eq182}
\begin{gathered}
    C_{f_a,f_b}^{(s)} \Biggl(s_{d,m}, s_{i,n} \mathlarger{\mathlarger{\vert}} \boldsymbol{\theta}_a^{(s)}, \boldsymbol{\theta}_b^{(s)} \Biggr) = 
    \begin{cases} 
    0 &\text{ if sites are not flow-connected } 
    \\[1.5ex] 
    \Biggl[ \mathlarger{\mathlarger{\prod}}_{\ k \in B_{s_{i,n},s_{d,m}}}  \Phi(\gamma_k) \Biggr] \conv_{f_a,f_b}^{(s)} \Biggl(
    \mathlarger{\sum}_{j \in \mathcal{T}_{s_{i,n},s_{d,m}}} \tau_j^2 
    \mathlarger{\mathlarger{\vert}} 
    \boldsymbol{\theta}_a^{(s)}, \boldsymbol{\theta}_b^{(s)} \Biggr) \text{ } \text{ } &\text{ if sites are flow-connected  }
    \end{cases}
    \end{gathered}
\end{equation}

\vspace{-0.65 cm}

\begin{equation}\label{Eq183}
\begin{split}
    \conv_{f_a,f_b}^{(s)} \Biggl(
    \mathlarger{\sum}_{j \in \mathcal{T}_{s_{i,n},s_{d,m}}} \tau_j^2
    \mathlarger{\mathlarger{\vert}} 
    \boldsymbol{\theta}_a^{(s)}, \boldsymbol{\theta}_b^{(s)} \Biggr)
    = \int\limits_{0}^{\infty} g_a(x + \sum_{j \in \mathcal{T}_{s_{i,n},s_{d,m}}} \tau_j^2  \vert \boldsymbol{\theta}_a^{(s)})g_b(x \vert \boldsymbol{\theta}_b^{(s)})  dx 
    \end{split}
\end{equation}

Note that, in general, the hydrological distance between two spatial locations $ s_{d,m} $ and $ s_{i,n} $ can be expressed as

\vspace{-0.15 cm}

\begin{equation}\label{Eq181}
\begin{gathered}
    d(s_{i,n},s_{d,m}) = \mathlarger{\sum}_{j \in \mathcal{T}_{s_{i,n},s_{d,m}}} \tau_j^2
    \end{gathered}
\end{equation}

Symbol $ \mathcal{T}_{s_{i,n},s_{d,m}} $ denotes the set containing all the $ \tau_j $ elements between spatial locations $ s_{i,n} $ and $ s_{d,m} $. Consequently, instead of performing inference directly over $ \boldsymbol{h} $ and $ \boldsymbol{w} $, inference is now performed over the alternative latent random variables $ \boldsymbol{\tau} $ and $ \boldsymbol{\gamma} $, respectively. 
Since the hydrological distances are either measured during the data collection procedure or estimated using a GIS, the estimated values can be used as a form of prior knowledge. More specifically, independent univariate Gaussian priors $\mathcal{N} \bigl(\tau_j \vert d_{\tau_{j}},\exp \{\eta_\tau \} \bigr)$ are used.
The symbol $ d_{\tau_{j}} $ denotes the measured/estimated (data) value for the hydrological distance under consideration, as mapped through the latent variable parameterisation, which is used as the mean for the univariate Gaussian density.
 The shared variance parameter is given by $ \exp \{\eta_\tau \} $. During the initial model development phase, a mutually independent Gaussian prior with shared variance parameter $ \sigma_\tau^2 $ was considered, however, empirical observations from the simulation-based case studies indicated that the $ \boldsymbol{\tau} $ variational posterior densities, and by association, the hydrological distance posterior densities, are sensitive to the magnitude of the shared prior variance parameter. Consequently, a hyperprior was placed on the shared prior variance parameter to circumvent this problem, with the prior variance $ \sigma_\tau^2 = \exp\{\eta_\tau\}$ (\citealp{Gredilla2011}).

Finally, independent, univariate Gaussian priors are used among the $ \boldsymbol{\gamma} $  vector components (which parameterise $ w_k = {\Phi^2(\gamma_k)} $) such that

\vspace{-0.35 cm}

\begin{equation}\label{Eq59}
    \gamma_k \sim \mathcal{N}(\gamma_k \vert d_{\gamma_k}, \sigma^2_{\gamma}) 
\end{equation}
The mean of the prior density is set to the estimated value $ d_{\gamma_k} $, which can be obtained from the measured flow rate data or a suitable proxy variable, as mapped through the latent variable parameterisation, whereas the shared prior variance $ \sigma^2_{\gamma} $ reflects the belief in the estimated value $ d_{\gamma_k} $. 

\vspace{-0.45 cm}

\subsection{Specifying The Inducing Variable Prior Density} \label{Section6.2}

To implement the BGP-LVM, it is necessary to introduce a set of auxiliary variables, $\boldsymbol{u}$, often referred to as the inducing variables. Following the work of \cite{Titsias2008,Titsias2009}, the inducing variables facilitate computational speedups and, when used as a mathematical tool, induce a variational lower bound on the log marginal likelihood of the developed probabilistic model (see Sects. \ref{Section4.1} and \ref{Section4.4}). Let the inducing variables $ \boldsymbol{u}$ represent $ M \ll N $ function points located on the underlying latent function. The joint density over $ \boldsymbol{f} $ and $ \boldsymbol{u} $, conditioned on the uncertain inputs, take the following functional form

\vspace{-0.45 cm}

\begin{equation}\label{Eq138}
    p(\boldsymbol{f}, \boldsymbol{u} \vert \boldsymbol{\tau},\boldsymbol{\gamma}) = \mathcal{N} \Biggl( \begin{bmatrix} \boldsymbol{f} \\ \boldsymbol{u} \end{bmatrix} \Biggl\vert \begin{bmatrix} \boldsymbol{0} \\ \boldsymbol{0} \end{bmatrix}; \begin{bmatrix} \boldsymbol{K}_{NN}(\boldsymbol{\tau},\boldsymbol{\gamma}) & \boldsymbol{K}_{NM}(\boldsymbol{\tau},\boldsymbol{\gamma},\boldsymbol{h}^{'},\boldsymbol{\alpha}) \\ \boldsymbol{K}_{MN}(\boldsymbol{\tau},\boldsymbol{\gamma},\boldsymbol{h}^{'},\boldsymbol{\alpha}) & \boldsymbol{K}_{MM}(\boldsymbol{h}^{'},\boldsymbol{\alpha}) \end{bmatrix} \Biggr)
\end{equation}

The quantities $ \boldsymbol{h}^{'} $ and $ \boldsymbol{\alpha} $, which will eventually be interpreted as variational parameters, will be introduced when constructing the covariance matrix $ \boldsymbol{K}_{MM}(\boldsymbol{h}^{'},\boldsymbol{\alpha}) $, as associated with the inducing variables $ \boldsymbol{u} $. Note that from here on the explicit dependence of any covariance matrix on $ 
\boldsymbol{\tau} $,$ 
\boldsymbol{\gamma} $, $ \boldsymbol{h}^{'} $, and $ \boldsymbol{\alpha} $ is dropped for national convenience. Using the properties of the multivariate Gaussian density, the conditional (joint) density in Eq. \eqref{Eq138} can be conditioned on the inducing variables $ \boldsymbol{u} $ such that

\vspace{-0.45 cm}

\begin{equation}\label{Eq139}
    p(\boldsymbol{f}, \boldsymbol{u} \vert \boldsymbol{\tau},\boldsymbol{\gamma}) = p(\boldsymbol{f} \vert \boldsymbol{u}, \boldsymbol{\tau},\boldsymbol{\gamma})p(\boldsymbol{u})
\end{equation}

The densities $ p(\boldsymbol{f} \vert \boldsymbol{u}, \boldsymbol{\tau},\boldsymbol{\gamma}) $ and $ p(\boldsymbol{u}) $ by definition corresponds to

\vspace{-0.25 cm}

\begin{equation}\label{Eq140}
    p(\boldsymbol{f} \vert \boldsymbol{u}, \boldsymbol{\tau},\boldsymbol{\gamma}) = \mathcal{N}(\boldsymbol{f} \vert \boldsymbol{K}_{NM}{\boldsymbol{K}^{-1}_{MM}}\boldsymbol{u},\boldsymbol{K}_{NN} - \boldsymbol{K}_{NM}{\boldsymbol{K}^{-1}_{MM}}\boldsymbol{K}_{MN})
\end{equation}
\begin{equation}\label{Eq141}
    p(\boldsymbol{u}) = \mathcal{N}(\boldsymbol{u} \vert \boldsymbol{0},\boldsymbol{K}_{MM})
\end{equation}
Using the proposed BGP-LVM frameworks results in a variational lower bound that requires inverting the matrix $ \boldsymbol{K}_{MM} $, instead of the matrix $ \boldsymbol{K}_{NN}$, which is computationally more efficient since $ M \ll N $. 
Note that there is no strict requirement for the one-to-one correspondence in the number of  underlying latent functions, $ K_f $, and the number of variational inducing functions, $ K_u $. Preferably, $ K_u \leq K_f $ to maintain computational efficiency, but the developments below use $K_u = 2$. In the original variational GP methodology proposed by \cite{Titsias2008,Titsias2009}, the inducing variables $ \boldsymbol{u} $ and $ \boldsymbol{f} $ share the same covariance, meaning $\boldsymbol{K}_{MM}$ would follow the same construction as the Co-Kriging prior outlined in section \ref{Section6.1}. However, this work follows a generalised framework (\citealp{Alvarez2010} and \citealp{Alvarez2011}) where the kernel functions associated with $ \boldsymbol{f} $ and $ \boldsymbol{u} $ can be different to improve approximation capacity. This is achieved by introducing additional covariance parameters that are estimated during gradient-based optimisation.

\vspace{-0.35 cm}
\begin{figure*}[!ht]
    \centering
    \includegraphics[scale = 0.90]{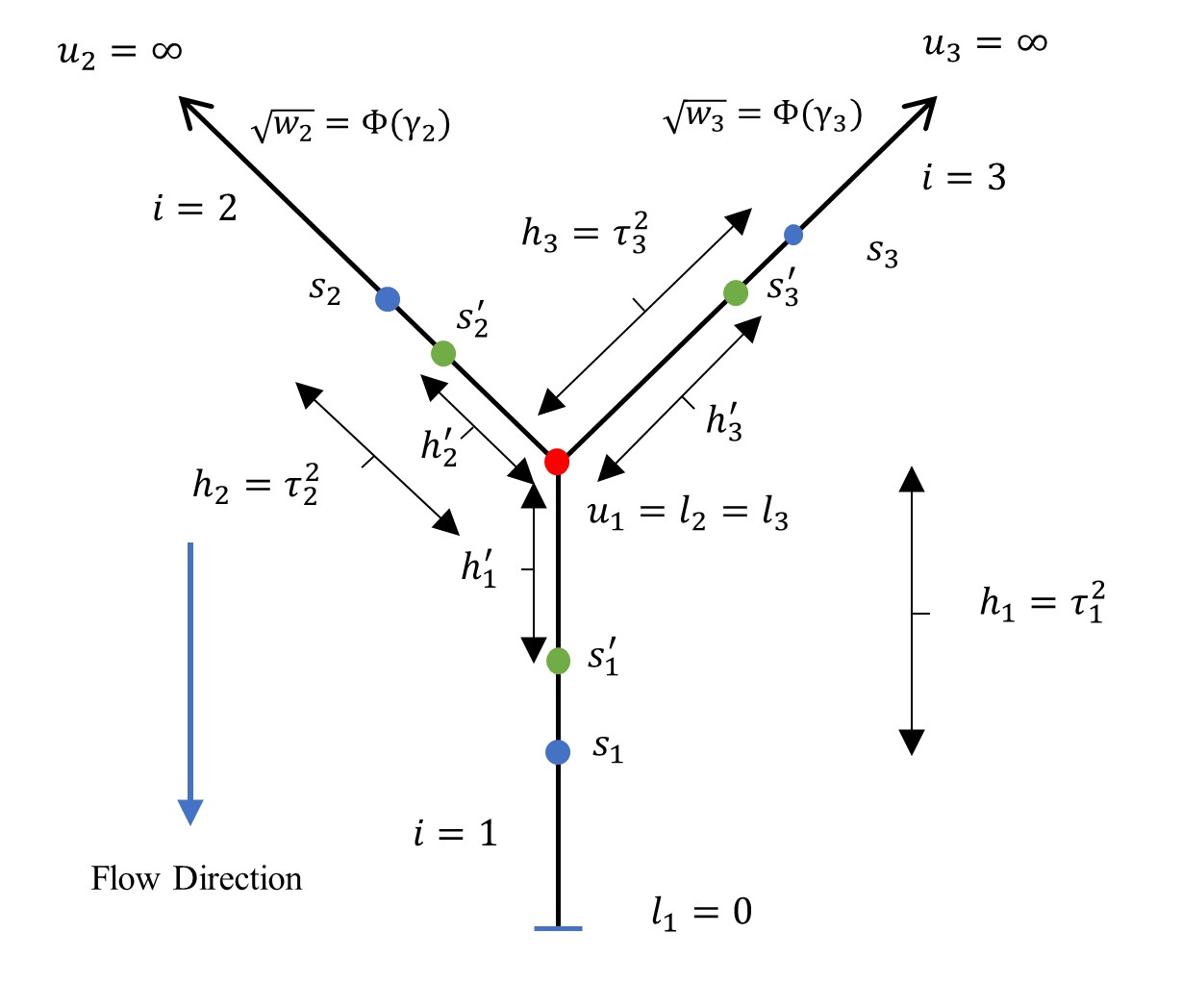}
    \caption{Hypothetical stream network consisting of three stream segments/branches labelled $ i = 1, i = 2 $, and $ i = 3 $, respectively, together with three sampled spatial locations denoted by $ s_1, s_2 $, and $ s_3 $ and three inducing spatial locations denoted by $ s_1^\prime, s_2^\prime $, and $ s_3^\prime $. Figure \ref{fig:Figure9} has been reproduced and adjusted from the work of \cite{VerHoef2006}. \vspace{-0.25 cm}}
    \label{fig:Figure9}
\end{figure*}

The stream network-based methodology of \cite{VerHoef2006} is again used to construct the spatial covariance component associated with $ \boldsymbol{u}$. However, some care must be taken to specify the spatial locations of these inducing points. Refer to Fig. \ref{fig:Figure9} and recall the three sampled spatial locations labelled as $ s_1 $, $ s_2 $, and $ s_3 $, respectively, as denoted by the blue dots. Associated with each of the sampled spatial locations is a spatial inducing input location labelled $ s_1^\prime $, $ s_2^\prime $, and $ s_3^\prime $, respectively, as denoted by the green dots. These locations can be adjusted by changing the variational hydrological distances $ h_1^\prime $, $ h_2^\prime $, and $ h_3^\prime $, respectively, such that the $ \mathcal{KL} $-divergence (see Eq. \eqref{Eq45}) is minimised. In this regard, the variational hydrological distances $ h_1^\prime $,$ h_2^\prime $, and $ h_3^\prime $ can be regarded as freely optimisable variational parameters.

In this work, the spatial inducing location $ s_1' $ is constrained to lie above the associated downstream sampled spatial location $ s_1 $, maintaining a minimum small distance $ \epsilon_{s_1,s_1'} $ away from spatial location $ s_1 $. The minimum distance requirement helps to avoid potential pitfalls that can arise when computing the required expected values that are necessary to obtain the variational lower bound in a closed-form solution. Furthermore, spatial inducing input location $ s_1' $ is not allowed to traverse the junction point $ u_1 $, in other words, the inducing input location is restricted to the stream segment/branch $ i = 1 $. All other spatial inducing input locations, for example, $ s_2' $ and $ s_3' $ in Fig. \ref{fig:Figure9}, are defined to be below the associated sampled spatial location, again  maintaining a small distance away from the actual sampled spatial location. Similar to before, these spatial inducing input locations are not allowed to traverse the junction point $ u_1 $.

In terms of constructing the spatial covariance component associated with the variational inducing function $ u_{1} $ at the spatial inducing input location $ s_1' $, the same procedure is followed as outlined in \cite{VerHoef2006} and Sect. \ref{Section6.1}. In other words, using the moving-average function construction procedure proposed by \cite{VerHoef2006}, the variational inducing function $ u_{1}(s_1') $ at spatial inducing input location $ s_1' $ can be constructed as 

\vspace{-0.15 cm}

\begin{equation}\label{Eq197}
\begin{gathered}
    u_{1}(s_1') = \int\limits_{s_1'}^{u_1} g_1'(x_1 - s_1')W(x_1)dx_1 + \Phi(\alpha_2) \int\limits_{u_1}^{\infty} g_1'(x_2 - s_1')W(x_2)dx_2  + \text{ } \Phi(\alpha_3) \int\limits_{u_1}^{\infty} g_1'(x_3 - s_1')W(x_3)dx_3
    \end{gathered}
\end{equation}

\vspace{-0.05 cm}

Notice from Eq. \eqref{Eq197} that the smoothing kernel $ g_1'(\cdot) $, similar to the construction procedure outlined in Sect. \ref{Section6.1}, splits at the junction point and is weighted by the variational weights $ \Phi(\alpha_2) $ and $ \Phi(\alpha_3) $, respectively. To maintain stationarity of the variances, it is also required that $ \Phi^2(\alpha_2) + \Phi^2(\alpha_3) = 1 $, which the authors impose using constrained-based numerical optimisation. Furthermore, the dependence of $ g_1'(\cdot) $ on any model parameters has been omitted for notational convenience. The freely optimisable parameters $ \alpha_2 $ and $ \alpha_3 $, collectively denoted as $ \boldsymbol{\alpha} = [{\alpha_2},{\alpha_3}]^T $, serve as the variational parameters which, when mapped through $ \Phi(\cdot) $, produce the weighting parameters that are required to produce the stationary random variable $ u_{1}(s_1') $ at spatial inducing input location $ s_1'$. In general, the variational inducing function random variable $ u_{a}(s'_{i,n}) $ can be constructed as

\vspace{-0.25 cm}

\begin{equation}\label{Eq198}
\begin{gathered}
    u_{a}(s'_{i,n}) = \mathlarger{\mathlarger{\sum}}_{j \in U_{s'_{i,n}}} \mathlarger{\Biggl(} \Biggl[ \mathlarger{\mathlarger{\prod}}_{\text{ } k \in B_{s'_{i,n},[j]}}  \Phi(\alpha_k) \Biggr] \int\limits_{l_j}^{u_j} g_a'(x_j - s'_{i,n})W(x_j)dx_j \mathlarger{\Biggr)}
    \end{gathered}
\end{equation}

\vspace{-0.05 cm}

\noindent Following the same arguments and procedure as outlined in Sect. \ref{Section6.1}, the separable spatio-temporal cross-covariance associated with the inducing variables can be constructed as follows

\vspace{-0.15 cm}

\begin{equation}\label{Eq209}
\begin{gathered}
    \mathbb{C}\text{ov}  \Bigl[ u_a(s'_{d,m}, t'_p), u_b(s'_{i.n}, t'_q) \mathlarger{\mathlarger{\vert}} \boldsymbol{\theta}'_a, \boldsymbol{\theta}'_b \Bigr] = \\[1.5ex] 
    \begin{cases} 
    0 
    &\text{ if sites are not flow-connected } \\[1.5ex] 
    \Biggl[ \mathlarger{\mathlarger{\prod}}_{\ k \in B_{s'_{i,n},s'_{d,m}}}  \Phi(\alpha_k) \Biggr] 
    \conv_{u_a, u_b}^{(s)\times(t)} \Biggl(\mathlarger{\sum}_{j \in \mathcal{T}_{s'_{i,n},s'_{d,m}}} h'_j , t'_q - t'_p \mathlarger{\mathlarger{\vert}} \boldsymbol{\theta}'_a, \boldsymbol{\theta}'_b \Biggr)
    &\text{ if sites are flow-connected }
    \end{cases}
    \end{gathered}
\end{equation}

\vspace{-0.15 cm}

\begin{equation}\label{Eq210}
\begin{split}
    \conv_{u_a, u_b}^{(s)\times(t)} \Biggl(\mathlarger{\sum}_{j \in \mathcal{T}_{s'_{i,n},s'_{d,m}}} h'_j , t'_q-t'_p \mathlarger{\mathlarger{\vert}} \boldsymbol{\theta}'_a, \boldsymbol{\theta}'_b \Biggr) 
    &= 
    \Biggl[ \ \int\limits_{0}^{\infty} 
    g'_a(x + \mathlarger{\sum}_{j \in \mathcal{T}_{s'_{i,n},s'_{d,m}}} h'_j  \vert \boldsymbol{\theta}_a^{'(s)}) 
    g'_b(x \vert \boldsymbol{\theta}_b^{'(s)})  dx \Biggr] 
    \\
    &\times \Biggl[ \ \int\limits_{-\infty}^{\infty} 
    G'_a(z - (t'_q - t'_p) \vert \boldsymbol{\theta}_a^{'(t)})
    G'_b(z\vert \boldsymbol{\theta}_b^{'(t)}) dz  \Biggr]  
    \ \forall \ a,b = 1,\cdots,K_u
    \end{split}
\end{equation}

\noindent Here, $ \boldsymbol{\theta}'_a = \{ \boldsymbol{\theta}_a^{'(s)}, \boldsymbol{\theta}_a^{'(t)} \} $  collectively denotes the spatial and temporal parameters associated with the smoothing kernel for the variational inducing function $a$. The cross-covariance parameterisation relies on the fact that the hydrological distance $ d(s'_{i,n},s'_{d,m}) $ between the spatial inducing input locations $ s'_{i,n} $ and $ s'_{d,m} $ can be expressed as 

\vspace{-0.10 cm}

\begin{equation}\label{Eq204}
\begin{gathered}
    d(s'_{i,n},s'_{d,m}) = \mathlarger{\sum}_{j \in \mathcal{T}_{s'_{i,n},s'_{d,m}}} h'_j
    \end{gathered}
\end{equation}

\vspace{-0.10 cm}

 Equations \eqref{Eq209} and \eqref{Eq210} completely specify the spatio-temporal cross-covariance associated with the variational inducing functions and, for a finite set of spatio-temporal inducing input locations, can be used to construct the covariance matrix $ \boldsymbol{K}_{MM} $, as associated with the inducing variable prior density given by Eqs. \eqref{Eq141}. The quantities $ \alpha_k, h'_j,t'_p $, and $ t'_q $ in Eqs. \eqref{Eq209} and \eqref{Eq210} are variational parameters that the practitioner can estimate using gradient-based optimisation such that the $ \mathcal{KL} $-divergence between the approximate and true posterior distribution in Eq. \eqref{Eq45} is minimised. 

For a computationally efficient lower bound implementation, and as part of the gradient-based optimisation routine specifications, the practitioner must select the number of spatio-temporal inducing variables $ M $ (as associated with covariance matrix $ \boldsymbol{K}_{MM} $) such that $ M < N $ (as associated with covariance matrix $ \boldsymbol{K}_{NN} $). From Eqs. \eqref{Eq209} and \eqref{Eq210}, observe that the spatio-temporal inducing variables $ \boldsymbol{u} $ are constructed from two different types of inducing input locations. The first type is the spatial inducing input locations, for example, $ s'_1, s'_2 $, and $ s'_3 $ depicted in Fig. \ref{fig:Figure9}. The second type is the temporal inducing input locations denoted by $ t'_p $ and $ t'_q $ in Eqs. \eqref{Eq209} and \eqref{Eq210}. The two different types of inducing input locations are then connected through the assumed separable (multiplicative) covariance to form the spatio-temporal inducing input locations. Since there is a spatial inducing input location defined for each physical river/stream network location, the spatial inducing input locations, as defined in this work, can not serve as a mechanism to induce computational efficiency. However, the temporal inducing input locations can be used as a mechanism to induce computational efficiency.

With the covariance matrices $ \boldsymbol{K}_{NN} $ (Sect. \ref{Section6.1}) and $ \boldsymbol{K}_{MM} $ (Sect. \ref{Section6.2}) specified, the only matrices that remain unspecified are the cross-covariance matrices $ \boldsymbol{K}_{NM} $ and $ \boldsymbol{K}_{MN} = \boldsymbol{K}_{NM}^T $. The spatial component of the cross-covariance $ \mathbb{C} $ov$[u_{a}(s'_{d,m}, t'_p),f_{b}(s_{i,n}, t_q)]$ can be computed for flow-connected sites with $ s_{i,n} > s'_{d,m} $ as follows

\vspace{-0.15 cm}

\begin{equation}\label{Eq212}
\begin{gathered}
    C_{u_a, f_b}^{(s)} (s'_{d,m},s_{i,n}) = 
    \mathlarger{\mathlarger{\sum}}_{j \in U_{s_{i,n}} } \mathlarger{\Biggl(} 
    \Biggl[ \mathlarger{\mathlarger{\prod}}_{\text{ } k' \in B_{s'_{d,m},[j]}}  \Phi(\alpha_{k'}) \Biggr] 
    \Biggl[ \mathlarger{\mathlarger{\prod}}_{\text{ } k \in B_{s_{i,n},[j]}}  \Phi(\gamma_k) \Biggr] 
    \int\limits_{l_j}^{u_j} g_a'(x_j - s'_{d,m})g_b(x_j - s_{i,n})  dx_j \mathlarger{\Biggr)} 
    \end{gathered}
\end{equation}

\vspace{-0.15 cm}

The spatial locations $ s_{i,n} $ and $ s'_{d,m} $ can not swap across the underlying latent and variational inducing functions as the locations are unique to the respective functions. In Eq. \eqref{Eq212} the downstream location $ s'_{d,m} $ was associated with $ u_a $. However, the downstream location can also be associated with the underlying latent function, for example, $ {f}_b $.  For the scenario where the physically sampled downstream location $ s_{d,m} $ is associated with the underlying latent function and the upstream spatial inducing input location $ s'_{i,n} $ is associated with the variational inducing function, the spatial component of the cross-covariance $ \mathbb{C}\text{ov}[u_a(s'_{i,n}, t'_p),f_b(s_{d,m}, t_q)]  $ can be computed as follows

\vspace{-0.15 cm}

\begin{equation}\label{Eq213}
\begin{gathered}
    C_{u_a, f_b}^{(s)} (s'_{i,n},s_{d,m}) = 
    \mathlarger{\mathlarger{\sum}}_{j \in U_{s'_{i,n}} } \mathlarger{\Biggl(} \Biggl[ \mathlarger{\mathlarger{\prod}}_{\ k' \in B_{s'_{i,n},[j]}}  \Phi(\alpha_k') \Biggr] 
    \Biggl[ \mathlarger{\mathlarger{\prod}}_{\ k \in B_{s_{d,m},[j]}}  \Phi(\gamma_{k}) \Biggr] 
    \int\limits_{l_j}^{u_j} 
    g'_a(x_j - s'_{i,n}) g_b(x_j - s_{d,m})  dx_j \mathlarger{\Biggr)} 
    \end{gathered}
\end{equation}

\vspace{-0.15 cm}

Finally, the general expression for the temporal cross-covariance component contribution can be computed as follows

\vspace{-0.45 cm}

\begin{equation}\label{Eq219}
\begin{gathered}
    C_{u_a,f_b}^{(t)} (t'_p, t_q \vert \boldsymbol{\theta}_a^{'(t)}, \boldsymbol{\theta}_b^{(t)}) 
    = \int\limits_{-\infty}^{\infty} 
    G'_a(t'_p - z \vert \boldsymbol{\theta}_a^{'(t)})
    G_b(t_q - z\vert \boldsymbol{\theta}_b^{(t)}) dz 
    \quad \forall \ a = 1,\cdots,K_u; 
    \quad \forall \ b = 1,\cdots,K_f
    \end{gathered}
\end{equation}

\vspace{-0.65 cm}

\subsection{Lower Bounding The Model Log Marginal Likelihood}{\label{Section4.1}}

\vspace{-0.15 cm}
With the specification of the joint Co-Kriging prior density and likelihood function (see Sect. \ref{SectionM}), as well as the uncertain input and inducing variable prior densities, it is now possible, using variational inference, to implement the BGP-LVM as a means of propagating the additional sources of input uncertainty. In the $ \mathcal{KL} $-divergence sense, variational inference aims to minimise the dissimilarity between the approximate variational posterior density and the actual underlying posterior density. In other words, the goal is to minimise

\vspace{-0.35 cm}

\begin{equation}\label{Eq45}
    \begin{gathered}
           \mathcal{KL}\bigl [q(\boldsymbol{f},\boldsymbol{u},\boldsymbol{\tau},\eta_\tau,\boldsymbol{\gamma}) \vert\vert p(\boldsymbol{f},\boldsymbol{u},\boldsymbol{\tau},\eta_\tau,\boldsymbol{\gamma} \vert \boldsymbol{y}) \bigr] = \int\limits_{\boldsymbol{f}} \int\limits_{\boldsymbol{u}} \int\limits_{\boldsymbol{\tau}} \int\limits_{\eta_\tau} \int\limits_{\boldsymbol{\gamma}} q(\boldsymbol{f},\boldsymbol{u},\boldsymbol{\tau},\eta_\tau,\boldsymbol{\gamma}) \ln\frac{q(\boldsymbol{f},\boldsymbol{u},\boldsymbol{\tau},\eta_\tau,\boldsymbol{\gamma})}{p(\boldsymbol{f},\boldsymbol{u},\boldsymbol{\tau},\eta_\tau,\boldsymbol{\gamma} \vert \boldsymbol{y})} d\boldsymbol{\gamma} d\eta_\tau d\boldsymbol{\tau} d\boldsymbol{u} d\boldsymbol{f}
    \end{gathered}       
\end{equation}       

\vspace{-0.15 cm}
 Using similar arguments to that outlined in Sect. \ref{Section3.2}, the variational lower bound $ F[q(\boldsymbol{f},\boldsymbol{u},\boldsymbol{\tau},\eta_\tau,\boldsymbol{\gamma})] $ can be computed as follows

\vspace{-0.40 cm}

\begin{equation}\label{Eq48}
    \begin{gathered}           
    \mathcal{F} \bigl [q(\boldsymbol{f},\boldsymbol{u},\boldsymbol{\tau},\eta_\tau,\boldsymbol{\gamma}) \bigr ] \text{ } = \int\limits_{\boldsymbol{f}} \int\limits_{\boldsymbol{u}} \int\limits_{\boldsymbol{\tau}} \int\limits_{\eta_\tau} \int\limits_{\boldsymbol{\gamma}} q(\boldsymbol{f},\boldsymbol{u},\boldsymbol{\tau},\eta_\tau,\boldsymbol{\gamma}) \ln\frac{p(\boldsymbol{y} \vert \boldsymbol{f})p(\boldsymbol{f} \vert \boldsymbol{u},\boldsymbol{\tau},\boldsymbol{\gamma})p(\boldsymbol{u})p(\boldsymbol{\tau} \vert \eta_\tau)p(\eta_\tau)p(\boldsymbol{\gamma})}{q(\boldsymbol{f},\boldsymbol{u},\boldsymbol{\tau},\eta_\tau,\boldsymbol{\gamma})} d\boldsymbol{\gamma} d\eta_\tau d\boldsymbol{\tau} d\boldsymbol{u} d\boldsymbol{f}
    \end{gathered}       
\end{equation} 

Following the ideas outlined in \cite{Titsias2008,Titsias2009}, \cite{Titsias2010}, and \cite{Damianou2016}, the following approximate variational posterior densities are selected

\vspace{-0.25 cm}

\begin{equation}\label{Eq49}    q(\boldsymbol{f},\boldsymbol{u},\boldsymbol{\tau},\eta_\tau,\boldsymbol{\gamma}) = p(\boldsymbol{f} \vert \boldsymbol{u},\boldsymbol{\tau},\boldsymbol{\gamma})q(\boldsymbol{u})q(\boldsymbol{\tau})q(\eta_\tau)q(\boldsymbol{\gamma})        
\end{equation} 

Note from Eq. \eqref{Eq49} that for the selected variational posterior densities, there are four free-form densities that can be optimised over since the density $ p(\boldsymbol{f} \vert \boldsymbol{u},\boldsymbol{\tau},\boldsymbol{\gamma}) $ corresponds to the conditional GP prior from the augmented model (see Eq. \eqref{Eq140} and \cite{Titsias2009} for more details). Using Eq. \eqref{Eq49}, the variational lower bound in Eq. \eqref{Eq48} can be expanded and rewritten to obtain 

\vspace{-0.35 cm}

\begin{equation}\label{Eq52}
    \begin{gathered}           
    \mathcal{F} \bigl[ q(\boldsymbol{u})q(\boldsymbol{\tau})q(\eta_\tau)q(\boldsymbol{\gamma}) \bigr] \text{ } = \text{ } \int\limits_{\boldsymbol{u}} q(\boldsymbol{u}) \biggl\{ \Omega\Bigl(\boldsymbol{u},q(\boldsymbol{\tau}),q(\boldsymbol{\gamma})\Bigr) \text{ } + \text{ } \mathbb{E}_{q(\eta_\tau)}\Bigl[-\mathcal{KL} \bigl[q(\boldsymbol{\tau}) \vert \vert p(\boldsymbol{\tau} \vert \eta_\tau) \bigr] \Bigr]  \text{ } \\[1.5ex] - \text{ } \mathcal{KL} \bigl[q(\boldsymbol{\gamma}) \vert \vert p(\boldsymbol{\gamma}) \bigr]  \text{ } - \text{ } \mathcal{KL} \bigl [q(\eta_\tau) \vert \vert p(\eta_\tau) \bigr] \text{ } + \text{ } \ln\frac{p(\boldsymbol{u})}{q(\boldsymbol{u})} \biggr\} d\boldsymbol{u}
    \end{gathered}       
\end{equation} 

The symbol $ \mathbb{E}_{q(\cdot)} $ denotes taking the expectation under the variational density $ q(\cdot) $ where $ \mathcal{KL}[q(\cdot) \vert \vert p(\cdot)] $ requires computing the $ \mathcal{KL} $-divergence between $ q(\cdot) $ and the corresponding prior density $ p(\cdot) $. Note that the quantities denoted by $ \Omega(\boldsymbol{u},q(\boldsymbol{\tau}),q(\boldsymbol{\gamma})) $ and $ \Psi(\boldsymbol{u},\boldsymbol{\tau},\boldsymbol{\gamma}) $ are defined as follows

\vspace{-0.15 cm}

\begin{equation}\label{Eq53}
    \begin{gathered}           \Omega\Bigl(\boldsymbol{u},q(\boldsymbol{\tau}),q(\boldsymbol{\gamma})\Bigr) = \mathbb{E}_{q(\boldsymbol{\tau})} \Bigl[ \mathbb{E}_{q(\boldsymbol{\gamma})} \bigl[ \Psi(\boldsymbol{u},\boldsymbol{\tau},\boldsymbol{\gamma}) \bigr] \Bigr] = \int\limits_{\boldsymbol{\tau}} q(\boldsymbol{\tau}) \biggl[ \int\limits_{\boldsymbol{\gamma}} q(\boldsymbol{\gamma}) \Psi(\boldsymbol{u},\boldsymbol{\tau},\boldsymbol{\gamma}) d\boldsymbol{\gamma} \biggr] d\boldsymbol{\tau}
    \end{gathered}       
\end{equation} 

\begin{equation}\label{Eq54}
    \begin{gathered}           
    \Psi(\boldsymbol{u},\boldsymbol{\tau},\boldsymbol{\gamma}) = \mathbb{E}_{p(\boldsymbol{f} \vert \boldsymbol{u},\boldsymbol{\tau},\boldsymbol{\gamma})} \Bigl[ \ln p(\boldsymbol{y} \vert \boldsymbol{f}) \Bigr] = \int\limits_{\boldsymbol{f}} p(\boldsymbol{f} \vert \boldsymbol{u},\boldsymbol{\tau},\boldsymbol{\gamma}) \ln p(\boldsymbol{y} \vert \boldsymbol{f})  d\boldsymbol{f}
    \end{gathered}       
\end{equation} 

\vspace{-0.25 cm}

Recall that it is important to select an approximate posterior density family that is flexible enough to model the true underlying posterior density while simultaneously resulting in a computationally efficient optimisation routine. In several applications, it is common to select the variational family to contain densities that factorise into subgroups of random variables, such as the variational densities associated with $ \boldsymbol{u},\boldsymbol{\tau}, \eta_\tau $, and $ \boldsymbol{\gamma} $ given by Eq. \eqref{Eq49}, or factorise completely in all of the random variables, or a combination of both. Consequently, variational methods based on these factorised approximations, unfortunately, ignore correlations between random variables that may be crucial in the process of learning model parameters (\citealp{Beal2003}; \citealp{Winn2004}; \citealp{Winn2005}; \citealp{Opper2009}). 

In this paper, the authors assume that the variational posterior densities associated with the latent variables $ \boldsymbol{\tau} $ and $ \boldsymbol{\gamma} $ factorise completely (i.e., a further independence assumption is made) whereas the density associated with $ \boldsymbol{u} $, for which an optimal functional form will be derived in subsequent sections (see Sect. \ref{Section4.4}), remains unspecified. In other words, Eq. \eqref{Eq49} can be rewritten as  

\vspace{-0.35 cm}

\begin{equation}\label{Eq63}    q(\boldsymbol{f},\boldsymbol{u},\boldsymbol{\tau},\eta_\tau,\boldsymbol{\gamma}) = p(\boldsymbol{f} \vert \boldsymbol{u},\boldsymbol{\tau},\boldsymbol{\gamma})q(\boldsymbol{u})\underbrace{\prod_{j}^{N_\tau}q(\tau_j)}_{q(\boldsymbol{\tau})}q(\eta_\tau)\underbrace{\prod_{k}^{N_\gamma}q(\gamma_k)}_{q(\boldsymbol{\gamma})}        
\end{equation} 

In Eq. \eqref{Eq63}, $ N_\tau $ is the number of $ \tau_j $ latent variables and $ N_\gamma $ is the number of $ \gamma_k $ latent variables. 
Next, the following functional forms are assumed for the variational posterior densities associated with $ \boldsymbol{\tau}, \eta_\tau $, and $ \boldsymbol{\gamma} $, respectively.

\vspace{-0.25 cm}

\begin{equation}\label{Eq67}         
   q(\tau_j) = \mathcal{N} \bigl(\tau_j \vert \mu_{\tau_j},\sigma^2_{\tau_j} \bigr) \;; \text{ } q(\eta_\tau) = \mathcal{N} \bigl(\eta_\tau \vert \mu_{\eta_\tau},\sigma^2_{\eta_\tau} \bigr)  \;; \text{ } q(\gamma_k) = \mathcal{N} \bigl(\gamma_k \vert \mu_{\gamma_k},\sigma^2_{\gamma_k} \bigr) 
\end{equation} 
The symbols $ \mu_{\tau_j}, \mu_{\eta_\tau} $, and $ \mu_{\gamma_k} $ denote the variational mean parameters whereas $ \sigma^2_{\tau_j}, \sigma^2_{\eta_\tau} $, and $ \sigma^2_{\gamma_k} $ denote the variational variance parameters. These variational parameters are selected such that the variational lower bound (Eq. \eqref{Eq52}) is maximised during gradient-based optimisation. 
With the specification of the latent variable prior and variational posterior densities, which all have an univariate Gaussian density functional form, the $ \mathcal{KL} $-divergence quantities associated with the variational lower bound in Eq. \eqref{Eq52} have closed-form solutions.

\vspace{-0.45 cm}

\subsection{Deriving The Optimal Variational Results}{\label{Section4.4}}

\vspace{-0.15 cm}
The first step to deriving the optimal density $ q(\boldsymbol{u}) $ is to note that $ q(\boldsymbol{u}) $ must be normalised. In other words, the optimal $ q(\boldsymbol{u}) $ is subject to the following integral constraint

\vspace{-0.35 cm}

\begin{equation}\label{Eq75}
    \int\limits_{\boldsymbol{u}} q(\boldsymbol{u}) d\boldsymbol{u} = 1 
\end{equation} 

Following the procedure outlined in \cite{Basson2023}, to analytically solve the constrained-based optimisation problem subject to the integral constraint given by Eq. \eqref{Eq75}, define the Lagrangian as follows (see \citealp{Logan2006} for more details) 

\vspace{-0.55 cm}

\begin{equation}\label{Eq76}
    \mathcal{L} \bigr[q(\boldsymbol{u}), \lambda \bigl] = q(\boldsymbol{u}) \Biggl[ \Lambda\Bigl(\boldsymbol{u},q(\boldsymbol{\tau}),q(\boldsymbol{\gamma})\Bigr) + \ln \frac{p(\boldsymbol{u})}{q(\boldsymbol{u})} \Biggr] + \lambda q(\boldsymbol{u})
\end{equation} 

The quantity $ \lambda $  denotes the Lagrange multiplier whereas $ \Lambda(\boldsymbol{u},q(\boldsymbol{\tau}),q(\boldsymbol{\gamma})) $, based on the variational lower bound in Eq. \eqref{Eq52}, is defined as follows

\vspace{-0.25 cm}

\begin{equation}\label{Eq77}
    \begin{gathered}         
\Lambda\Bigl(\boldsymbol{u},q(\boldsymbol{\tau}),q(\boldsymbol{\gamma})\Bigr) \text{ } = \text{ } \Omega\Bigl(\boldsymbol{u},q(\boldsymbol{\tau}),q(\boldsymbol{\gamma})\Bigr) \text{ } + \text{ } \mathbb{E}_{q(\eta_\tau)}\Bigl[-\mathcal{KL} \bigl[q(\boldsymbol{\tau}) \vert \vert p(\boldsymbol{\tau} \vert \eta_\tau) \bigr] \Bigr] \\[1.5ex] \text{ } \text{ } \text{ } \text{ } \text{ } \text{ } \text{ } \text{ } \text{ } - \text{ } \mathcal{KL} \bigl[q(\boldsymbol{\gamma}) \vert \vert p(\boldsymbol{\gamma}) \bigr] \text{ } - \text{ } \mathcal{KL} \bigl [q(\eta_\tau) \vert \vert p(\eta_\tau) \bigr]
    \end{gathered}       
\end{equation} 

From the Euler-Lagrange equation, the optimal variational density $ q(\boldsymbol{u}) $ that satisfies the stationarity condition corresponds to

\vspace{-0.45 cm}

\begin{equation}\label{Eq79}
    q(\boldsymbol{u}; \{ q(\boldsymbol{\tau}),q(\boldsymbol{\gamma}) \}) = \cfrac{p(\boldsymbol{u})\exp \Bigl\{\Omega\bigl(\boldsymbol{u},q(\boldsymbol{\tau}),q(\boldsymbol{\gamma})\bigr) \Bigr\} }{ \mathlarger{ \int\limits_{\boldsymbol{u}} } p(\boldsymbol{u})\exp \Bigl\{\Omega\bigl(\boldsymbol{u},q(\boldsymbol{\tau}),q(\boldsymbol{\gamma})\bigr) \Bigr\} d\boldsymbol{u}} 
\end{equation} 

Note that the additional arguments $ \{ q(\boldsymbol{\tau}),q(\boldsymbol{\gamma}) \} $ in Eq. \eqref{Eq79} denote that the optimal variational density $ q(\boldsymbol{u}; \{ \cdot \}) $ depends on being able to compute the analytical (i.e., closed-form) expectations with respect to the variational densities $ q(\boldsymbol{\tau}) $ and $ q(\boldsymbol{\gamma}) $ via the quantity $ \Omega\bigl(\boldsymbol{u},q(\boldsymbol{\tau}),q(\boldsymbol{\gamma})\bigr) $ which Eqs. \eqref{Eq53} and \eqref{Eq54} define. Back substituting Eq. \eqref{Eq79} into Eq. \eqref{Eq52} results in the following general solution for the collapsed variational lower bound

\vspace{-0.15 cm}

\begin{equation}\label{Eq80}
    \begin{gathered}           
    \mathcal{F}^\ast \bigl( \boldsymbol{\theta};\{ q(\boldsymbol{\tau}),q(\eta_\tau),q(\boldsymbol{\gamma}) \} \bigr) \text{ } = \text{ }  \ln \int\limits_{\boldsymbol{u}} p(\boldsymbol{u})\exp \Bigl\{\Omega\bigl(\boldsymbol{u},q(\boldsymbol{\tau}),q(\boldsymbol{\gamma})\bigr) \Bigr\} d\boldsymbol{u} \text{ } + \text{ } \\[1ex] \mathbb{E}_{q(\eta_\tau)}\Bigl[-\mathcal{KL} \bigl[q(\boldsymbol{\tau}) \vert \vert p(\boldsymbol{\tau} \vert \eta_\tau) \bigr] \Bigr] \text{ } - \text{ } \mathcal{KL} \bigl[q(\boldsymbol{\gamma}) \vert \vert p(\boldsymbol{\gamma}) \bigr] \text{ } - \text{ } \mathcal{KL} \bigl [q(\eta_\tau) \vert \vert p(\eta_\tau) \bigr]
    \end{gathered}       
\end{equation} 

Since $ \mathcal{F}^\ast ( \boldsymbol{\theta};\{ q(\boldsymbol{\tau}),q(\eta_\tau),q(\boldsymbol{\gamma}) \} )  $ lower bounds the model log marginal likelihood, a point estimate for the parameter vector $ \boldsymbol{\theta} $ can be obtained by maximising the variational lower bound using gradient-based optimisation (see Sects. \ref{Section2.1} and \ref{Section2.2}). 
For the mixed Co-Kriging likelihood function associated with censored, and potentially missing, observational data, Eqs. \eqref{Eq79} and \eqref{Eq80} are analytically intractable and not available in closed-form solution due to the presence of the cdf terms. However, analytical tractability can be induced by considering local variational methods which aim to locally lower (or upper) bound functions defined on groups or individual random variables (\citealp{Jaakkola1996}; \citealp{Jordan1999}; \citealp{Gibbs2000}; \citealp{Nickisch2008}; \citealp{Bishop2009}). Following the procedure outlined in \cite{Basson2023}, it is possible to derive a locally lower bounded mixed Co-Kriging likelihood function, denoted as $ p_l(\boldsymbol{y} \vert \boldsymbol{f} ; \{\boldsymbol{\zeta}, \boldsymbol{\sigma} \}) $, for censored stream network-based problems. Using the locally lower bounded likelihood function, Eqs. \eqref{Eq53}, \eqref{Eq54}, \eqref{Eq79}, and \eqref{Eq80} still holds true, however, $  \ln p(\boldsymbol{y} \vert \boldsymbol{f}) $ is replaced with $ \ln p_l(\boldsymbol{y} \vert \boldsymbol{f} ; \{\boldsymbol{\zeta}, \boldsymbol{\sigma} \}) $. In doing so, the use of the locally lower bounded mixed Co-Kriging likelihood function induces an analytically tractable secondary variational lower bound by noting that

\vspace{-0.25 cm}

\begin{equation}\label{Eq132}
    \ln p(\boldsymbol{y}) \geq \mathcal{F}^\ast \bigl( \boldsymbol{\theta};\{ q(\boldsymbol{\tau}),q(\eta_\tau),q(\boldsymbol{\gamma}) \} \bigr) \geq \mathcal{F}_l^\ast \bigl( \boldsymbol{\theta};\{ q(\boldsymbol{\tau}),q(\eta_\tau),q(\boldsymbol{\gamma}) \} ; \{\boldsymbol{\zeta}, \boldsymbol{\sigma} \} \bigr)
\end{equation}

\noindent The secondary variational lower bound itself can be computed as follows

\vspace{-0.15 cm}

\begin{equation}\label{Eq131}
    \begin{gathered}           
    \mathcal{F}_l^\ast \bigl( \boldsymbol{\theta};\{ q(\boldsymbol{\tau}),q(\eta_\tau),q(\boldsymbol{\gamma}) \} ; \{\boldsymbol{\zeta}, \boldsymbol{\sigma} \}\}\bigr) \text{ } = \text{ }  \ln \int\limits_{\boldsymbol{u}} p(\boldsymbol{u})\exp \Bigl\{\Omega_{l}\bigl(\boldsymbol{u},q(\boldsymbol{\tau}),q(\boldsymbol{\gamma}) ; \{\boldsymbol{\zeta}, \boldsymbol{\sigma} \}\bigr) \Bigr\} d\boldsymbol{u} \text{ } + \text{ } \\[1.5ex] \mathbb{E}_{q(\eta_\tau)}\Bigl[-\mathcal{KL} \bigl[q(\boldsymbol{\tau}) \vert \vert p(\boldsymbol{\tau} \vert \eta_\tau) \bigr] \Bigr] \text{ } - \text{ } \mathcal{KL} \bigl[q(\boldsymbol{\gamma}) \vert \vert p(\boldsymbol{\gamma}) \bigr] \text{ } - \text{ } \mathcal{KL} \bigl [q(\eta_\tau) \vert \vert p(\eta_\tau) \bigr]
    \end{gathered}       
\end{equation} 

The parameter vector $ \boldsymbol{\zeta} $ indicates that the locally lower bounded Co-Kriging likelihood function depends on these additional parameters which govern the tightness of the local lower bounds (see \citealp{Basson2023} for more details). Furthermore, the parameter vector $ \boldsymbol{\sigma} $ makes explicit the heteroskedastic-based parameterisation that was introduced in Eq. \eqref{Eq107}.

With the likelihood function construction procedure for the various data-generating mechanisms, as introduced in Sects. \ref{Specifying The Joint Co-Kriging Likelihood} and \ref{Section5.2}, and the latent/variational inducing function prior construction procedure from Sects. \ref{Section6.1} and \ref{Specifying The Uncertain Input Prior Densities}, in conjunction with the variational results, it is now possible to derive an analytical expression for the optimal inducing variable density $ q(\boldsymbol{u}) $ as well as a closed-form solution for the variational lower bound $ \mathcal{F}^{\ast}(\boldsymbol{\theta};\cdot) $. In this section, the optimal inducing variable density and secondary variational lower bound results will be provided for the case of censored observational data (see Sect. \ref{Section5.2}), however, following the same derivation procedure as outlined in the Supplementary Information (see Sect. SI.1), it will also be possible to derive the optimal variational results for the fully observed or missing observational data scenario - definitions for the vectors and matrices used below can also be obtained in Sect. SI.1. The variational results for the optimal inducing variable $ \boldsymbol{u} $ posterior density and the secondary variational lower bound, all subject to censored observational data, are given by Eqs. \eqref{Eq268} and \eqref{Eq280}, respectively. For $ q_l(\boldsymbol{u}; \{ q(\boldsymbol{\tau}),q(\boldsymbol{\gamma})\} ; \{\boldsymbol{\zeta}, \boldsymbol{\sigma} \}) $ it can be shown that 

\vspace{-0.25 cm}

\begin{equation}\label{Eq268}
\begin{gathered}
    q_l(\boldsymbol{u}; \{ q(\boldsymbol{\tau}),q(\boldsymbol{\gamma})\} ; \{\boldsymbol{\zeta}, \boldsymbol{\sigma} \}) = \mathcal{N}(\boldsymbol{u} \vert \boldsymbol{\mu}_{\boldsymbol{u}},\boldsymbol{\Sigma}_{\boldsymbol{u}})
    \end{gathered}
\end{equation} 

\noindent The covariance matrix $ \boldsymbol{\Sigma}_{\boldsymbol{u}} $ and mean vector $ \boldsymbol{\mu}_{\boldsymbol{u}} $ associated with $ q_l(\boldsymbol{u}; \{ q(\boldsymbol{\tau}),q(\boldsymbol{\gamma})\} ; \{\boldsymbol{\zeta}, \boldsymbol{\sigma} \}) $ can be computed as follows

\vspace{-0.25 cm}

\begin{equation}\label{Eq269}
\begin{gathered}
    \boldsymbol{\Sigma}_{\boldsymbol{u}} = \boldsymbol{K}_{MM}\boldsymbol{Q}^{-1}\boldsymbol{K}_{MM}
    \end{gathered}
\end{equation} 

\vspace{-0.30 cm}

\begin{equation}\label{Eq270}
\begin{gathered}
    \boldsymbol{\mu}_{\boldsymbol{u}} = \boldsymbol{\Sigma}_{\boldsymbol{u}}\boldsymbol{K}^{-1}_{MM}\overline{\boldsymbol{\Psi}}^T_1\boldsymbol{\Sigma}^{-1}_{l}\boldsymbol{y}_{l} 
    \end{gathered}
\end{equation} 

\vspace{0.15 cm}

\noindent The collapsed secondary variational lower bound corresponds to

\begin{equation}\label{Eq280}
    \begin{gathered}           
    \mathcal{F}_l^\ast \bigl( \boldsymbol{\theta};\{ q(\boldsymbol{\tau}),q(\eta_\tau),q(\boldsymbol{\gamma}) \} ; \{\boldsymbol{\zeta}, \boldsymbol{\sigma} \}\}\bigr) \text{ } 
    = \ln \mathlarger{\Biggl\{} \frac{\vert \boldsymbol{K}_{MM} \vert^{\frac{1}{2}}}{ (2\pi\sigma^2_{1})^{\frac{N_{1,[>q_1]}}{2}} (2\pi\sigma^2_{2})^{\frac{N_{2,[>q_2]}}{2}}  \vert \boldsymbol{Q} \vert^{\frac{1}{2}}} \exp \bigl\{\mathcal{A}_{\mathcal{F}_l^\ast} \bigr\} \mathlarger{\Biggr\}}
    \\[1.0ex]
    - \frac{{\overline{\psi}_0}}{2} + \frac{1}{2}\text{tr$ \Bigl\{ \boldsymbol{K}^{-1}_{MM}\overline{\boldsymbol{\Psi}}_2 \Bigr\} $} 
    -\mathlarger{\sum}_{j=1}^{N_{\tau}} \text{ } \mathcal{KL} \Bigl[\mathcal{N} \Bigl(\tau_j \vert \mu_{\tau_j},\sigma^2_{\tau_j} \Bigr) \Bigr\vert \Bigr\vert \mathcal{N} \Bigl(\tau_j \vert d_{\tau_j},\exp \Bigl\{\mu_{\eta_\tau} - \frac{\sigma^2_{\eta_\tau}}{2} \Bigr\} \Bigr) \Bigr] 
    \\[1.0ex]
    - \mathlarger{\sum}_{k=1}^{N_{\gamma}} \text{ } \mathcal{KL} \Bigl[\mathcal{N} \Bigl(\gamma_k \vert \mu_{\gamma_k},\sigma^2_{\gamma_k} \Bigr) \Bigr\vert \Bigr\vert \mathcal{N}(\gamma_k \vert d_{\gamma_k}, \sigma^2_{\gamma}) \Bigr] 
    - \mathcal{KL} \Bigl [\mathcal{N} \bigl(\eta_\tau \vert \mu_{\eta_\tau},\sigma^2_{\eta_\tau} \bigr)  \Bigr\vert \Bigr\vert \mathcal{N} \bigl(\eta_\tau \vert d_{\eta_{\tau{o}}},\sigma^2_{\eta_{\tau{o}}} \bigr) \Bigr] \\
    \mathcal{A}_{\mathcal{F}_l^\ast} = - \frac{1}{2}\boldsymbol{y}^T_{l}\boldsymbol{A}\boldsymbol{y}_{l} + \Biggl[\frac{1}{2}\boldsymbol{b}^T\boldsymbol{\Sigma}^{-1}_c\boldsymbol{b} + \boldsymbol{c}^T\boldsymbol{\Sigma}^{-1}_c\boldsymbol{1}^{\ast} - \boldsymbol{b}^T\boldsymbol{\Sigma}^{-1}_c\boldsymbol{d} \Biggr]
\end{gathered}       
\end{equation}

\noindent Another common step after deriving the variational lower bound is obtaining the latent function predictive equations given the approximate variational posterior densities. The authors have deferred this derivation to Sect. SI. 2 in the Supplementary Information.

\vspace{-0.45 cm}
\section{Simulation-Based Experiments}\label{Section9}

To demonstrate the BGP-LVM framework for river/stream networks, consider its application to two synthetically produced case studies that have been developed for the river/stream network configuration depicted in Fig \ref{fig:Figure9}. The case studies will illustrate both the benefits and the limitations associated with the developed BGP-LVM framework for river/stream networks.

\vspace{-0.55 cm}

\subsection{Synthetic Data: Case Study 1}\label{Section9.2}

\vspace{-0.10 cm}

The first simulation-based case study focuses on a quantitative assessment of the multi-output BGP-LVM (MO-BGP-LVM) for river/stream networks, relative to the competing benchmarks, in the absence of any missing and censored observational data. The benchmarks considered in this case study include \{1\} the exact/deterministic input multi-output spatio-temporal GPR framework (Exact-GPR), \{2\} the measured/estimated input multi-output spatio-temporal GPR framework (Uncertain-GPR), and \{3\} the independent BGP-LVM (In-BGP-LVM) framework for river/stream networks. The independent BGP-LVM framework can be recovered from the multi-output BGP-LVM by setting the sub-cross-covariance matrices
 \text{ }to be the zero-matrix. 

To implement the various multi-output spatio-temporal GPR frameworks, the unified model representation from Sect. \ref{Section2.2}, in conjunction with the latent function log-based transformation, can be used. Recall from Sect. \ref{Section2.2} that throughout this work the authors assumed a zero-mean centred latent function GP prior. The covariance matrix $ \boldsymbol{K}_{NN} $, as associated with the zero-mean centred latent function GP prior, was constructed from the separable spatio-temporal covariance given by Eqs. \eqref{Eq192} and \eqref{Eq193}. For the Uncertain-GPR framework the measured/estimated inputs for $ \boldsymbol{\tau} $ and $ \boldsymbol{\gamma} $, as outlined in Table SI.2, were used to construct matrix $ \boldsymbol{K}_{NN} $. 
For the Exact-GPR framework the corresponding deterministic values for $ \boldsymbol{\tau} $ and $ \boldsymbol{\gamma} \text{ }  $
were used to construct matrix $ \boldsymbol{K}_{NN} $. For this case study, the Exact-GPR framework purely serves as a mechanism to establish how predictions obtained from the remaining frameworks (with the uncertain inputs) hold up when compared to predictions obtained from the GPR framework where the covariance matrix $ \boldsymbol{K}_{NN} $ inputs, in other words, $ \boldsymbol{\tau} $ and $ \boldsymbol{\gamma} $, are known exactly.

To generate observational data, a single ground truth latent function sample was generated from the conditional GP prior (see Eq. \eqref{Eq143}) using the spatial and temporal moving-average function parametric values outlined in Table SI.1, as well as the deterministic input values for $ \boldsymbol{\tau} $ and $ \boldsymbol{\gamma} $ from Table SI.2. The ground truth latent function sample was then partitioned into the two sub-group latent functions, in other words, $ \boldsymbol{f} = [\boldsymbol{f}^T_{1},\boldsymbol{f}^T_{2}]^T $, for a total of $ K_f = 2 $ latent functions. 
The sub-group latent functions were then further partitioned, as a function of time, into three spatial latent function contributions following the river/stream network configuration depicted in Fig. \ref{fig:Figure9}. 
Each spatial partitioning consisted of a thousand latent function points, equally spaced over the temporal range $ 0 \leq t \leq 10 $, which serves as the ground truth latent function contribution at a specific spatial location, viewed as a function of time. To generate a single noise-corrupted observational data set,
each spatial latent function was sub-sampled at 50 temporal points equally spaced between $ 0 \leq t \leq 10 $,
followed by artificially corrupting the sub-sampled function values by adding zero-mean Gaussian distributed noise with a known standard deviation parameter corresponding to either $ \xi_{1} $ or $ \xi_{2} $ (see Table SI.2) depending on which latent function is considered. This process was repeated to generate a total of 100 noise-corrupted observational data sets. 
For the case study under consideration, the authors set the number of temporal inducing input locations to $ M_t = 20 $ (see Sect. SI.4 in the Supplementary Information for more details).

\begin{figure*}[h!tb]
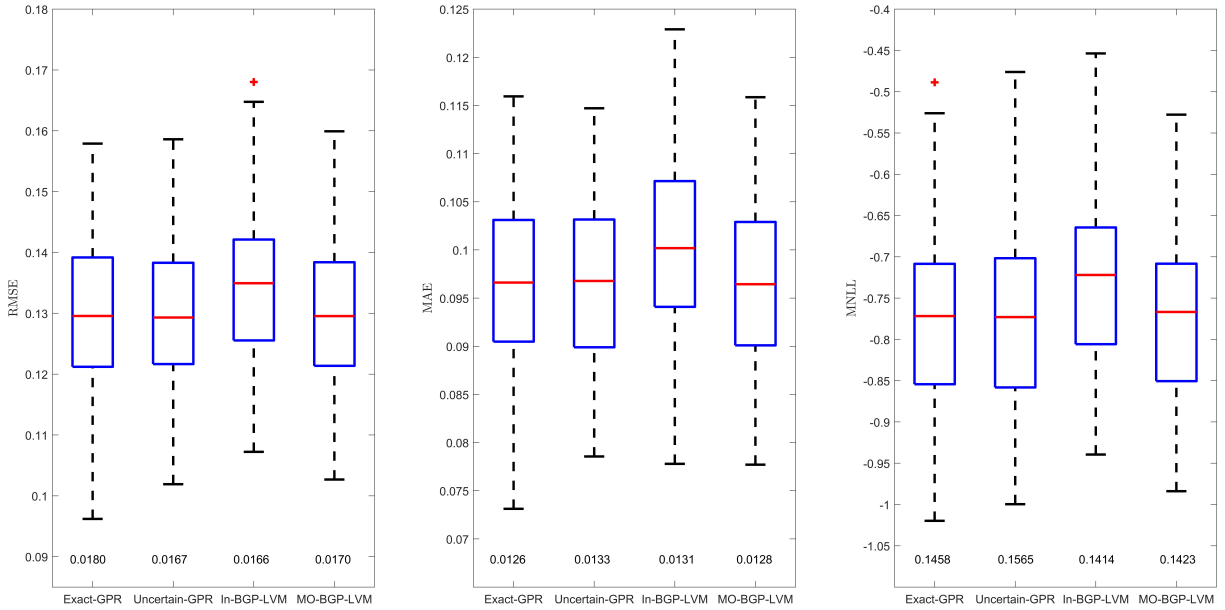

   \vspace{-0.25 cm}
   \hspace{-0.50in}  
   \sbox0{\includegraphics{Fig13.jpeg}}
   \includegraphics[scale = 0.044]{Fig13.jpeg}
   \caption{Box plot visualisation for the distribution of the RMSE (left panel), MAE (middle panel), and MNLL (right panel) performance metric results across the remaining 96 noise-corrupted observational data sets, respectively, for each framework considered. The interquartile range is denoted below the bottom whisker of each framework box plot. The red '+'-signs denote the outliers, as associated with the remaining 96 data sets, which are defined as values that are more than 1.5 times the interquartile range. \vspace{-0.45 cm}}
   \label{fig:Figure15}
\end{figure*}

The multi-output BGP-LVM, as well as the competing benchmarks, were trained on each of the 100 noise-corrupted observational data sets using gradient-based optimisation, with $ M_t = 20 $ and initialised to evenly spaced input points across the function temporal domain for the BGP-LVM frameworks, followed by computing the RMSE, MAE, and MNLL performance metrics (see Sect. SI. 3.4) for each framework and data set considered. 
Due to the non-convex nature of the model objective function, some optimisation routines failed to converge. This problem manifests itself as calculated RMSE, MAE, and MNLL values that are significantly larger when compared to the remaining computed values, skewing the overall performance metric results. To circumvent the skewed performance metric results, the authors identified each framework's RMSE, MAE, and MNLL outliers. Outliers were defined as performance metric results that are more than 1.5 times the interquartile range. 
The identified outliers were then consistently removed across all frameworks as a means to fairly compare the performance metric results. 
After removing the outliers, RMSE, MAE, and MNLL results for 96 of the 100 noise-corrupted observational data sets remained. Figure \ref{fig:Figure15} depicts and compares the distribution of the RMSE, MAE, and MNLL results for the remaining 96 observational data sets using box plots. 

Qualitatively, from the RMSE (left panel), MAE (middle panel), and MNLL (right panel) results in Fig. \ref{fig:Figure15}, there is no significant difference in the predictive performance results for the various frameworks across the 96 noise-corrupted observational data sets. This result is further emphasised when considering the mean RMSE, MAE, and MNLL results for each framework, as outlined in Table \ref{Table6}, across the 96 data sets. The mean performance metric results are practically the same in numeric value. Arguably, the independent BGP-LVM framework, the only framework where the latent functions are a priori independent, performs marginally worse. 

%\vspace{-0.25 cm}
\bgroup
\def\arraystretch{1.3} 
\begin{table}[h!tb]
    \centering
    \begin{tabular}{|c|l|c|c|c|}
    \cline{3-5}
         \multicolumn{2}{c|}{} & \multicolumn{3}{|c|}{\textbf{Performance Metric}}  \\ 
         \cline{3-5}
         \multicolumn{2}{c|}{} & Mean RMSE & Mean MAE & Mean MNLL \\ \hline
         \parbox[t]{2mm}{\multirow{4}{*}{\rotatebox[origin=c]{90}{\textbf{Framework}}}}  & Exact-GPR & 0.1295 & 0.0966 & {-0.7736} \\ \cline{2-5}
         & Uncertain-GPR & 0.1295  & 0.0966 & -0.7735 \\ \cline{2-5}
         & In-BGP-LVM & 0.1350 & 0.1009 & -0.7223 \\ \cline{2-5}
         & MO-BGP-LVM & {0.1293} & {0.0965} & -0.7733 \\ \hline
    \end{tabular}
    \caption{Summary of the mean RMSE, MAE, and MNLL performance metric results for each framework considered in the first simulation-based case study across the remaining 96 noise-corrupted observational data sets.}
    \label{Table6}
\end{table}

The authors expected the Exact-GPR model, which uses the true $\boldsymbol{\tau}$ and $\boldsymbol{\gamma}$ values, to outperform the Uncertain-GPR model, which uses measured/estimated $\boldsymbol{\tau}$ and $\boldsymbol{\gamma}$ values, but this was not the case.
One potential explanation for this unexpected behaviour is that the Uncertain-GPR model might not be not sensitive to the magnitude of the measured/estimated values for $ \boldsymbol{\tau} $ and $ \boldsymbol{\gamma} $, as outlined in Table SI.2, for the case study under consideration. 
A second reason might be the simplicity of the river/stream network configuration used in the case study. The river/stream network configuration depicted in Fig. \ref{fig:Figure9} is not particularly large in terms of the selected network branching structure, and only 5 measured/estimated inputs are required to construct the Uncertain-GPR model covariance matrix $ \boldsymbol{K}_{NN} $. Consequently, the selected small river/stream network configuration can result in the Uncertain-GPR model performing similarly to the Exact-GPR model, especially if the Uncertain-GPR model is already insensitive to the magnitude of the measured/estimated values for $ \boldsymbol{\tau} $ and $ \boldsymbol{\gamma} $. 

When comparing the RMSE, MAE, and MNLL performance metric results for the MO-BGP-LVM and both the Exact and Uncertain-GPR frameworks, observe that all frameworks have very similar performance metric results. This indicates that the MO-BGP-LVM framework infers underlying latent function results that are comparable to the results obtained from the simpler frameworks. 
One reason to prefer the MO-BGP-LVM framework is that, amongst the four options considered, it most closely follows the Bayesian philosophy to model all the known sources of uncertainty. 
From a more practical point of view, the MO-BGP-LVM also has computational run time benefits over the simpler GPR approaches.  To empirically demonstrate this, the authors trained both the Exact-GPR and the MO-BGP-LVM frameworks on data sets of various sizes. First, 15 data sets of increasing size were generated.  
To capture the sensitivity of the run-times to the initialisation of the parameters, different starting values were generated by first generating 10 unique random number generator seed values and then, for each seed, generating a batch of 40 random parameter starting points. 
The MO-BGP-LVM framework was then trained on each combination of data set and random parameter starting values. A similar process was carried out for the Exact-GPR framework. 
The optimisation run time for those starting points that converged was then averaged within each of the 10 batches to obtain 10 average computational run times per data set size and framework.

Figure \ref{fig:Figure16} shows the average optimisation run time for the MO-BGP-LVM (blue) and Exact-GPR (black) frameworks, across the replicates per selected data set size, together with one standard deviation error bars. Figure \ref{fig:Figure16} empirically demonstrates that as the data set size increases, the Exact-GPR framework becomes more computationally demanding despite having significantly fewer model parameters to optimise over. Recall from Sect. \ref{Section2.1} that the GPR framework requires inverting an $ N \times N $ covariance matrix $ \boldsymbol{\Sigma}_K $, where $ N $ corresponds to the total number of stream network observations, during the gradient-based optimisation routine. The (numerical) inversion process can become prohibitively slow since the GPR model time complexity scales as $ \mathcal{O}(N^3) $, rendering the GPR framework computationally intractable for larger data sets. However, the MO-BGP-LVM framework requires inverting a smaller $ M \times M $ covariance matrix $ \boldsymbol{K}_{MM} $, which facilitates computational speedups for larger data sets  while maintaining prediction accuracy. 
\begin{figure*}[h!tb]
   \hspace{-1in}  
   \sbox0{\includegraphics{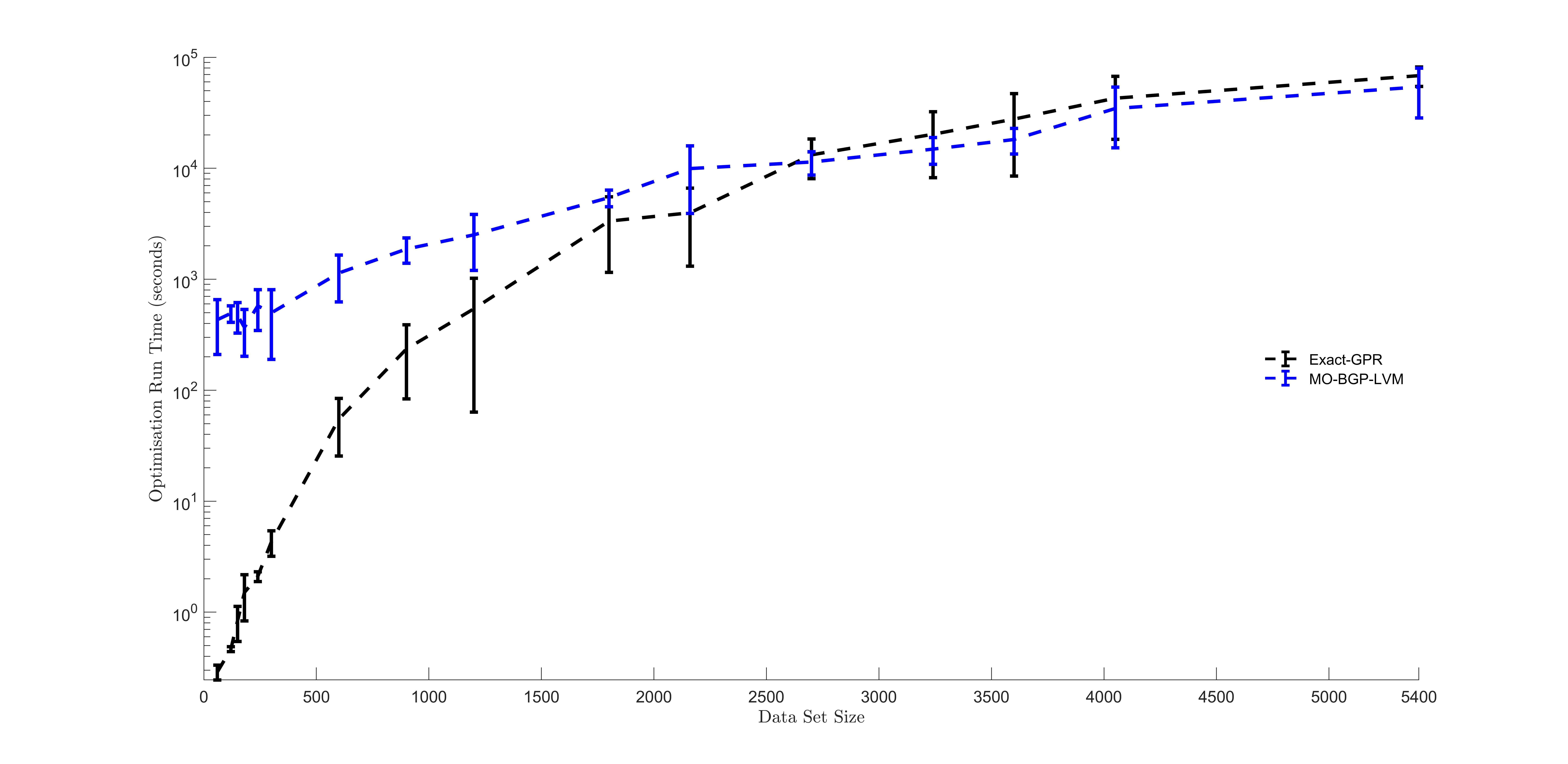}}
   \includegraphics[scale = 0.060]{Fig15.jpeg}
   \vspace{-0.65 cm}
   \caption{Average optimisation run time for \{1\} the MO-BGP-LVM (blue) and \{2\} the Exact-GPR (black) frameworks. The curves depict and compare the average optimisation run time for the two frameworks considered, across the 10 replicates, together with one standard deviation error bars. \vspace{- 0.35 cm}}
   \label{fig:Figure16}
\end{figure*}

Consequently, for the simulation-based case study under consideration, despite the MO-BGP-LVM yielding similar latent function prediction results, for larger data sets there is a clear computational advantage associated with the MO-BGP-LVM framework relative to the competing GPR framework. 

\vspace{-0.5 cm}

\subsection{Synthetic Data: Case Study 2}\label{Section9.4}

\vspace{-0.05 cm}

The second simulation-based case study focuses on the quantitative assessment of the multi-output BGP-LVM framework (MO-BGP-LVM) for river/stream networks for data with censored and missing values. The same ground truth latent function and spatial partitioning procedure from Case Study 1 was used to generate a total of 100 noise-corrupted observational data sets. For each data set, censoring was artificially introduced, with censoring and detection limits for $ \boldsymbol{f}_{1} $ set to the $25^{\text{th}}$ and the $15^{\text{th}}$ percentile, respectively. 
Similarly, for $ \boldsymbol{f}_{2} $, the censoring and detection limits  were set to the $35^{\text{th}}$ and $20^{\text{th}}$ percentile, respectively. This results in a data set that contains the remaining noise-corrupted latent function values as well as the censored observational data. For each spatial partitioned data set, across the 100 uniquely generated observational data sets, the authors, at random, removed samples to simulate missing observational data. The number of removed samples, per generated data set, is outlined in Table \ref{Table7}. This process results in 100 spatial partitioned data sets that consisted of noise-corrupted and censored observational data subject to missing values. Recall that the BGP-LVM frameworks also require the practitioner to specify the number of temporal inducing input locations $ M_t $. Using the same procedure as discussed in Sect. SI.4, $ M_t = 20 $ temporal inducing input locations was sufficient for a tight secondary variational lower bound.

%\vspace{-0.25 cm}
\bgroup
\def\arraystretch{1.5} 
\begin{table}[ht!]
    \centering
\begin{tabular}{c|ccc|}
\cline{2-4}
                                                                                   & \multicolumn{3}{c|}{\textbf{Sampled Spatial Locations}}                                                                           \\ \cline{2-4} 
                                                                                   & \multicolumn{1}{c|}{$ s_1 $}                     & \multicolumn{1}{c|}{$ s_2 $}                     & $ s_3 $                     \\ \hline
\multicolumn{1}{|c|}{\multirow{2}{*}{Fully Observed}}                              & \multicolumn{1}{c|}{$ \boldsymbol{f}_{1} $ - 32} & \multicolumn{1}{c|}{$ \boldsymbol{f}_{1} $ - 32} & $ \boldsymbol{f}_{1} $ - 32 \\ \cline{2-4} 
\multicolumn{1}{|c|}{}                                                             & \multicolumn{1}{c|}{$ \boldsymbol{f}_{2} $ - 28} & \multicolumn{1}{c|}{$ \boldsymbol{f}_{2} $ - 28} & $ \boldsymbol{f}_{2} $ - 28 \\ \hline
\multicolumn{1}{|c|}{\multirow{2}{*}{Between Quantification and Detection Limits}} & \multicolumn{1}{c|}{$ \boldsymbol{f}_{1} $ - 1}  & \multicolumn{1}{c|}{$ \boldsymbol{f}_{1} $ - 3}  & $ \boldsymbol{f}_{1} $ - 0  \\ \cline{2-4} 
\multicolumn{1}{|c|}{}                                                             & \multicolumn{1}{c|}{$ \boldsymbol{f}_{2} $ - 6}  & \multicolumn{1}{c|}{$ \boldsymbol{f}_{2} $ - 5}  & $ \boldsymbol{f}_{2} $ - 2  \\ \hline
\multicolumn{1}{|c|}{\multirow{2}{*}{Below Detection Limit}}                       & \multicolumn{1}{c|}{$ \boldsymbol{f}_{1} $ - 0}  & \multicolumn{1}{c|}{$ \boldsymbol{f}_{1} $ - 3}  & $ \boldsymbol{f}_{1} $ - 3  \\ \cline{2-4} 
\multicolumn{1}{|c|}{}                                                             & \multicolumn{1}{c|}{$ \boldsymbol{f}_{2} $ - 2}  & \multicolumn{1}{c|}{$ \boldsymbol{f}_{2} $ - 2}  & $ \boldsymbol{f}_{2} $ - 10 \\ \hline
\end{tabular}
    \caption{Summary of the number of samples that were removed at random, across the 100 generated data sets, as associated with underlying latent functions $ \boldsymbol{f}_{1} $ and $ \boldsymbol{f}_{2} $, to simulate noise-corrupted and censored observational data subject to missing values.}
    \label{Table7}
\end{table}

The MO-BGP-LVM, as well as the competing approaches, were trained on each of the 100 noise-corrupted and censored observational data sets with missing values, and the RMSE, MAE, and MNLL performance metrics were calculated for each framework and data set. 
Similar to before, the outliers associated with each framework's RMSE, MAE, and MNLL results were removed based on 1.5 times the interquartile range. 
Figure \ref{fig:Figure18} depicts and compares the distribution of the RMSE, MAE, and MNLL results for the remaining 83 data sets using box plots.  

Qualitatively, from the RMSE (left panel), MAE (middle panel), and MNLL (right panel) results, there is indeed a difference in the predictive performance results for the various frameworks. 
The RMSE results for the MO-BGP-LVM framework are better, relative to all three competing frameworks, though this improvement is small when considering the y-axis RMSE scale. When considering the MAE results in Fig. \ref{fig:Figure18}, the Exact-GPR and the Uncertain-GPR frameworks produce quite similar results, whereas the In-BGP-LVM framework MAE results are marginally better. The MAE results for the MO-BGP-LVM framework show a clear small improvement. 
The comparability of the results from the Exact-GPR, Uncertain-GPR, and In-BGP-LVM potentially indicates that explicitly modelling censored observational data and the additional sources of input uncertainty (as in the In-BGP-LVM) gives a competitive edge even in the absence of an explicit correlation structure between latent functions. The MO-BGP-LVM, which differs from the In-BGP-LVM framework only in the sense that the former jointly models the underlying latent functions $ \boldsymbol{f}_{1} $ and $ \boldsymbol{f}_{2}$, further improves on the RMSE and MAE results.

%\vspace{-0.75 cm}
\begin{figure*}[h!tb]
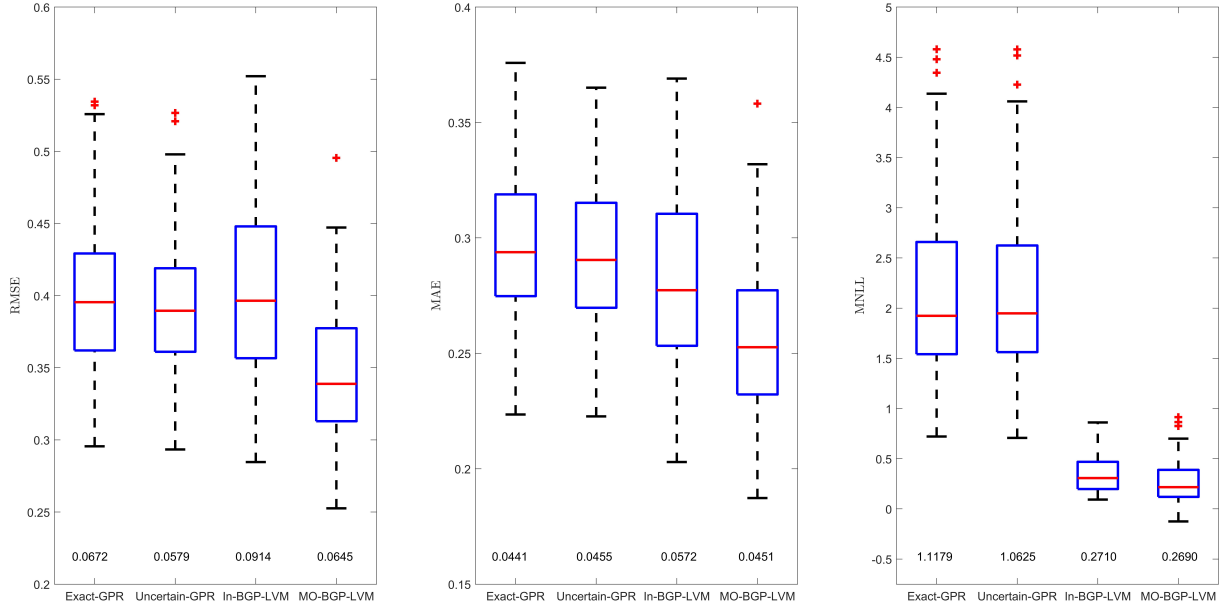

   \hspace{-0.50in}  
   \sbox0{\includegraphics{Fig17.jpeg}}
   \includegraphics[scale = 0.044]{Fig17.jpeg}
   \vspace{-0.55 cm}
   \caption{Box plot visualisation for the distribution of the RMSE (left panel), MAE (middle panel), and MNLL (right panel) performance metric results across the remaining 83 observational data sets, respectively, for each framework considered. The interquartile range is denoted below the bottom whisker of each framework box plot. The red '+'-signs denote the outliers, as associated with the remaining 83 data sets, which are defined as values that are more than 1.5 times the interquartile range. \vspace{-0.40 cm}}
   \label{fig:Figure18}
\end{figure*}

Further evidence for the superior performance of the BGP-LVM frameworks can be seen when considering the MNLL performance metric results in Fig. \ref{fig:Figure18}. From Fig. \ref{fig:Figure18}, the MNLL (right panel) results for both BGP-LVM frameworks are significantly better when compared to the GPR framework counterparts. From an inference-based perspective, the MNLL results indicate that the BGP-LVM frameworks provide a mechanism for improved predictive latent function uncertainty quantification. 
When considering the performance metrics as a whole, the MO-BGP-LVM framework clearly gives the best predictive performance results for noise-corrupted and censored observational data subject to missing values. The improved predictive performance associated with the MO-BGP-LVM framework is further emphasised when considering the mean metrics across the 83 remaining data sets, shown in Table \ref{Table8}, with bold values indicating the best-performing framework.

%\vspace{-0.25 cm}
\bgroup
\def\arraystretch{1.3} 
\begin{table}[h!tb]
    \centering
    \begin{tabular}{|c|l|c|c|c|}
    \cline{3-5}
         \multicolumn{2}{c|}{} & \multicolumn{3}{|c|}{\textbf{Performance Metric}}  \\ \cline{3-5}
         \multicolumn{2}{c|}{} & Mean RMSE & Mean MAE & Mean MNLL \\ \hline
         \parbox[t]{2mm}{\multirow{4}{*}{\rotatebox[origin=c]{90}{\textbf{Framework}}}}  & Exact-GPR & 0.4031 & 0.2993 & 2.1994 \\ \cline{2-5}
         & Uncertain-GPR & 0.3963  & 0.2938 & 2.1997 \\ \cline{2-5}
         & In-BGP-LVM & 0.4006 & 0.2846 & 0.3416 \\ \cline{2-5}
         & MO-BGP-LVM & \textbf{0.3443} & \textbf{0.2561} & \textbf{0.2818} \\ \hline
    \end{tabular}
    \caption{Summary of the mean RMSE, MAE, and MNLL performance metric results for each framework considered in the second simulation-based case study across the remaining 83 observational data sets. For each performance metric, the bold value indicates, on average, the best-performing framework.}
    \label{Table8}
\end{table}

\begin{figure*}[h!tb]
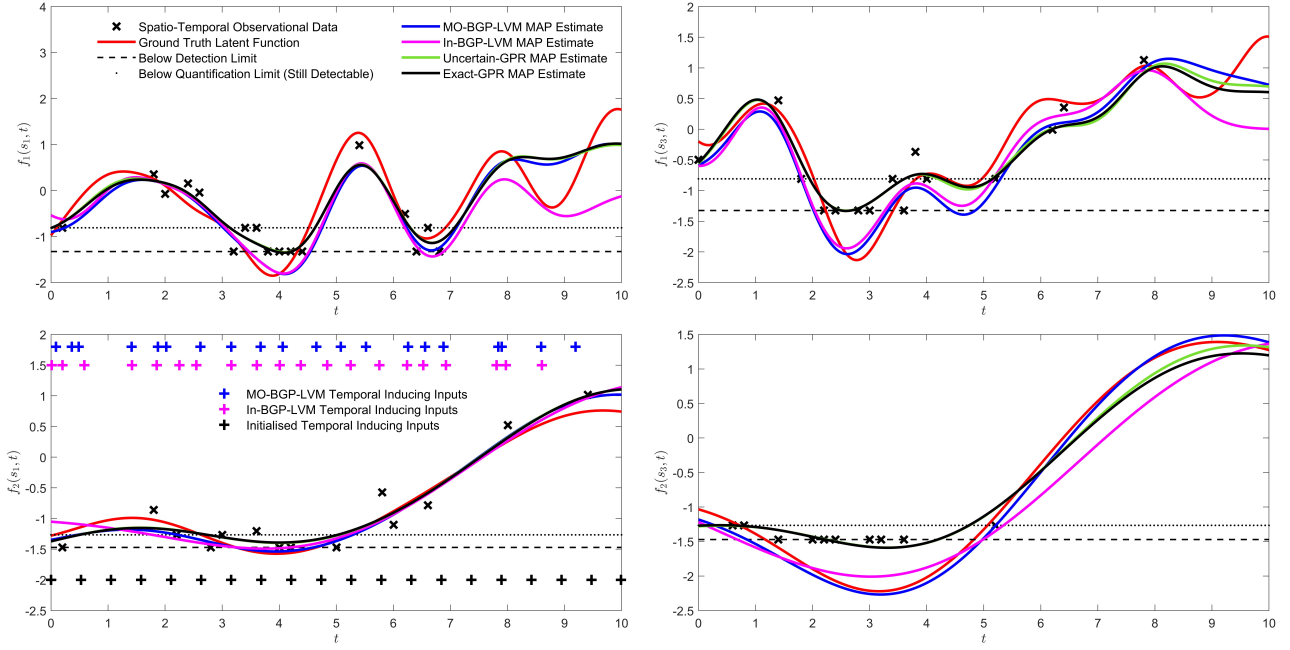

   \hspace{-0.70in}  
   \sbox0{\includegraphics{Fig18.jpeg}}
   \includegraphics[scale = 0.046]{Fig18.jpeg}
   \vspace{-0.65 cm}
   \caption{Latent function prediction results for the various frameworks considered in this case study, as pertaining to underlying latent functions $ \boldsymbol{f}_{1} $ (top row) and $ \boldsymbol{f}_{2} $ (bottom row), respectively, at sampled spatial locations $ s_1 $ (left panels) and $ s_3 $ (right panels), viewed as a function of time. The black 'x'-sign denotes the spatio-temporal observational data (noisy or censored). Note that the ground truth latent function is denoted in red whereas predictions obtained from \{1\} the MO-BGP-LVM, \{2\} the In-BGP-LVM, \{3\} the Uncertain-GPR, and \{4\} the Exact-GPR frameworks are depicted in dark blue, magenta, green, and black, respectively. The maximum a posteriori (MAP) estimate for each framework has been used for the qualitative comparative analysis. The black '+'-sign denotes the initial temporal inducing input locations, which were initialised to evenly spaced input points across the function temporal domain, whereas the dark blue and magenta '+'-signs denote the optimised temporal inducing input locations for the MO-BGP-LVM and In-BGP-LVM frameworks, respectively. Note that the temporal inducing input locations are common to all the latent function profiles. \vspace{-0.55 cm}}
   \label{fig:Figure19}
\end{figure*}

%\vspace{-0.25 cm}

To qualitatively investigate the predictive performance of the various frameworks considered in this case study, an additional random data set was generated. The MO-BGP-LVM, In-BGP-LVM, Uncertain-GPR, and Exact-GPR frameworks were trained on this data set and the latent function MAP estimate obtained from each framework (see Sect. SI. 2). These estimates and the ground truth latent function used to generate the observational data are depicted in Fig. \ref{fig:Figure19}. Qualitatively, the BGP-LVM frameworks are both able to infer a reasonable estimate for the underlying latent function in regions where the practitioner has access to censored observational data. However, 
the GPR framework counterparts seem to directly interpolate the censored observational data, which is undesirable from a latent function inference-based perspective.

Figure \ref{fig:Figure20} depicts the approximate variational posterior densities obtained from the MO-BGP-LVM (blue) and the In-BGP-LVM (magenta) frameworks for the additional randomly generated data set. The shared variance parameter $ \sigma^2_{\tau} $, hydrological distance, and weighting parameter posterior densities were constructed by inverting the alternative latent variable parameterisations introduced in Sect. \ref{Section4.1} (also see SI. 3.3). The black curves in the bottom row of Fig. \ref{fig:Figure20} depict the specified prior densities for the weighting parameters and the shared variance parameter that were used for both frameworks during the gradient-based optimisation procedure. Since a hyperprior was placed on $ \eta_{\tau} $, which governs the variance associated with $ p(\tau_i \vert \eta_{\tau}) $, the multiple black curves in the top row of Fig. \ref{fig:Figure20} depict sample prior densities that are based on 10 randomly generated samples for $ \eta_{\tau} $ from the specified prior density $ p(\eta_{\tau}) $ (last panel, bottom row). From the inferred approximate variational posterior densities depicted in Fig. \ref{fig:Figure20}, observe that both the MO-BGP-LVM and In-BGP-LVM frameworks produce overlapping results. This observation further emphasises that the improved MO-BGP-LVM framework RMSE, MAE, and MNLL performance metric results, as outlined in Fig. \ref{fig:Figure18}, can be attributed to the frameworks' capacity to jointly model the underlying latent functions $ \boldsymbol{f}_{1} $ and $ \boldsymbol{f}_{2} $.

\begin{figure*}[h!tb]
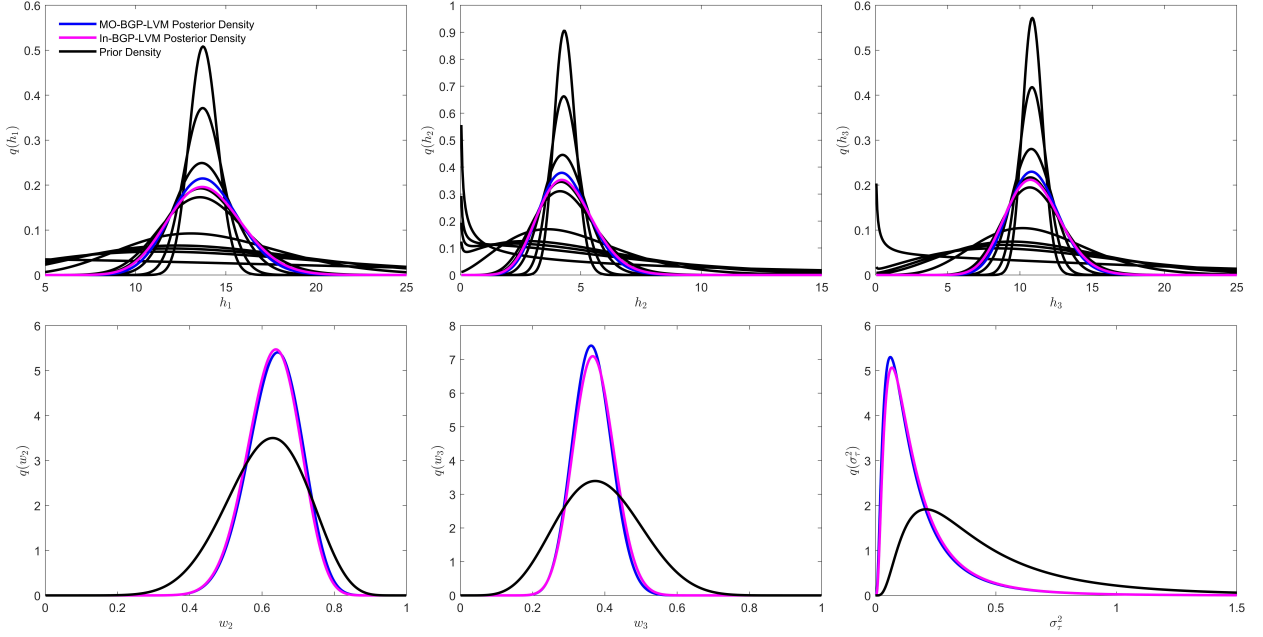

   \hspace{-0.70in}  
   \sbox0{\includegraphics{Fig19.jpeg}}
   \includegraphics[scale = 0.045]{Fig19.jpeg}
   \vspace{-0.65 cm}
   \caption{Approximate variational posterior densities for the MO-BGP-LVM (blue) and In-BGP-LVM (magenta) frameworks for the additional randomly generated data set. The top row depicts the approximate posterior densities associated with the hydrological distance $ h_i $ whereas the bottom row depicts the weighting parameter $ w_i $ (first two panels) and shared variance parameter (last panel) approximate posterior densities. The black curves in the bottom row depict the specified prior densities whereas the black curves in the top row depict 10 sample prior densities generated from the hyperprior over $ \eta_{\tau} $. \vspace{- 0.55 cm}}
   \label{fig:Figure20}
\end{figure*}

When considering the repeated experimental results (see Fig. \ref{fig:Figure18}), the average performance metric results outlined in Table \ref{Table8}, the MAP latent function results from Fig. \ref{fig:Figure19}, as well as the approximate variational posterior densities in Fig. \ref{fig:Figure20}, it is clear that the MO-BGP-LVM framework outperforms the remaining frameworks for noise-corrupted and censored observational data subject to missing values. Consequently, for the case study under consideration, it was beneficial to model the censored observational data and the additional sources of input uncertainty associated with $ \boldsymbol{\tau} $ and $ \boldsymbol{\gamma} $ within a correlated latent function structure, in other words, jointly modelling $ \boldsymbol{f}_{1} $ and $ \boldsymbol{f}_{2} $. Arguably, one can claim that, when considering all the results of the case study under consideration, there is not enough evidence to truly support the use of the MO-BGP-LVM over the In-BGP-LVM. Therefore, the practitioner can consider using the In-BGP-LVM framework instead of the MO-BGP-LVM framework since the mathematical derivation and implementation associated with the former is easier.

\vspace{-0.50 cm}

\section{Discussion And Model Limitations}\label{Section10}

\vspace{-0.15 cm}

In this paper, a variational inference-based framework for training a multi-output Gaussian process latent variable model, specifically tailored to the tails-up spatio-temporal river/stream network, was developed. The proposed framework relies on the variational sparse GP regression framework, local variational methods, and the Bayesian Gaussian process latent variable model which allows the authors to variationally integrate over the additional sources of input uncertainty associated with the measured/estimated hydrological distances and weighting parameters. 

The developed BGP-LVM framework for river/stream networks can also account for \{1\} the spatio-temporal evolution underpinning the passive downstream movement of materials like CECs or other waterborne chemicals in stream networks via a multi-output (i.e., Co-Kriging) dependency structure using separable kernel functions, \{2\} impose a positivity constraint on the CEC concentration profiles such that the inferred results are consistent with the physics underpinning the problem, and \{3\} account for limitations associated with censoring in spatio-temporal data sets. The authors demonstrated the proposed multi-output BGP-LVM framework for river/stream networks on synthetically produced data, subject to different data-generating mechanisms (noisy, missing, and censored) and found that the developed framework can produce improved latent function results when compared to the other benchmarks considered in the case studies. 

The aforementioned result was especially noticeable in the case where the practitioner had access to a noise-corrupted observational data set that was subject to missing and censored data entries (Case Study 2). In cases where the multi-output BGP-LVM framework produced comparable latent function prediction results to the competing benchmarks (Case Study 1), an empirical computational run time advantage was observed for the multi-output BGP-LVM framework when compared to the exact GPR model. However, given the marginal improvement in the RMSE, MAE, and MNLL performance metrics associated with the multi-output BGP-LVM framework, relative to the independent BGP-LVM framework, there is not enough evidence to truly support the use of the multi-output BGP-LVM framework over the independent BGP-LVM framework. Therefore, the practitioner can consider using the independent BGP-LVM framework instead of the multi-output BGP-LVM framework since the mathematical derivation and implementation associated with the former are easier than that of the latter.

One of the biggest limitations of the proposed BGP-LVM framework is its ability to scale to arbitrary river/stream network configurations. Due to the mathematical complexity associated with the tails-up construction procedure (especially when deriving and computing expected values of the cross-covariance matrix entries), and the propagation of input uncertainty through this tails-up construction procedure, it becomes practically impossible to derive the functional form of the vector/matrix quantities necessary to compute the variational lower bound (see Sect. SI. 1). 
To preserve the upstream construction procedure and stationarity of the variances during the gradient-based optimisation routine, it is necessary to derive and impose a variational analogue constraint for $ w_2 + w_3 = 1$. Further, additional constraints are required to `fix' the location of the spatial inducing input locations, which depend on the placement of the optimisable parameters associated with the uncertain spatial locations. For the stream network configuration considered in this work, the constraints were developed ‘by hand’ and checked by visually inspecting properties of the collapsed variational lower bound, but this is an inherently difficult problem to solve in full generality. New constraints are required every time (i) the river network configuration changes, (ii) a new spatial and/or temporal moving-average function is selected, (iii) the variational approximation is changed, or (iv) a new spatial inducing input configuration is explored. Therefore, further work to automate the derivation associated with the functional form of the covariance functions, the vector/matrix quantities, and the constraint necessary to implement the variational lower bound is needed to realise the full practical benefits of the proposed framework. Of course, practitioners can consider using symbolic integration and differentiation techniques to bypass some of the mathematical complexities. However, the authors do anticipate that the prohibitive step will be the human element required to come up with a suitable functional form of the spatial-based constraints. A potential remedy here can be to focus on one specific choice of moving-average function, derive the associated constraints and implement these into software for arbitrary stream network configurations.

Several other minor limitations also arise during the theoretical development of the variational lower bound. First, several quantities that appear in the lower bound calculations ($\overline{\boldsymbol{\Psi}}_0, \overline{\boldsymbol{\Psi}}_1 $, and $ \overline{\boldsymbol{\Psi}}_2 $ derived in Sect. SI.1) are only available in closed-form for specific covariance/cross-covariance functions. Practitioners can consider using numerical-based integration techniques for more complex covariance/cross-covariance functions at the expense of computational efficiency. Second, assuming that the latent function (marginal) predictive posterior density can be well approximated with the Gaussian distribution (see Sect. SI.2) may lead to predictive inaccuracies, especially if the true underlying latent function predictive density is skewed. Here, MCMC-based integration techniques could be considered as an alternative to gain access to the (marginal) latent function predictive posterior density.}

A final limitation stems from the use of the constant weighting parameters that are required to maintain the stationarity of the variances. From a practical and implementation perspective, it is, as a first-trial-attempt, reasonable to assume constant weighting parameters. However, the weighting parameters should ideally be computed using the flow rate contribution of each river/stream branch/segment. Consequently, there will be temporal, and potentially also spatial, variation associated with the weighting parameters. For example, seasonal variations in weather patterns (winter rain and summer drought) directly influence the flow of water in the river/stream branch/segment which, at least, implies temporal variation in the weighting parameters. Therefore, a more practical version of the proposed BGP-LVM framework would be one that allows for temporal variation in the weighting parameters. This would require introducing the notion of temporal stationarity of the variances, i.e., to maintain stationarity of the variances per temporal instance. Interestingly enough, practitioners can also considering using temporal variation in the weighting parameters to introduce non-stationarity into the covariance functions.

For future work, the authors would like to explore and address the limitations outlined above. Having the capacity to automate the proposed BGP-LVM construction procedure can greatly benefit practitioners as it will allow the proposed framework to be applied to any arbitrary river/stream network. Consequently, practitioners will be able to perform Co-Kriging for river/stream networks in a more computationally efficient manner while simultaneously propagating all the sources of input uncertainty and addressing censoring/missing data in spatio-temporal data sets. Allowing the weighting parameters to be a function of time would also align the proposed model more with the physics underpinning the flow characteristics of the river/stream network under consideration. Introducing temporal variation in the weighting parameters can also potentially improve the underlying latent function predictions as the water flow rate has a direct impact, both physically and in terms of the model construction procedure, on the passive downstream movement of materials like CECs or other waterborne chemicals in the river/stream network under consideration.

\vspace{-0.35 cm}
\begin{acknowledgements}
This work was supported in part by the Engineering and Physical Sciences Research Council
(EPSRC) (EP/P028403/1) and the School of Data Science and Computational Thinking (Stellenbosch University).
\end{acknowledgements}

\vspace{-0.65 cm}

\section*{Declarations}
\begin{itemize}
\item Competing Interests - The authors have no competing interests to declare relevant to this article's content.
\item Funding - This work was supported in part by the Engineering and Physical Sciences Research Council (EPSRC) (EP/P028403/1) and the School of Data Science and Computational Thinking (Stellenbosch University).
\item Ethics approval - Not Applicable
\item Consent to participate - Not Applicable
\item Consent for publication - All authors agreed with the content of this article and have given explicit consent for submission/publication
\item Availability of data and materials - No novel data was produced during this study.
\item Code availability - No code has been made publicly available.
\item Authors' contributions - See below.
\end{itemize}

\vspace{-0.55 cm}
\section*{Author Contributions}

Conceptualisation, Methodology, Formal analysis, and investigation, Writing - original draft preparation: Marno Basson; Writing - review and editing, Supervision: Tobias M. Louw and Theresa R. Smith; Funding acquisition: Tobias M. Louw

\vspace{-0.45 cm}
\bibliographystyle{MG}       
{\footnotesize
\bibliography{bibliography}}  

\newpage

\begin{center}
   \vspace*{3mm} {\LARGE Supplementary Material}
\end{center}
\vspace{1mm}

\setcounter{figure}{0}
\setcounter{table}{0}
\setcounter{equation}{0}
\setcounter{section}{0}
\setcounter{subsection}{0}

\renewcommand{\thesubsection}{SI.\arabic{section}.\arabic{subsection}}
\renewcommand{\thesection}{SI.\arabic{section}}
\renewcommand{\theequation}{SI-\arabic{equation}}
\renewcommand{\thetable}{SI.\arabic{table}}
\renewcommand{\thefigure}{SI.\arabic{figure}}

% hyperref anchor names
\renewcommand{\theHequation}{SI.\arabic{equation}}
\renewcommand{\theHtable}{SI.\arabic{table}}
\renewcommand{\theHfigure}{SI.\arabic{figure}}
\renewcommand{\theHsection}{SI.\arabic{section}}
\renewcommand{\theHsubsection}{SI.\arabic{section}.\arabic{subsection}}

\section{Derivation - Analytically Computing The Optimal Variational Results}\label{SI_1}

\subsection{Analytically Computing The Optimal Variational Inducing Variable Posterior Density}\label{SI_1_1}

For the case of censored observational data, the first step in deriving the analytical results associated with the optimal variational posterior density $ q_l(\boldsymbol{u}; \{ q(\boldsymbol{\tau}),q(\boldsymbol{\gamma})\} ; \{\boldsymbol{\zeta}, \boldsymbol{\sigma} \}) $ (see Eqs. (72) to (74)) would be to compute the quantity $ \Omega_{l}\bigl(\boldsymbol{u},q(\boldsymbol{\tau}),q(\boldsymbol{\gamma}) ; \{\boldsymbol{\zeta}, \boldsymbol{\sigma} \}\bigr)  $. Recall from Eq. (61), where $  \ln p(\boldsymbol{y} \vert \boldsymbol{f}) $ has been substituted with $ \ln p_l(\boldsymbol{y} \vert \boldsymbol{f} ; \{\boldsymbol{\zeta}, \boldsymbol{\sigma} \}) $, that

\begin{equation}\label{Eq222}
    \begin{gathered}           \Omega_{l}\Bigl(\boldsymbol{u},q(\boldsymbol{\tau}),q(\boldsymbol{\gamma}) ; \{\boldsymbol{\zeta}, \boldsymbol{\sigma} \} \Bigr) = \int\limits_{\boldsymbol{\tau}} q(\boldsymbol{\tau}) \biggl[ \int\limits_{\boldsymbol{\gamma}} q(\boldsymbol{\gamma}) \Psi_{l}(\boldsymbol{u},\boldsymbol{\tau},\boldsymbol{\gamma} ; \{\boldsymbol{\zeta}, \boldsymbol{\sigma} \}) d\boldsymbol{\gamma} \biggr] d\boldsymbol{\tau}
    \end{gathered}       
\end{equation} 

 Observe that $ \Omega_{l}\bigl(\boldsymbol{u},q(\boldsymbol{\tau}),q(\boldsymbol{\gamma}) ; \{\boldsymbol{\zeta}, \boldsymbol{\sigma} \}\bigr)  $ in Eq. \eqref{Eq222} itself requires the quantity $ \Psi_{l}(\boldsymbol{u},\boldsymbol{\tau},\boldsymbol{\gamma} ; \{\boldsymbol{\zeta}, \boldsymbol{\sigma} \}) $ which, from Eq. (62), is computed as follows

\begin{equation}\label{Eq223}
    \begin{gathered}           
    \Psi_{l}(\boldsymbol{u},\boldsymbol{\tau},\boldsymbol{\gamma} ; \{\boldsymbol{\zeta}, \boldsymbol{\sigma} \}) = \int\limits_{\boldsymbol{f}} p(\boldsymbol{f} \vert \boldsymbol{u},\boldsymbol{\tau},\boldsymbol{\gamma}) \ln p_l(\boldsymbol{y} \vert \boldsymbol{f} ; \{\boldsymbol{\zeta}, \boldsymbol{\sigma} \})  d\boldsymbol{f} = \\[1.5ex] \mathbb{E}_{p(\boldsymbol{f} \vert \boldsymbol{u},\boldsymbol{\tau},\boldsymbol{\gamma})} \bigl[\ln p_l(\boldsymbol{y} \vert \boldsymbol{f} ; \{\boldsymbol{\zeta}, \boldsymbol{\sigma} \}) \bigr]
    \end{gathered}       
\end{equation}

\noindent Following the notation and nomenclature of \cite{Basson2023}, the locally lower bounded mixed Co-Kriging log-likelihood function for censored observational data (see Sect. 3.3) can be expressed as follows

\begin{equation}\label{Eq224}
\begin{gathered}    \
    \ln p_l(\boldsymbol{y} \vert \boldsymbol{f} ; \{\boldsymbol{\zeta}, \boldsymbol{\sigma} \}) = \\[1.5ex] 
    \mathlarger{\mathlarger{\sum}}_{a = 1}^{K_f = 2} \text{ } \ln \mathlarger{\Biggl\{} \mathlarger{\prod_{s \in \sample_a^{(s)}}} 
    {\color{black}\Biggl [} \
    \mathlarger{\prod_{t \in \{\sample_{a,s}^{(t)} | y_a(s,t) > l_{q_a} \}}} \ \mathcal{N} \bigl(y_a(s,t) \vert f_{a}(s,t),\sigma_a^2 \bigr) \times \\[1.5ex] \mathlarger{\prod_{t \in \{\sample_{a,s}^{(t)} | y_a(s,t) = l_{q_a} \}}}  g \bigl(f_{a}(s,t) \vert \zeta_{q_ad_a}(s,t), l_{d_a}, l_{q_a},\sigma_a^2,\sigma_{q_ad_a}^2 \bigr) \times \\[1.5ex] \mathlarger{\prod_{t \in \{\sample_{a,s}^{(t)} | y_a(s,t) = l_{d_a} \}}} \text{ } g \bigl( f_{a}(s,t) \vert \zeta_{d_a}(s,t), l_{d_a},\sigma_a^2,\sigma_{d_a}^2 \bigr)  {\color{black}\Biggr]} \mathlarger{\Biggl\}} 
\end{gathered}  
%\hspace{0.5cm}\raisebox{-25mm}{\includegraphics[scale = 0.35]{Figure100.jpeg}}
\end{equation} 

The symbol $ g(\cdot) $ denotes the quadratic locally lower bounded functions that allow the authors to compute closed-form variational results for the case of censored observational data (see \citealp{Basson2023} for more details). Furthermore, note that the locally lower bounded mixed Co-Kriging log-likelihood function in Eq. \eqref{Eq224} has been constructed using the running example corresponding to $ K_f = 2 $ underlying latent functions with the assumed river/stream network configuration depicted in Figs. \ref{fig:Figure7} and 4, in other words, $ \sample_1^{(s)} = \sample_2^{(s)} = \{s_1,s_2,s_3\} $, and $\sample_{a,s}^{(t)}$ is the set of all sample times for component $a=1,2$ at sampled spatial locations $s = s_1, s_2, $ or $ s_3 $. For derivation purposes, it is also assumed that all sampled spatial locations are observed, however, this assumption can easily be relaxed. Due to the complexity associated with deriving the variational closed-form solutions, the authors find it easier and more instructive to solve the variational results using an explicit example. 
From Eq. \eqref{Eq223}, using the running example, the results from Eq. \eqref{Eq224}, and linearity of the expectation operator, it can be shown that

\begin{equation}\label{Eq225}
    \begin{gathered}           
    \Psi_{l}(\boldsymbol{u},\boldsymbol{\tau},\boldsymbol{\gamma} ; \{\boldsymbol{\zeta}, \boldsymbol{\sigma} \}) = \\[1.5ex] \mathlarger{\mathlarger{\sum}}_{a = 1}^{K_f = 2} \text{ } \mathbb{E}_{p(\boldsymbol{f} \vert \boldsymbol{u},\boldsymbol{\tau},\boldsymbol{\gamma})} \mathlarger{\Biggl[}\ln \mathlarger{\Biggl\{} \mathlarger{\prod_{s \in \sample_a^{(s)}}} {\color{black}\Biggl [} \text{ } \mathlarger{\prod_{t \in \{\sample_{a,s}^{(t)} | y_a(s,t) > l_{q_a} \}}} \text{ } \mathcal{N} \bigl(y_a(s,t) \vert f_{a}(s,t),\sigma_a^2 \bigr) \times \\[1.5ex] 
    \mathlarger{\prod_{t \in \{\sample_{a,s}^{(t)} | y_a(s,t) = l_{q_a} \}}} \text{ } g \bigl(f_{a}(s,t) \vert \zeta_{q_ad_a}(s,t), l_{d_a}, l_{q_a},\sigma_a^2,\sigma_{q_ad_a}^2 \bigr) \times \\[1.5ex] 
    \mathlarger{\prod_{t \in \{\sample_{a,s}^{(t)} | y_a(s,t) > l_{d_a} \}}} \text{ } g \bigl( f_{a}(s,t) \vert \zeta_{d_a}(s,t), l_{d_a},\sigma_a^2,\sigma_{d_a}^2 \bigr)  {\color{black}\Biggr]} \mathlarger{\Biggl\}} \mathlarger{\Biggr]}
    \end{gathered}       
\end{equation}

\noindent From Eq. \eqref{Eq225}, note that the quantity $ \Psi_{l}(\boldsymbol{u},\boldsymbol{\tau},\boldsymbol{\gamma} ; \{\boldsymbol{\zeta}, \boldsymbol{\sigma} \}) $ can be decomposed as a sum over underlying latent function contributions, with $ K_f = 2 $, such that

%\noindent Next, the sum over $ K_f = 2 $ in Eq. \eqref{Eq225} can be decomposed such that

\begin{equation}\label{Eq226}
    \begin{gathered}           
    \Psi_{l}(\boldsymbol{u},\boldsymbol{\tau},\boldsymbol{\gamma} ; \{\boldsymbol{\zeta}, \boldsymbol{\sigma} \}) = \underbrace{\Psi_{l_1}(\boldsymbol{u},\boldsymbol{\tau},\boldsymbol{\gamma} ; \{\boldsymbol{\zeta}, \boldsymbol{\sigma} \})}_{ a = 1} + \underbrace{\Psi_{l_2}(\boldsymbol{u},\boldsymbol{\tau},\boldsymbol{\gamma} ; \{\boldsymbol{\zeta}, \boldsymbol{\sigma} \})}_{a = 2}
    \end{gathered}       
\end{equation}

The factor $ \Psi_{l_1}(\boldsymbol{u},\boldsymbol{\tau},\boldsymbol{\gamma} ; \{\boldsymbol{\zeta}, \boldsymbol{\sigma} \}) $, as associated with underlying latent function $ {f}_{1} $ and Eq. \eqref{Eq225}, can be computed as follows

\begin{equation}\label{Eq227}
    \begin{gathered}           
    \Psi_{l_1}(\boldsymbol{u},\boldsymbol{\tau},\boldsymbol{\gamma} ; \{\boldsymbol{\zeta}, \boldsymbol{\sigma} \}) = \\[1.5ex] \mathbb{E}_{p(\boldsymbol{f} \vert \boldsymbol{u},\boldsymbol{\tau},\boldsymbol{\gamma})} \mathlarger{\Biggl[}\ln \mathlarger{\Biggl\{} \mathlarger{\prod_{s \in \sample_1^{(s)}}} {\color{black}\Biggl [} \text{ } 
    \mathlarger{\prod_{t \in \{\sample_{1,s}^{(t)} | y_1(s,t) > l_{q_1} \}}} \text{ } \mathcal{N} \bigl(y_{{1}}(s,t) \vert f_{{{1}}}(s,t),\sigma_{{1}}^2 \bigr) \times \\[1.5ex] 
    \mathlarger{\prod_{t \in \{\sample_{1,s}^{(t)} | y_1(s,t) = l_{q_1} \}}} \text{ } g \bigl(f_{{{1}}}(s,t) \vert \zeta_{q_1d_1}(s,t), l_{d_1}, l_{q_1},\sigma_{{1}}^2,\sigma_{q_1d_1}^2 \bigr) \times \\[1.5ex] 
    \mathlarger{\prod_{t \in \{\sample_{1,s}^{(t)} | y_1(s,t) = l_{d_1} \}}} \text{ } g \bigl( f_{{{1}}}(s,t) \vert \zeta_{d_1}(s,t), l_{d_1},\sigma_{{1}}^2,\sigma_{d_1}^2 \bigr)  {\color{black}\Biggr]} \mathlarger{\Biggl\}} \mathlarger{\Biggr]}
    \end{gathered}       
\end{equation}

\noindent A similar arguments holds for the factor $ \Psi_{l_2}(\boldsymbol{u},\boldsymbol{\tau},\boldsymbol{\gamma} ; \{\boldsymbol{\zeta}, \boldsymbol{\sigma} \}) $, as associated with underlying latent function $ {f}_{2} $, by setting the index $ i = 2 $. Next, observe from Eq. \eqref{Eq227}, and by linearity of the expectation operator, that

\begin{equation}\label{Eq228}
    \begin{gathered}           
    \Psi_{l_1}(\boldsymbol{u},\boldsymbol{\tau},\boldsymbol{\gamma} ; \{\boldsymbol{\zeta}, \boldsymbol{\sigma} \}) = \\[1.5ex] 
    \mathlarger{\sum_{s \in \sample_1^{(s)}}} \mathbb{E}_{p(\boldsymbol{f} \vert \boldsymbol{u},\boldsymbol{\tau},\boldsymbol{\gamma})} \Biggl[\ln {\color{black}\Biggl \{} \text{ } 
    \mathlarger{\prod_{t \in \{\sample_{1,s}^{(t)} | y_1(s,t) > l_{q_1} \}}} \text{ } \mathcal{N} \bigl(y_{{1}}(s,t) \vert f_{{{1}}}(s,t),\sigma_{{1}}^2 \bigr) \times \\[1.5ex] 
    \mathlarger{\prod_{t \in \{\sample_{1,s}^{(t)} | y_1(s,t) = l_{q_1} \}}} \text{ } g \bigl(f_{{{1}}}(s,t) \vert \zeta_{q_1d_1}(s,t), l_{d_1}, l_{q_1},\sigma_{{1}}^2,\sigma_{q_1d_1}^2 \bigr) \times \\[1.5ex] 
    \mathlarger{\prod_{t \in \{\sample_{1,s}^{(t)} | y_1(s,t) > l_{d_1} \}}} \text{ } g \bigl( f_{{{1}}}(s,t) \vert \zeta_{d_1}(s,t), l_{d_1},\sigma_{{1}}^2,\sigma_{d_1}^2 \bigr)  {\color{black}\Biggr\}} \Biggr] 
    \end{gathered}       
\end{equation}

\noindent From Eq. \eqref{Eq228}, observe that the quantity $ \Psi_{l_1}(\boldsymbol{u},\boldsymbol{\tau},\boldsymbol{\gamma} ; \{\boldsymbol{\zeta}, \boldsymbol{\sigma} \}) $ can be further decomposed as a sum over sampled spatial location contributions such that

\begin{equation}\label{Eq229}
    \begin{gathered}           
    \Psi_{l_1}(\boldsymbol{u},\boldsymbol{\tau},\boldsymbol{\gamma} ; \{\boldsymbol{\zeta}, \boldsymbol{\sigma} \}) = \underbrace{\Psi^{s_1}_{l_1}(\boldsymbol{u},\boldsymbol{\tau},\boldsymbol{\gamma} ; \{\boldsymbol{\zeta}, \boldsymbol{\sigma} \})}_{\text{sampled site $ s_1 $}} + \underbrace{\Psi^{s_2}_{l_1}(\boldsymbol{u},\boldsymbol{\tau},\boldsymbol{\gamma} ; \{\boldsymbol{\zeta}, \boldsymbol{\sigma} \})}_{\text{sampled site $ s_2 $}} + \underbrace{\Psi^{s_3}_{l_1}(\boldsymbol{u},\boldsymbol{\tau},\boldsymbol{\gamma} ; \{\boldsymbol{\zeta}, \boldsymbol{\sigma} \})}_{\text{sampled site $ s_3 $}}
    \end{gathered}       
\end{equation}

\noindent The factor $ \Psi^{s_1}_{l_1}(\boldsymbol{u},\boldsymbol{\tau},\boldsymbol{\gamma} ; \{\boldsymbol{\zeta}, \boldsymbol{\sigma} \}) $, as associated with the sampled spatial location $ s_1 $ in Figs. \ref{fig:Figure7} and 4, and Eq. \eqref{Eq229}, can be computed as follows

\begin{equation}\label{Eq230}
    \begin{gathered}           
     \Psi^{s_1}_{l_1}(\boldsymbol{u},\boldsymbol{\tau},\boldsymbol{\gamma} ; \{\boldsymbol{\zeta}, \boldsymbol{\sigma} \}) = \\[1.5ex] \mathbb{E}_{p(\boldsymbol{f} \vert \boldsymbol{u},\boldsymbol{\tau},\boldsymbol{\gamma})}\Biggl[ \ln {\color{black}\Biggl \{} \text{ } 
     \mathlarger{\prod_{t \in \{\sample_{1,s_1}^{(t)} | y_1(s_1,t) > l_{q_1} \}}} \text{ } \mathcal{N} \bigl(y_{{1}}(s_1,t) \vert f_{{{1}}}(s_1,t),\sigma_{{1}}^2 \bigr) \times \\[1.5ex] 
     \mathlarger{\prod_{t \in \{\sample_{1,s_1}^{(t)} | y_1(s_1,t) = l_{q_1} \}}} \text{ } g \bigl(f_{{{1}}}(s_1,t) \vert \zeta_{q_1d_1}(s_1,t), l_{d_1}, l_{q_1},\sigma_{{1}}^2,\sigma_{q_1d_1}^2 \bigr) \times \\[1.5ex] 
     \mathlarger{\prod_{t \in \{\sample_{1,s_1}^{(t)} | y_1(s_1,t) = l_{d_1} \}}} \text{ } g \bigl( f_{{{1}}}(s_1,t) \vert \zeta_{d_1}(s_1,t), l_{d_1},\sigma_{{1}}^2,\sigma_{d_1}^2 \bigr)  {\color{black}\Biggr\}} \Biggr]
    \end{gathered}       
\end{equation}

\noindent Observe from Eqs. \eqref{Eq226} and \eqref{Eq229} that Eq. \eqref{Eq223} can alternative be expressed as 

\begin{equation}\label{Eq231}
    \begin{gathered}           
    \Psi_{l}(\boldsymbol{u},\boldsymbol{\tau},\boldsymbol{\gamma} ; \{\boldsymbol{\zeta}, \boldsymbol{\sigma} \}) = \\[1.5ex] \underbrace{\underbrace{\Psi^{s_1}_{l_1}(\boldsymbol{u},\boldsymbol{\tau},\boldsymbol{\gamma} ; \{\boldsymbol{\zeta}, \boldsymbol{\sigma} \})}_{\text{sampled site $ s_1 $}} + \underbrace{\Psi^{s_2}_{l_1}(\boldsymbol{u},\boldsymbol{\tau},\boldsymbol{\gamma} ; \{\boldsymbol{\zeta}, \boldsymbol{\sigma} \})}_{\text{sampled site $ s_2 $}} + \underbrace{\Psi^{s_3}_{l_1}(\boldsymbol{u},\boldsymbol{\tau},\boldsymbol{\gamma} ; \{\boldsymbol{\zeta}, \boldsymbol{\sigma} \})}_{\text{sampled site $ s_3 $}}}_{\text{Associated with underlying latent function $ \boldsymbol{f}_{1} $}} + \\[1.5ex]
    \underbrace{\underbrace{\Psi^{s_1}_{l_2}(\boldsymbol{u},\boldsymbol{\tau},\boldsymbol{\gamma} ; \{\boldsymbol{\zeta}, \boldsymbol{\sigma} \})}_{\text{sampled site $ s_1 $}} + \underbrace{\Psi^{s_2}_{l_2}(\boldsymbol{u},\boldsymbol{\tau},\boldsymbol{\gamma} ; \{\boldsymbol{\zeta}, \boldsymbol{\sigma} \})}_{\text{sampled site $ s_2 $}} + \underbrace{\Psi^{s_3}_{l_2}(\boldsymbol{u},\boldsymbol{\tau},\boldsymbol{\gamma} ; \{\boldsymbol{\zeta}, \boldsymbol{\sigma} \})}_{\text{sampled site $ s_3 $}}}_{\text{Associated with underlying latent function $ \boldsymbol{f}_{2} $}}
    \end{gathered}       
\end{equation}

\noindent Consequently, from Eq. \eqref{Eq231}, it is sufficient to only consider a single factor $ \Psi^{s}_{l_a}(\boldsymbol{u},\boldsymbol{\tau},\boldsymbol{\gamma} ; \{\boldsymbol{\zeta}, \boldsymbol{\sigma} \}) $ when computing the overall quantity $ \Psi_{l}(\boldsymbol{u},\boldsymbol{\tau},\boldsymbol{\gamma} ; \{\boldsymbol{\zeta}, \boldsymbol{\sigma} \}) $ as all other factors will have the same mathematical functional form and contribution. Next, consider the contributing factor $ \Psi^{s_1}_{l_1}(\boldsymbol{u},\boldsymbol{\tau},\boldsymbol{\gamma} ; \{\boldsymbol{\zeta}, \boldsymbol{\sigma} \})  $ in Eq. \eqref{Eq230}. Recall from Eq. (47) that the density $ p(\boldsymbol{f} \vert \boldsymbol{u},\boldsymbol{\tau},\boldsymbol{\gamma}) $ corresponds to a (conditional) multivariate Gaussian density. Furthermore, the vector of latent function values $ \boldsymbol{f}_{1} $ is, for mathematical convenience, defined to have the ordering $ \boldsymbol{f}_{1}  = [\boldsymbol{f}^T_{1}(s_1),\boldsymbol{f}^T_{1}(s_2),\boldsymbol{f}^T_{1}(s_3)]^T $. Since the factor $ \Psi^{s_1}_{l_1}(\boldsymbol{u},\boldsymbol{\tau},\boldsymbol{\gamma} ; \{\boldsymbol{\zeta}, \boldsymbol{\sigma} \}) $ pertains to only a subset of the latent function values, as associated with the physically sampled spatial location $ s_1 $, with the multivariate Gaussian density $ p(\boldsymbol{f} \vert \boldsymbol{u},\boldsymbol{\tau},\boldsymbol{\gamma}) $ jointly defined over the underlying latent function values, the marginalisation property of the multivariate Gaussian density can be used to simplify Eq. \eqref{Eq230} such that

\begin{equation}\label{Eq232}
    \begin{gathered}           
     \Psi^{s_1}_{l_1}(\boldsymbol{u},\boldsymbol{\tau},\boldsymbol{\gamma} ; \{\boldsymbol{\zeta}, \boldsymbol{\sigma} \}) = \\[1.5ex] \mathbb{E}_{p(\boldsymbol{f}_{1}(s_1) \vert \boldsymbol{u},\boldsymbol{\tau},\boldsymbol{\gamma})}\Biggl[ \ln {\color{black}\Biggl \{} \text{ } 
     \mathlarger{\prod_{t \in \{\sample_{1,s_1}^{(t)} | y_1(s_1,t) > l_{q_1} \}}} \text{ } \mathcal{N} \bigl(y_{{1}}(s_1,t)  \vert f_{{{1}}}(s_1,t),\sigma_{{1}}^2 \bigr) \Biggl\} \Biggr] \text{ } + \\[1.5ex] \mathbb{E}_{p(\boldsymbol{f}_{1}(s_1) \vert \boldsymbol{u},\boldsymbol{\tau},\boldsymbol{\gamma})} \Biggl[ \Biggr\{ 
     \mathlarger{\prod_{t \in \{\sample_{1,s_1}^{(t)} | y_1(s_1,t) = l_{q_1} \}}} \text{ } g \bigl(f_{{{1}}}(s_1,t) \vert \zeta_{q_1d_1}(s_1,t), l_{d_1}, l_{q_1},\sigma_{{1}}^2,\sigma_{q_1d_1}^2 \bigr) \Biggr\} \Biggr]  \text{ } + \\[1.5ex] \mathbb{E}_{p(\boldsymbol{f}_{1}(s_1) \vert \boldsymbol{u},\boldsymbol{\tau},\boldsymbol{\gamma})} \Biggl[ \Biggl\{ 
     \mathlarger{\prod_{t \in \{\sample_{1,s_1}^{(t)} | y_1(s_1,t) = l_{d_1} \}}} \text{ } g \bigl( f_{{{1}}}(s_1,t) \vert \zeta_{d_1}(s_1,t), l_{d_1},\sigma_{{1}}^2,\sigma_{d_1}^2 \bigr)   {\color{black}\Biggr\}} \Biggr ]
    \end{gathered}       
\end{equation}

\noindent Note that to arrive at Eq. \eqref{Eq232}, the product rule for the natural logarithm and linearity of the expectation operator has been used. To calculate the expectations associated with the quantity $ \Psi^{s_1}_{l_1}(\boldsymbol{u},\boldsymbol{\tau},\boldsymbol{\gamma} ; \{\boldsymbol{\zeta}, \boldsymbol{\sigma} \}) $, the log-likelihood factors can be rewritten into a more amenable form. Observe that the first product of factor associated with $ y_{{1}}(s_1,t) > l_{q_1} $ in Eq. \eqref{Eq232} can be compactly written as a multivariate Gaussian density across temporal observations such that

\begin{equation}\label{Eq233}
    \begin{gathered}           
      \mathlarger{\prod_{t \in \{\sample_{1,s_1}^{(t)} | y_1(s_1,t) > l_{q_1} \}}} \text{ } \mathcal{N} \bigl(y_{{1}}(s_1,t)  \vert f_{{{1}}}(s_1,t),\sigma_{{1}}^2 \bigr) =  \mathcal{N} \bigl(\boldsymbol{y}_{{1}}(s_1;[> l_{q_1}])  \vert \boldsymbol{f}_{{{1}}}(s_1;[> l_{q_1}]), \boldsymbol{\Sigma}_{s_1,[>q_1]} \bigr)
    \end{gathered}       
\end{equation}

In Eq. \eqref{Eq233}, the symbol $ \boldsymbol{y}_{{1}}(s_1;[> l_{q_1}]) $ collectively denotes the noise-corrupted temporal observations for the sampled spatial location $ s_1 $, as associated with the latent function values $ \boldsymbol{f}_{1}(s_1;\cdot) $, that are uncensored. Note that $ [>q_1] $ will be used as shorthand notation for $ [> l_{q_1}] $, in other words, $ [>q_1] $ serves as shorthand notation for the quantities associated with uncensored temporal observations. The covariance matrix $ \boldsymbol{\Sigma}_{s_1,[>q_1]} $ is defined as 

\begin{equation}\label{Eq240}
    \begin{gathered}           
      \boldsymbol{\Sigma}_{s_1,[>q_1]} = \sigma_{{1}}^2 \boldsymbol{I_{[> q_1]}}
    \end{gathered}       
\end{equation}

The symbol $ \boldsymbol{I_{[> q_1]}} $ denotes the identity matrix of appropriate size that is associated with the uncensored observations. Using the local likelihood lower bound, as introduced by \cite{Basson2023}, the product of factors associated with $ y_{{1}}(s_1,t) = l_{q_1} $ in Eq. \eqref{Eq232} can be compactly written as a multivariate quadratic function across temporal instances such that

\begin{equation}\label{Eq234}
    \begin{gathered}           
      \mathlarger{\prod_{t \in \{\sample_{1,s_1}^{(t)} | y_1(s_1,t) = l_{q_1} \}}} \text{ } g \bigl(f_{{{1}}}(s_1,t) \vert \zeta_{q_1d_1}(s_1,t), l_{d_1}, l_{q_1},\sigma_{{1}}^2,\sigma_{q_1d_1}^2 \bigr) = \\[1.5ex] - \frac{1}{2} \Biggl[\boldsymbol{f}^T_{1}(s_1;l_{q_1},l_{d_1}) \boldsymbol{\Sigma}^{-1}_{s_1,q_1d_1}\boldsymbol{f}_{1}(s_1;l_{q_1},l_{d_1}) \text{ } + \\[1.5ex] \boldsymbol{b}^T_{s_1,q_1d_1}\boldsymbol{\Sigma}^{-1}_{s_1,q_1d_1}\bigl(\boldsymbol{f}_{1}(s_1;l_{q_1},l_{d_1}) \text{ } - \text{ } [l_{q_1} + l_{d_1}]\boldsymbol{1}_{s_1,q_1d_1} \bigr) \text{ } + \text{ } \boldsymbol{c}^T_{s_1,q_1d_1}\boldsymbol{\Sigma}^{-1}_{s_1,q_1d_1}\boldsymbol{1}_{s_1,q_1d_1} \Biggr]
    \end{gathered}       
\end{equation}

In Eq. \eqref{Eq234}, the symbol $ \boldsymbol{f}_{1}(s_1;l_{q_1},l_{d_1}) $ denotes the latent function values, as associated with sampled spatial location $ s_1 $, that correspond to temporal observations that are below the quantification limit $ l_{q_1} $ but is still detectable (with detection limit $ l_{d_1} $). Note that the $ q_1d_1 $ subscript/superscript notation in Eq. \eqref{Eq234} again serves as a shorthand notation for the quantities associated with the below quantification (but still detectable) limit. The symbol $ \boldsymbol{1}_{s_1,q_1d_1} $ denotes the column vector of ones. The matrix $ \boldsymbol{\Sigma}_{s_1,q_1d_1} $ is defined as follows

\begin{equation}\label{Eq235}
    \begin{gathered}           
      \boldsymbol{\Sigma}_{s_1,q_1d_1} = (\sigma_{{1}}^2 + \sigma_{q_1d_1}^2)\boldsymbol{I}_{s_1,q_1d_1}
    \end{gathered}       
\end{equation}

\noindent Symbol $ \boldsymbol{I}_{s_1,q_1d_1} $ in Eq. \eqref{Eq235} denotes the identity matrix of appropriate size as pertaining to sampled spatial location $ s_1 $. Furthermore, for each censored temporal observation, the element-wise entries of the column vectors $ \boldsymbol{b}_{s_1,q_1d_1} $ and $ \boldsymbol{c}_{s_1,q_1d_1} $ can be computed following the procedure outlined in \cite{Basson2023}. Consequently, from Eq. \eqref{Eq234}, there is a column vector $ \boldsymbol{\zeta}_{s_1,q_1d_1} $ that can be adjusted during the gradient-based optimisation routine to keep the local lower bounds tight, as associated with vectors $ \boldsymbol{b}_{s_1,q_1d_1} $ and $ \boldsymbol{c}_{s_1,q_1d_1} $, such that

\begin{equation}\label{Eq236}
    \begin{gathered}           
      \boldsymbol{\zeta}_{s_1,q_1d_1} = \begin{bmatrix} \zeta_{q_1d_1}(s_1,t_2) \\ \zeta_{q_1d_1}(s_1,t_6) \\ \zeta_{q_1d_1}(s_1,t_7) \\ \vdots \\ \zeta_{q_1d_1}(s_1,t_{13}) \end{bmatrix}; \text{ }\boldsymbol{b}_{s_1,q_1d_1} = \begin{bmatrix} b_{q_1d_1}(s_1,t_2) \\ b_{q_1d_1}(s_1,t_6) \\ b_{q_1d_1}(s_1,t_7) \\ \vdots \\ b_{q_1d_1}(s_1,t_{13}) \end{bmatrix}; \text{ } \boldsymbol{c}_{s_1,q_1d_1} = \begin{bmatrix} c_{q_1d_1}(s_1,t_2) \\ c_{q_1d_1}(s_1,t_6) \\ c_{q_1d_1}(s_1,t_7) \\ \vdots \\ c_{q_1d_1}(s_1,t_{13}) \end{bmatrix};
    \end{gathered}       
\end{equation}

Note that in Eq. \eqref{Eq236} the temporal indices $ t_2,t_6,t_7, \cdots, t_{13} $ are purely suggestive for illustrative purposes. The true temporal indices will depend on the data set available to the practitioner. The results for the product of factors associated with the condition $ y_{{1}}(s_1,t) = l_{d_1} $ can be derived in a similar manner as outlined above for $ y_{{1}}(s_1,t) = l_{q_1} $.

For sampled spatial location $ s_1 $, note from Eq. \eqref{Eq232} that each contributing product of factors in the locally lower bounded mixed Co-Kriging log-likelihood function is defined over a subset of the temporal indices, as associated with $ \boldsymbol{f}_{1}(s_1) $. Since the density $ p(\boldsymbol{f}_{1}(s_1) \vert \boldsymbol{u},\boldsymbol{\tau},\boldsymbol{\gamma}) $ is jointly defined over $ \boldsymbol{f}_{1}(s_1) $, and also follows a multivariate Gaussian density, the expectation of the first locally lower bounded mixed Co-Kriging log-likelihood function product of factors, using the marginalisation property of the Gaussian density and Eq. \eqref{Eq233}, can be computed as 

\begin{equation}\label{Eq241}
    \begin{gathered}           
      \mathbb{E}_{p(\boldsymbol{f}_{1}(s_1;[> l_{q_1}]) \vert \boldsymbol{u},\boldsymbol{\tau},\boldsymbol{\gamma})}\Biggl[ \ln \mathcal{N} \bigl(\boldsymbol{y}_{{1}}(s_1;[> l_{q_1}])  \vert \boldsymbol{f}_{{{1}}}(s_1;[> l_{q_1}]), \boldsymbol{\Sigma}_{s_1,[>q_1]} \bigr) \Biggr] = \\[1.5ex] -\frac{N_{s_1,[>q_1]}}{2} \ln(2\pi\sigma_{{1}}^2) \text{ } - \text{ } \frac{1}{2} \text{tr$ \Bigl\{ \boldsymbol{\Sigma}^{-1}_{s_1,[>q_1]} \Bigl[ \boldsymbol{K}^{s_1,[>q_1]}_{NN} - \boldsymbol{K}^{s_1,[>q_1]}_{NM}\boldsymbol{K}^{-1}_{MM}\boldsymbol{K}^{s_1,[>q_1]}_{MN} \Bigr] \Bigr\} $} \text{ } - \text{ } \\[1.5ex] \frac{1}{2} \Bigl[\boldsymbol{y}^T_{{1}}(s_1;[>q_1])\boldsymbol{\Sigma}^{-1}_{s_1,[>q_1]}\boldsymbol{y}_{{1}}(s_1;[>q_1]) - 2\boldsymbol{y}^T_{{1}}(s_1;[> q_1])\boldsymbol{\Sigma}^{-1}_{s_1,[>q_1]}\boldsymbol{K}^{s_1,[>q_1]}_{NM}\boldsymbol{K}^{-1}_{MM}\boldsymbol{u} \text{ } + \text{ } \\[1.5ex] (\boldsymbol{K}^{s_1,[>q_1]}_{NM}\boldsymbol{K}^{-1}_{MM}\boldsymbol{u})^T\boldsymbol{\Sigma}^{-1}_{s_1,[>q_1]}(\boldsymbol{K}^{s_1,[>q_1]}_{NM}\boldsymbol{K}^{-1}_{MM}\boldsymbol{u}) \Bigr]
    \end{gathered}       
\end{equation}

In Eq. \eqref{Eq241}, the symbol $ N_{s_1,[>q_1]} $ denotes the total number of uncensored temporal observations associated with sampled spatial location $ s_1 $ and underlying latent function $ \boldsymbol{f}_{1} $. The symbol $ \text{tr}\{\cdot\} $ denotes the matrix trace operator. The symbol $ \boldsymbol{K}^{s_1,[>q_1]}_{NN} $ denotes the spatio-temporal covariance matrix contribution associated with sampled spatial location $ s_1 $, as computed on the spatio-temporal input locations pertaining to the uncensored observations, and can be obtained from evaluating the separable spatio-temporal covariance given by Eqs. (25) and (26). The symbol $ \boldsymbol{K}^{s_1,[>q_1]}_{NM} $ denotes the cross-covariance contribution between the spatio-temporal input locations associated with $ s_1 $ and the spatio-temporal inducing input locations. The contribution to the cross-covariance matrix can be obtained by evaluating the spatio-temporal cross-covariance given by Eqs. (54) to (56). Lastly, the symbol $ \boldsymbol{K}_{MM} $ denotes the covariance between the spatio-temporal inducing input locations and can be obtained by evaluating the spatio-temporal covariance given by Eqs. (51) and (52). 

The expectation of the second product of factors, which are associated with the site $ s_1 $ temporal observations that are below the quantification limit $ l_{q_1} $ (but still detectable), in a similar manner to the first locally lower bounded mixed Co-Kriging log-likelihood function product of factors (see Eqs. \eqref{Eq232} and \eqref{Eq241}), using the results from Eq. \eqref{Eq234} and the marginalisation property of the multivariate Gaussian density, can be computed as follows

\begin{equation}\label{Eq242}
    \begin{gathered}           
      \mathbb{E}_{p(\boldsymbol{f}_{1}(s_1;l_{q_1}) \vert \boldsymbol{u},\boldsymbol{\tau},\boldsymbol{\gamma})} \Biggl[ \Biggr\{ 
      \mathlarger{\prod_{t \in \{\sample_{1,s_1}^{(t)} | y_1(s_1,t) = l_{q_1} \}}} \text{ } g \bigl(f_{{{1}}}(s_1,t) \vert \zeta_{q_1d_1}(s_1,t), l_{d_1}, l_{q_1},\sigma_{{1}}^2,\sigma_{q_1d_1}^2 \bigr) \Biggr\} \Biggr] = \\[1.5ex] - \frac{1}{2} \Biggl[(\boldsymbol{K}^{s_1,q_1d_1}_{NM}\boldsymbol{K}^{-1}_{MM}\boldsymbol{u})^T \boldsymbol{\Sigma}^{-1}_{s_1,q_1d_1}(\boldsymbol{K}^{s_1,q_1d_1}_{NM}\boldsymbol{K}^{-1}_{MM}\boldsymbol{u}) - 2\boldsymbol{b}^T_{s_1,q_1d_1}\boldsymbol{\Sigma}^{-1}_{s_1,q_1d_1}\boldsymbol{K}^{s_1,q_1d_1}_{NM}\boldsymbol{K}^{-1}_{MM}\boldsymbol{u} \Biggr] - \\[1.5ex] \frac{1}{2} \text{tr$ \Bigl\{ \boldsymbol{\Sigma}^{-1}_{s_1,q_1d_1} \Bigl[ \boldsymbol{K}^{s_1,q_1d_1}_{NN} - \boldsymbol{K}^{s_1,q_1d_1}_{NM}\boldsymbol{K}^{-1}_{MM}\boldsymbol{K}^{s_1,q_1d_1}_{MN} \Bigr] \Bigr\} $} \text{ } + \text{ }  \\[1.5ex] \boldsymbol{c}^T_{s_1,q_1d_1}\boldsymbol{\Sigma}^{-1}_{s_1,q_1d_1}\boldsymbol{1}_{s_1,q_1d_1} - \boldsymbol{b}^T_{s_1,q_1d_1}\boldsymbol{\Sigma}^{-1}_{s_1,q_1d_1}\boldsymbol{d}_{s_1,q_1d_1}
    \end{gathered}       
\end{equation}

\noindent In Eq. \eqref{Eq242}, the column vector $ \boldsymbol{d}_{s_1,q_1d_1} $ is defined as follows

\begin{equation}\label{Eq243}
    \begin{gathered}           
      \boldsymbol{d}_{s_1,q_1d_1} = [l_{d_1} +  l_{q_1}]\boldsymbol{1}_{s_1,q_1d_1}
    \end{gathered}       
\end{equation}

\noindent The matrices $ \boldsymbol{K}^{s_1,q_1d_1}_{NN}$ and $ \boldsymbol{K}^{s_1,q_1d_1}_{NM} $ are calculated in a similar manner to $ \boldsymbol{K}^{s_1,[>q_1]}_{NN} $ and $ \boldsymbol{K}^{s_1,[>q_1]}_{NM} $, respectively, however, now the input locations corresponding to the indices where temporal observations are below the quantification limit (but still detectable) are used. The results for the product of factors associated with the condition $ y_{{1}}(s_1,t) = l_{d_1} $ can be derived in a similar manner as outlined above for $ y_{{1}}(s_1,t) = l_{q_1} $.

Recall that the goal is to compute the quantity $ \Psi_{l}( \boldsymbol{u},\boldsymbol{\tau},\boldsymbol{\gamma} ; \{\boldsymbol{\zeta}, \boldsymbol{\sigma} \}) $ in Eq. \eqref{Eq231}. From the preceding discussion, the quantity $ {\Psi^{s_1}_{l_1}(\boldsymbol{u},\boldsymbol{\tau},\boldsymbol{\gamma}; \{\boldsymbol{\zeta}, \boldsymbol{\sigma} \})} $, as associated with sampled spatial location $ s_1 $ and underlying latent function $ \boldsymbol{f}_{1} $, has been computed. The expected values associated with sampled spatial locations $ s_2 $ and $ s_3 $, pertaining to the underlying latent function $ \boldsymbol{f}_{1} $, can also be computed in the exact same manner as outlined in the preceding discussion. The same procedure can then be repeated for underlying latent function $ \boldsymbol{f}_{2} $, as associated with sampled spatial locations $ s_1, s_2 $, and $ s_3 $. Observe from Eqs. \eqref{Eq241} and \eqref{Eq242}, as well as for the condition $ y_{{1}}(s_1,t) = l_{d_1} $, that after taking the expectation with respect to the density $ p(\boldsymbol{f} \vert \boldsymbol{u},\boldsymbol{\tau},\boldsymbol{\gamma}) $, the resulting expressions are multivariate quadratic functions in the inducing variables $ \boldsymbol{u} $. Once all the quantities associated with $ \Psi_{l}(\boldsymbol{u},\boldsymbol{\tau},\boldsymbol{\gamma} ; \{\boldsymbol{\zeta}, \boldsymbol{\sigma} \}) $ in Eq. \eqref{Eq231} are computed, the multivariate quadratic property of the resulting expressions can be used to compactly express $ \Psi_{l}(\boldsymbol{u},\boldsymbol{\tau},\boldsymbol{\gamma} ; \{\boldsymbol{\zeta}, \boldsymbol{\sigma} \}) $ as follows

\begin{equation}\label{Eq246}
    \begin{gathered}           
      \Psi_{l}(\boldsymbol{u},\boldsymbol{\tau},\boldsymbol{\gamma} ; \{\boldsymbol{\zeta}, \boldsymbol{\sigma} \}) = c_1 - \frac{1}{2} \Biggl[ \bigl(\boldsymbol{y}_{l} - \boldsymbol{K}^{l}_{NM}\boldsymbol{K}^{-1}_{MM}\boldsymbol{u} \bigr)^T \boldsymbol{\Sigma}^{-1}_{l}(\boldsymbol{y}_{l} - \boldsymbol{K}^{l}_{NM}\boldsymbol{K}^{-1}_{MM}\boldsymbol{u} \bigr) \Biggr] \\[1.5ex] \text{ } \text{ } \text{ } \text{ } \text{ } \text{ } \text{ } \text{ } \text{ } \text{ }  - \frac{1}{2}\text{tr$ \Bigl\{ \boldsymbol{\Sigma}^{-1}_{l} \Bigl[ \boldsymbol{K}^{l}_{NN} - \boldsymbol{K}^{l}_{NM}\boldsymbol{K}^{-1}_{MM}\boldsymbol{K}^{l}_{MN} \Bigr] \Bigr\} $}
    \end{gathered}       
\end{equation}

\noindent The quantity $ c_1 $ in Eq. \eqref{Eq246} is defined as follows

\begin{equation}\label{Eq247}
    \begin{gathered}           
       c_1 = - \frac{1}{2} \Biggl[N_{1,[>q_1]} \ln(2\pi\sigma^2_{1}) + N_{2,[>q_2]} \ln(2\pi\sigma^2_{2}) \Biggr] + \Biggl[\frac{1}{2}\boldsymbol{b}^T\boldsymbol{\Sigma}^{-1}_c\boldsymbol{b} + \boldsymbol{c}^T\boldsymbol{\Sigma}^{-1}_c\boldsymbol{1}^{\ast} - \boldsymbol{b}^T\boldsymbol{\Sigma}^{-1}_c\boldsymbol{d} \Biggr] 
    \end{gathered}       
\end{equation}

\noindent In Eq. \eqref{Eq247}, the symbols $ N_{1,[>q_1]} $ and $ N_{2,[>q_2]} $ denote the total number of uncensored temporal observations associated with underlying latent functions $ \boldsymbol{f}_{1} $ and $ \boldsymbol{f}_{2} $, respectively. The column vector $ \boldsymbol{y}_{l} $, which can be interpreted as a pseudo-data vector that stems from the locally lower bounded likelihood methodology, is defined as follows

\begin{equation}\label{Eq248}
    \begin{gathered}           
       \boldsymbol{y}_{l} = \begin{bmatrix} \boldsymbol{y}_{{1}}([> q_1]) \\ \boldsymbol{y}_{{2}}([> q_2]) \\ \boldsymbol{b}_{q1d1} \\ \boldsymbol{b}_{q2d2}  \\ \boldsymbol{b}_{d1} \\ \boldsymbol{b}_{d2} \end{bmatrix}; \text{ } \boldsymbol{y}_{{1}}([> q_1]) = \begin{bmatrix} \boldsymbol{y}_{{1}}(s_1;[> q_1]) \\ \boldsymbol{y}_{{1}}(s_2;[> q_1]) \\ \boldsymbol{y}_{{1}}(s_3;[> q_1]) \end{bmatrix} ; \boldsymbol{y}_{{2}}([> q_2]) = \begin{bmatrix} \boldsymbol{y}_{{2}}(s_1;[> q_2]) \\ \boldsymbol{y}_{{2}}(s_2;[> q_2]) \\ \boldsymbol{y}_{{2}}(s_3;[> q_2]) \end{bmatrix} ; \text{ } \\[1.5ex] \boldsymbol{b}_{q1d1} = \begin{bmatrix} \boldsymbol{b}_{s_1,q_1d_1} \\ \boldsymbol{b}_{s_2,q_1d_1} \\ \boldsymbol{b}_{s_3,q_1d_1} \end{bmatrix} ; \text{ } \boldsymbol{b}_{q2d2} = \begin{bmatrix} \boldsymbol{b}_{s_1,q_2d_2} \\ \boldsymbol{b}_{s_2,q_2d_2} \\ \boldsymbol{b}_{s_3,q_2d_2} \end{bmatrix};  \boldsymbol{b}_{d_1} = \begin{bmatrix} \boldsymbol{b}_{s_1,d_1} \\ \boldsymbol{b}_{s_2,d_1} \\ \boldsymbol{b}_{s_3,d_1} \end{bmatrix} ; \text{ } \boldsymbol{b}_{d2} = \begin{bmatrix} \boldsymbol{b}_{s_1,d_2} \\ \boldsymbol{b}_{s_2,d_2} \\ \boldsymbol{b}_{s_3,d_2} \end{bmatrix}
    \end{gathered}       
\end{equation}

The remaining vectors $ \boldsymbol{b},\boldsymbol{c},\boldsymbol{1}^{\ast} $, and $ \boldsymbol{d} $ are defined as follows 

\begin{equation}\label{Eq249}
    \begin{gathered}           
       \boldsymbol{b} = \begin{bmatrix} \boldsymbol{b}_{qd} \\ \boldsymbol{b}_{d} \end{bmatrix}; \text{ } \boldsymbol{b}_{qd} = \begin{bmatrix} \boldsymbol{b}_{q1d1} \\ \boldsymbol{b}_{q2d2} \end{bmatrix}; \text{ } \boldsymbol{b}_{d} = \begin{bmatrix} \boldsymbol{b}_{d1} \\ \boldsymbol{b}_{d2} \end{bmatrix}; %\\[1.5ex] 
       \boldsymbol{c} = \begin{bmatrix} \boldsymbol{c}_{qd} \\ \boldsymbol{c}_{d} \end{bmatrix}; \text{ } \boldsymbol{c}_{qd} = \begin{bmatrix} \boldsymbol{c}_{q1d1} \\ \boldsymbol{c}_{q2d2} \end{bmatrix}; \text{ } \boldsymbol{c}_{d} = \begin{bmatrix} \boldsymbol{c}_{d1} \\ \boldsymbol{c}_{d2} \end{bmatrix}; \\[1.5ex] \text{ } \boldsymbol{c}_{q1d1} = \begin{bmatrix} \boldsymbol{c}_{s_1,q_1d_1} \\ \boldsymbol{c}_{s_2,q_1d_1} \\ \boldsymbol{c}_{s_3,q_1d_1} \end{bmatrix} ; \text{ } \boldsymbol{c}_{q2d2} = \begin{bmatrix} \boldsymbol{c}_{s_1,q_2d_2} \\ \boldsymbol{c}_{s_2,q_2d_2} \\ \boldsymbol{c}_{s_3,q_2d_2} \end{bmatrix} \boldsymbol{c}_{d_1} = \begin{bmatrix} \boldsymbol{c}_{s_1,d_1} \\ \boldsymbol{c}_{s_2,d_1} \\ \boldsymbol{c}_{s_3,d_1} \end{bmatrix} ; \text{ } \boldsymbol{c}_{d2} = \begin{bmatrix} \boldsymbol{c}_{s_1,d_2} \\ \boldsymbol{c}_{s_2,d_2} \\ \boldsymbol{c}_{s_3,d_2} \end{bmatrix}; \\[1.5ex] \boldsymbol{1}^{\ast} =  \begin{bmatrix} \boldsymbol{1}_{qd} \\ \boldsymbol{1}_{d} \end{bmatrix}; \text{ } \boldsymbol{1}_{qd} = \begin{bmatrix} \boldsymbol{1}_{q_1d_1} \\ \boldsymbol{1}_{q_2d_2} \end{bmatrix}; \text{ } \boldsymbol{1}_{d} = \begin{bmatrix} \boldsymbol{1}_{d_1} \\ \boldsymbol{1}_{d_2} \end{bmatrix}; \\[1.5ex] \boldsymbol{1}_{q1d1} = \begin{bmatrix} \boldsymbol{1}_{s_1,q_1d_1} \\ \boldsymbol{1}_{s_2,q_1d_1} \\ \boldsymbol{1}_{s_3,q_1d_1} \end{bmatrix} ; \text{ } \boldsymbol{1}_{q2d2} = \begin{bmatrix} \boldsymbol{1}_{s_1,q_2d_2} \\ \boldsymbol{1}_{s_2,q_2d_2} \\ \boldsymbol{1}_{s_3,q_2d_2} \end{bmatrix} \boldsymbol{1}_{d_1} = \begin{bmatrix} \boldsymbol{1}_{s_1,d_1} \\ \boldsymbol{1}_{s_2,d_1} \\ \boldsymbol{1}_{s_3,d_1} \end{bmatrix} ; \text{ } \boldsymbol{1}_{d2} = \begin{bmatrix} \boldsymbol{1}_{s_1,d_2} \\ \boldsymbol{1}_{s_2,d_2} \\ \boldsymbol{1}_{s_3,d_2} \end{bmatrix}; \\[1.5ex] \boldsymbol{d} = \begin{bmatrix} \boldsymbol{d}_{qd} \\ \boldsymbol{d}_{d} \end{bmatrix}; \text{ } \boldsymbol{d}_{qd} = \begin{bmatrix} \boldsymbol{d}_{q1d1} \\ \boldsymbol{d}_{q2d2} \end{bmatrix}; \text{ } \boldsymbol{d}_{d} = \begin{bmatrix} \boldsymbol{d}_{d1} \\ \boldsymbol{d}_{d2} \end{bmatrix}; \\[1.5ex] \text{ } \boldsymbol{d}_{q1d1} = \begin{bmatrix} \boldsymbol{d}_{s_1,q_1d_1} \\ \boldsymbol{d}_{s_2,q_1d_1} \\ \boldsymbol{d}_{s_3,q_1d_1} \end{bmatrix} ; \text{ } \boldsymbol{d}_{q2d2} = \begin{bmatrix} \boldsymbol{d}_{s_1,q_2d_2} \\ \boldsymbol{d}_{s_2,q_2d_2} \\ \boldsymbol{d}_{s_3,q_2d_2} \end{bmatrix} \boldsymbol{d}_{d_1} = \begin{bmatrix} \boldsymbol{d}_{s_1,d_1} \\ \boldsymbol{d}_{s_2,d_1} \\ \boldsymbol{d}_{s_3,d_1} \end{bmatrix} ; \text{ } \boldsymbol{d}_{d2} = \begin{bmatrix} \boldsymbol{d}_{s_1,d_2} \\ \boldsymbol{d}_{s_2,d_2} \\ \boldsymbol{d}_{s_3,d_2} \end{bmatrix}
    \end{gathered}       
\end{equation}

\noindent The matrices associated with Eqs. \eqref{Eq246} and \eqref{Eq247} are defined as follows

\begin{equation}\label{Eq250}
    \begin{gathered}           
       \boldsymbol{\Sigma}_c = \begin{bmatrix} \boldsymbol{\Sigma}_{qd} & \boldsymbol{0} \\ \boldsymbol{0} & \boldsymbol{\Sigma}_{d} \end{bmatrix}; \text{ } \boldsymbol{\Sigma}_{qd} = \begin{bmatrix} \boldsymbol{\Sigma}_{q_1d_1} & \boldsymbol{0} \\ \boldsymbol{0} & \boldsymbol{\Sigma}_{q_2d_2} \end{bmatrix} ; \text{ } \boldsymbol{\Sigma}_{d} = \begin{bmatrix} \boldsymbol{\Sigma}_{d_1} & \boldsymbol{0} \\ \boldsymbol{0} & \boldsymbol{\Sigma}_{d_2} \end{bmatrix} ; \\[1.5ex] \boldsymbol{\Sigma}_{q_1d_1} = \begin{bmatrix} \boldsymbol{\Sigma}_{s_1,q_1d_1} & \boldsymbol{0} & \boldsymbol{0} \\ \boldsymbol{0} & \boldsymbol{\Sigma}_{s_2,q_1d_1} & \boldsymbol{0} \\ \boldsymbol{0} & \boldsymbol{0} & \boldsymbol{\Sigma}_{s_2,q_1d_1}  \end{bmatrix}; \text{ } \boldsymbol{\Sigma}_{q_2d_2} = \begin{bmatrix} \boldsymbol{\Sigma}_{s_1,q_2d_2} & \boldsymbol{0} & \boldsymbol{0} \\ \boldsymbol{0} & \boldsymbol{\Sigma}_{s_2,q_2d_2} & \boldsymbol{0} \\ \boldsymbol{0} & \boldsymbol{0} & \boldsymbol{\Sigma}_{s_3,q_2d_2}  \end{bmatrix}; \\[1.5ex] \boldsymbol{\Sigma}_{d_1} = \begin{bmatrix} \boldsymbol{\Sigma}_{s_1,d_1} & \boldsymbol{0} & \boldsymbol{0} \\ \boldsymbol{0} & \boldsymbol{\Sigma}_{s_2,d_1} & \boldsymbol{0} \\ \boldsymbol{0} & \boldsymbol{0} & \boldsymbol{\Sigma}_{s_3,d_1}  \end{bmatrix}; \text{ } \boldsymbol{\Sigma}_{d_2} = \begin{bmatrix} \boldsymbol{\Sigma}_{s_1,d_2} & \boldsymbol{0} & \boldsymbol{0} \\ \boldsymbol{0} & \boldsymbol{\Sigma}_{s_2,d_2} & \boldsymbol{0} \\ \boldsymbol{0} & \boldsymbol{0} & \boldsymbol{\Sigma}_{s_3,d_2}  \end{bmatrix};
       \end{gathered}
\end{equation}

\begin{equation}\label{Eq251}
    \begin{gathered}           
       \boldsymbol{\Sigma}_l = \begin{bmatrix} \boldsymbol{\Sigma}_{N_{[>q]}} & \boldsymbol{0} \\ \boldsymbol{0} & \boldsymbol{\Sigma}_{c} \end{bmatrix}; \text{ } \boldsymbol{\Sigma}_{N_{[>q]}} = \begin{bmatrix} \boldsymbol{\Sigma}_{N_1,[> q_1]} & \boldsymbol{0} \\ \boldsymbol{0} & \boldsymbol{\Sigma}_{N_2,[> q_2]} \end{bmatrix}; \\[1.5ex] \boldsymbol{\Sigma}_{N_1,[> q_1]} = \begin{bmatrix} \boldsymbol{\Sigma}_{s_1,[>q_1]} & \boldsymbol{0} & \boldsymbol{0} \\ \boldsymbol{0} & \boldsymbol{\Sigma}_{s_2,[>q_1]} & \boldsymbol{0} \\ \boldsymbol{0} & \boldsymbol{0} & \boldsymbol{\Sigma}_{s_3,[>q_1]}  \end{bmatrix}; \text{ } \boldsymbol{\Sigma}_{N_2,[> q_2]} = \begin{bmatrix} \boldsymbol{\Sigma}_{s_1,[>q_2]} & \boldsymbol{0} & \boldsymbol{0} \\ \boldsymbol{0} & \boldsymbol{\Sigma}_{s_2,[>q_2]} & \boldsymbol{0} \\ \boldsymbol{0} & \boldsymbol{0} & \boldsymbol{\Sigma}_{s_3,[>q_2]}  \end{bmatrix}
       \end{gathered}
\end{equation}

\begin{equation}\label{Eq252}
    \begin{gathered}           
       \boldsymbol{K}^{l}_{NM} = \begin{bmatrix} \boldsymbol{K}^{[>q]}_{NM} \\ \boldsymbol{K}^{{qd}}_{NM} \\ \boldsymbol{K}^{{d}}_{NM} \end{bmatrix}; \text{ } \boldsymbol{K}^{l}_{MN} = [\boldsymbol{K}^{l}_{NM}]^T ; \text{ } \boldsymbol{K}^{[>q]}_{NM} = \begin{bmatrix} \boldsymbol{K}^{[>q_1]}_{N_{1}M} \\ \boldsymbol{K}^{[>q_2]}_{N_{2}M} \end{bmatrix}; \text{ } \boldsymbol{K}^{[>q_1]}_{N_{1}M} = \begin{bmatrix} \boldsymbol{K}^{s_1,[>q_1]}_{N_1M} \\ \boldsymbol{K}^{s_2,[>q_1]}_{N_1M} \\ \boldsymbol{K}^{s_3,[>q_1]}_{N_1M} \end{bmatrix} ; \\[1.5ex] \boldsymbol{K}^{[>q_2]}_{N_{2}M} = \begin{bmatrix} \boldsymbol{K}^{s_1,[>q_2]}_{N_2M} \\ \boldsymbol{K}^{s_2,[>q_2]}_{N_2M} \\ \boldsymbol{K}^{s_3,[>q_2]}_{N_2M} \end{bmatrix}; \text{ }  \boldsymbol{K}^{qd}_{NM} = \begin{bmatrix} \boldsymbol{K}^{q_1d_1}_{N_{1}M} \\ \boldsymbol{K}^{q_2d_2}_{N_{2}M} \end{bmatrix}; \text{ } \boldsymbol{K}^{q_1d_1}_{N_{1}M} = \begin{bmatrix} \boldsymbol{K}^{s_1,q_1d_1}_{N_1M} \\ \boldsymbol{K}^{s_2,q_1d_1}_{N_1M} \\ \boldsymbol{K}^{s_3,q_1d_1}_{N_1M} \end{bmatrix} ; \text{ } \boldsymbol{K}^{q_2d_2}_{N_{2}M} = \begin{bmatrix} \boldsymbol{K}^{s_1,q_2d_2}_{N_2M} \\ \boldsymbol{K}^{s_2,q_2d_2}_{N_2M} \\ \boldsymbol{K}^{s_3,q_2d_2}_{N_2M} \end{bmatrix}; \\[1.5ex] \boldsymbol{K}^{d}_{NM} = \begin{bmatrix} \boldsymbol{K}^{d_1}_{N_{1}M} \\ \boldsymbol{K}^{d_2}_{N_{2}M} \end{bmatrix}; \text{ } \boldsymbol{K}^{d_1}_{N_{1}M} = \begin{bmatrix} \boldsymbol{K}^{s_1,d_1}_{N_1M} \\ \boldsymbol{K}^{s_2,d_1}_{N_1M} \\ \boldsymbol{K}^{s_3,d_1}_{N_1M} \end{bmatrix} ; \text{ } \boldsymbol{K}^{d_2}_{N_{2}M} = \begin{bmatrix} \boldsymbol{K}^{s_1,d_2}_{N_2M} \\ \boldsymbol{K}^{s_2,d_2}_{N_2M} \\ \boldsymbol{K}^{s_3,d_2}_{N_2M} \end{bmatrix}
       \end{gathered}
\end{equation}

\begin{equation}\label{Eq253}
    \begin{gathered}           
       \boldsymbol{K}^{l}_{NN} = \begin{bmatrix} \boldsymbol{K}^{[>q]}_{NN} & \boldsymbol{0} & \boldsymbol{0} \\ 
        \boldsymbol{0} & \boldsymbol{K}^{qd}_{NN} & \boldsymbol{0} \\
        \boldsymbol{0} & \boldsymbol{0} & \boldsymbol{K}^{d}_{NN} \\
        \end{bmatrix}; \text{ } \boldsymbol{K}^{[>q]}_{NN} = \begin{bmatrix} \boldsymbol{K}^{[>q_1]}_{N_1N_1} & \boldsymbol{0} \\ \boldsymbol{0} & \boldsymbol{K}^{[>q_2]}_{N_2N_2}  \end{bmatrix} ;
        \boldsymbol{K}^{[>q_1]}_{N_1N_1} = \begin{bmatrix} \boldsymbol{K}^{s_1,[>q_1]}_{N_1N_1} & \boldsymbol{0} & \boldsymbol{0} \\ 
        \boldsymbol{0} & \boldsymbol{K}^{s_2,[>q_1]}_{N_1N_1} & \boldsymbol{0} \\
        \boldsymbol{0} & \boldsymbol{0} & \boldsymbol{K}^{s_3,[>q_1]}_{N_1N_1} \\
        \end{bmatrix}; \\[1.5ex] \text{ }  \boldsymbol{K}^{[>q_2]}_{N_2N_2} = \begin{bmatrix} \boldsymbol{K}^{s_1,[>q_2]}_{N_2N_2} & \boldsymbol{0} & \boldsymbol{0} \\ 
        \boldsymbol{0} & \boldsymbol{K}^{s_2,[>q_2]}_{N_2N_2} & \boldsymbol{0} \\
        \boldsymbol{0} & \boldsymbol{0} & \boldsymbol{K}^{s_3,[>q_2]}_{N_2N_2}\end{bmatrix} ; \boldsymbol{K}^{qd}_{NN} = \begin{bmatrix} \boldsymbol{K}^{q_1d_1}_{N_1N_1} & \boldsymbol{0} \\ \boldsymbol{0} & \boldsymbol{K}^{q_2d_2}_{N_2N_2} \end{bmatrix}; \\[1.5ex] \text{ } \boldsymbol{K}^{q_1d_1}_{N_1N_1} = \begin{bmatrix} \boldsymbol{K}^{s_1,q_1d_1}_{N_1N_1} & \boldsymbol{0} & \boldsymbol{0} \\ 
        \boldsymbol{0} & \boldsymbol{K}^{s_2,q_1d_1}_{N_1N_1} & \boldsymbol{0} \\
        \boldsymbol{0} & \boldsymbol{0} & \boldsymbol{K}^{s_3,q_1d_1}_{N_1N_1} \\   
        \end{bmatrix};  \boldsymbol{K}^{q_2d_2}_{N_2N_2} = \begin{bmatrix} \boldsymbol{K}^{s_1,q_2d_2}_{N_2N_2} & \boldsymbol{0} & \boldsymbol{0} \\ 
        \boldsymbol{0} & \boldsymbol{K}^{s_2,q_2d_2}_{N_2N_2} & \boldsymbol{0} \\
        \boldsymbol{0} & \boldsymbol{0} & \boldsymbol{K}^{s_3,q_2d_2}_{N_2N_2} \\   
        \end{bmatrix} ; \\[1.5ex] \text{ } \boldsymbol{K}^{d}_{NN} = \begin{bmatrix} \boldsymbol{K}^{d_1}_{N_1N_1} & \boldsymbol{0} \\ \boldsymbol{0} & \boldsymbol{K}^{d_2}_{N_2N_2} \end{bmatrix} ; \boldsymbol{K}^{d_1}_{N_1N_1} = \begin{bmatrix} \boldsymbol{K}^{s_1,d_1}_{N_1N_1} & \boldsymbol{0} & \boldsymbol{0} \\ 
        \boldsymbol{0} & \boldsymbol{K}^{s_2,d_1}_{N_1N_1} & \boldsymbol{0} \\
        \boldsymbol{0} & \boldsymbol{0} & \boldsymbol{K}^{s_3,d_1}_{N_1N_1} \\   
        \end{bmatrix} ; \text{ } \boldsymbol{K}^{d_2}_{N_2N_2} = \begin{bmatrix} \boldsymbol{K}^{s_1,d_2}_{N_2N_2} & \boldsymbol{0} & \boldsymbol{0} \\ 
        \boldsymbol{0} & \boldsymbol{K}^{s_2,d_2}_{N_2N_2} & \boldsymbol{0} \\
        \boldsymbol{0} & \boldsymbol{0} & \boldsymbol{K}^{s_3,d_2}_{N_2N_2} \\   
        \end{bmatrix}
       \end{gathered}
\end{equation}

Since the vector and matrix quantities associated with Eqs. \eqref{Eq246} and \eqref{Eq247} have been fully specified, the quantity $ \Omega_{l}\bigl(\boldsymbol{u},q(\boldsymbol{\tau}),q(\boldsymbol{\gamma}) ; \{\boldsymbol{\zeta}, \boldsymbol{\sigma} \} \bigr) $ given by Eq. \eqref{Eq222} can be computed. From Eqs. \eqref{Eq222} and \eqref{Eq246}, and using the cyclic property of the trace operator (\citealp{Bishop2009}), it can be shown that 

\begin{equation}\label{Eq254}
    \begin{gathered}           \Omega_{l}\Bigl(\boldsymbol{u},q(\boldsymbol{\tau}),q(\boldsymbol{\gamma}) ; \{\boldsymbol{\zeta}, \boldsymbol{\sigma} \} \Bigr) = c_1 - \frac{1}{2} \Biggl[\boldsymbol{y}^T_{l}\boldsymbol{\Sigma}^{-1}_{l}\boldsymbol{y}_{l} - 2\boldsymbol{u}^T\boldsymbol{K}^{-1}_{MM}\underbrace{\mathbb{E}_{q(\boldsymbol{\tau})} \bigl[\mathbb{E}_{q(\boldsymbol{\gamma})} \bigl[ (\boldsymbol{K}^{l}_{NM})^T \bigr] \bigr]}_{\overline{\boldsymbol{\Psi}}^T_1}\boldsymbol{\Sigma}^{-1}_{l}\boldsymbol{y}_{l} \text{ } + \\[1.5ex] \boldsymbol{u}^T\boldsymbol{K}^{-1}_{MM}\underbrace{\mathbb{E}_{q(\boldsymbol{\tau})} \bigl[\mathbb{E}_{q(\boldsymbol{\gamma})}\bigl[\boldsymbol{K}^{l}_{MN}\boldsymbol{\Sigma}^{-1}_{l}\boldsymbol{K}^{l}_{NM} \bigr] \bigr]}_{\overline{\boldsymbol{\Psi}}_2}\boldsymbol{K}^{-1}_{MM}\boldsymbol{u} \Biggr] - \frac{1}{2}\underbrace{\text{tr$ \Bigl\{ \underbrace{\mathbb{E}_{q(\boldsymbol{\tau})} \bigl[\mathbb{E}_{q(\boldsymbol{\gamma})}\bigl[\boldsymbol{K}^{l}_{NN} \bigr] \bigr]}_{\overline{\boldsymbol{\Psi}}_0} \boldsymbol{\Sigma}^{-1}_{l}\Bigr\} $}}_{\overline{{\psi}}_0} \text{ } + 
    \\[1.5ex] \frac{1}{2}\text{tr$ \Bigl\{ \boldsymbol{K}^{-1}_{MM}\underbrace{\mathbb{E}_{q(\boldsymbol{\tau})} \bigl[\mathbb{E}_{q(\boldsymbol{\gamma})}\bigl[\boldsymbol{K}^{l}_{MN}\boldsymbol{\Sigma}^{-1}_{l}\boldsymbol{K}^{l}_{NM} \bigr] \bigr]}_{\overline{\boldsymbol{\Psi}}_2} \Bigr\} $}
    \end{gathered}       
\end{equation}

\noindent Note that after taking the expected values with respect to the approximate variational posterior densities $ q(\boldsymbol{\gamma}) $ and $ q(\boldsymbol{\tau}) $, respectively, the quantity $ \Omega_{l}\bigl(\boldsymbol{u},q(\boldsymbol{\tau}),q(\boldsymbol{\gamma}) ; \{\boldsymbol{\zeta}, \boldsymbol{\sigma} \} \bigr) $ can be compactly rewritten as 

\begin{equation}\label{Eq255}
    \begin{gathered}           \Omega_{l}\Bigl(\boldsymbol{u},q(\boldsymbol{\tau}),q(\boldsymbol{\gamma}) ; \{\boldsymbol{\zeta}, \boldsymbol{\sigma} \} \Bigr) =  c_2 - \frac{1}{2} \Biggl[\boldsymbol{y}^T_{l}\boldsymbol{\Sigma}^{-1}_{l}\boldsymbol{y}_{l} - 2\boldsymbol{u}^T\boldsymbol{K}^{-1}_{MM}\overline{\boldsymbol{\Psi}}^T_1\boldsymbol{\Sigma}^{-1}_{l}\boldsymbol{y}_{l} \text{ } + \boldsymbol{u}^T\boldsymbol{K}^{-1}_{MM}\overline{\boldsymbol{\Psi}}_2\boldsymbol{K}^{-1}_{MM}\boldsymbol{u} \Biggr]
    \end{gathered}       
\end{equation}

\noindent The quantity $ c_2 $ in Eq. \eqref{Eq255} is defined as follows

\begin{equation}\label{Eq256}
    \begin{gathered}          
    c_2 = c_1 - \frac{{\overline{\psi}_0}}{2} + \frac{1}{2}\text{tr$ \Bigl\{ \boldsymbol{K}^{-1}_{MM}\overline{\boldsymbol{\Psi}}_2 \Bigr\} $}
    \end{gathered}       
\end{equation}

Notice from Eq. \eqref{Eq254} that, under the variational inference-based framework, the uncertain inputs associated with $ \boldsymbol{\tau} $ and $ \boldsymbol{\gamma} $ are replaced by the expected values under the approximate variational posterior densities $ q(\boldsymbol{\tau}) $ and $ q(\boldsymbol{\gamma}) $, respectively. Consequently, the practitioner can think of the quantities $ \overline{\psi}_0, \overline{\boldsymbol{\Psi}}_0, \overline{\boldsymbol{\Psi}}_1 $, and $ \overline{\boldsymbol{\Psi}}_2 $ as statistics that must be computed in order the apply the BGP-LVM for river/stream networks. Unfortunately, the statistics associated with Eq. \eqref{Eq254} are only available in a closed-form solution for a few covariance/cross-covariance functions which limits the application of the BGP-LVM for river/stream networks. Recall that the inducing variable prior density (see Sect. 5.2) has the following assumed multivariate Gaussian density functional form

\begin{equation}\label{Eq258}
    p(\boldsymbol{u}) = \mathcal{N}(\boldsymbol{u} \vert \boldsymbol{0},\boldsymbol{K}_{MM})
\end{equation}

To compute the optimal approximate inducing variable posterior density for the case of censored observational data, the following posterior density normalisation constant is required

\begin{equation}\label{Eq259}
    \mathcal{Z}_{l} \bigl(q(\boldsymbol{\tau}),q(\boldsymbol{\gamma}) ; \{\boldsymbol{\zeta}, \boldsymbol{\sigma} \} \bigr) = \mathlarger{ \int\limits_{\boldsymbol{u}} } p(\boldsymbol{u})\exp \Bigl\{\Omega_{l}\bigl(\boldsymbol{u},q(\boldsymbol{\tau}),q(\boldsymbol{\gamma}) ; \{\boldsymbol{\zeta}, \boldsymbol{\sigma} \}\bigr) \Bigr\} d\boldsymbol{u}
\end{equation}

Since both the inducing variable prior density $ p(\boldsymbol{u}) $ and the quantity $ \Omega_{l}\bigl(\boldsymbol{u},q(\boldsymbol{\tau}),q(\boldsymbol{\gamma}) ; \{\boldsymbol{\zeta}, \boldsymbol{\sigma} \} \bigr) $ are multivariate quadratic in $ \boldsymbol{u} $, the normalisation constant in Eq. \eqref{Eq259}, by design of the proposed BGP-LVM methodology, is available in a closed-form solution. More specifically, consider the integrand in Eq. \eqref{Eq259} that can be rewritten, using Eqs. \eqref{Eq255} and \eqref{Eq258}, to obtain

\begin{equation}\label{Eq260}
\begin{gathered}
     p(\boldsymbol{u})\exp \Bigl\{\Omega_{l}\bigl(\boldsymbol{u},q(\boldsymbol{\tau}),q(\boldsymbol{\gamma}) ; \{\boldsymbol{\zeta}, \boldsymbol{\sigma} \}\bigr) \Bigr\} = \\[1.5ex] c_3 \exp \Biggl\{-\frac{1}{2} \Biggl[\boldsymbol{u}^T\underbrace{\boldsymbol{K}^{-1}_{MM}\boldsymbol{K}_{MM}\boldsymbol{K}^{-1}_{MM}}_{\boldsymbol{K}^{-1}_{MM}}\boldsymbol{u} + \boldsymbol{y}^T_{l}\boldsymbol{\Sigma}^{-1}_{l}\boldsymbol{y}_{l} - 2\boldsymbol{u}^T\boldsymbol{K}^{-1}_{MM}\overline{\boldsymbol{\Psi}}^T_1\boldsymbol{\Sigma}^{-1}_{l}\boldsymbol{y}_{l} \text{ } + \boldsymbol{u}^T\boldsymbol{K}^{-1}_{MM}\overline{\boldsymbol{\Psi}}_2\boldsymbol{K}^{-1}_{MM}\boldsymbol{u} \Biggr] \Biggr\} = \\[1.5ex] 
     c_3 \exp \Biggl\{-\frac{1}{2} \Biggl[\boldsymbol{u}^T\boldsymbol{K}^{-1}_{MM}\underbrace{\bigl(\boldsymbol{K}_{MM} + \overline{\boldsymbol{\Psi}}_2 \bigr)}_{\boldsymbol{Q}}\boldsymbol{K}^{-1}_{MM}\boldsymbol{u} - 2\boldsymbol{u}^T\boldsymbol{K}^{-1}_{MM}\overline{\boldsymbol{\Psi}}^T_1\boldsymbol{\Sigma}^{-1}_{l}\boldsymbol{y}_{l} + \boldsymbol{y}^T_{l}\boldsymbol{\Sigma}^{-1}_{l}\boldsymbol{y}_{l} \Biggr] \Biggr\} =\\[1.5ex]
     c_3 \exp \Biggl\{-\frac{1}{2} \Biggl[\boldsymbol{u}^T\boldsymbol{K}^{-1}_{MM}\boldsymbol{Q}\boldsymbol{K}^{-1}_{MM}\boldsymbol{u} - 2\boldsymbol{u}^T\boldsymbol{K}^{-1}_{MM}\overline{\boldsymbol{\Psi}}^T_1\boldsymbol{\Sigma}^{-1}_{l}\boldsymbol{y}_{l} + \boldsymbol{y}^T_{l}\boldsymbol{\Sigma}^{-1}_{l}\boldsymbol{y}_{l} \Biggr] \Biggr\}
     \end{gathered}
\end{equation}

% \noindent The matrix $ \boldsymbol{Q} $ and quantity $ c_3 $ in Eq. \eqref{Eq260} is defined as follows

\noindent The quantity $ c_3 $ in Eq. \eqref{Eq260} is defined as follows

\begin{equation}\label{Eq261}
\begin{gathered}
     %\boldsymbol{Q} = \boldsymbol{K}_{MM} + \overline{\boldsymbol{\Psi}}_2 \\[1.5ex]
     c_3 = \frac{1}{(2\pi)^{\frac{M}{2}}} \frac{1}{\vert \boldsymbol{K}_{MM} \vert^{\frac{1}{2}}} \exp\{ c_2 \}
     \end{gathered}
\end{equation}

\noindent In Eq. \eqref{Eq261} the symbol $ \vert \cdot \vert $ denotes the matrix determinant. For Eq. \eqref{Eq260}, the following definitions, which are based on the multivariate quadratic functional form associated with the Gaussian density, are imposed

\begin{equation}\label{Eq262}
\begin{gathered}
     p(\boldsymbol{u})\exp \Bigl\{\Omega_{l}\bigl(\boldsymbol{u},q(\boldsymbol{\tau}),q(\boldsymbol{\gamma}) ; \{\boldsymbol{\zeta}, \boldsymbol{\sigma} \}\bigr) \Bigr\} = \\[1.5ex]
     c_3 \exp \Biggl\{-\frac{1}{2} \Biggl[\boldsymbol{u}^T\underbrace{\boldsymbol{K}^{-1}_{MM}\boldsymbol{Q}\boldsymbol{K}^{-1}_{MM}}_{\boldsymbol{\Sigma}^{-1}_{\boldsymbol{u}}}\boldsymbol{u} - 2\boldsymbol{u}^T\underbrace{\boldsymbol{K}^{-1}_{MM}\overline{\boldsymbol{\Psi}}^T_1\boldsymbol{\Sigma}^{-1}_{l}\boldsymbol{y}_{l}}_{\boldsymbol{\Sigma}^{-1}_{\boldsymbol{u}}\boldsymbol{\mu}_{\boldsymbol{u}}} + \boldsymbol{y}^T_{l}\boldsymbol{\Sigma}^{-1}_{l}\boldsymbol{y}_{l} \Biggr] \Biggr\} = \\[1.5ex]
     c_3 \exp \Biggl\{-\frac{1}{2} \Biggl[\boldsymbol{u}^T\boldsymbol{\Sigma}^{-1}_{\boldsymbol{u}}\boldsymbol{u} - 2\boldsymbol{u}^T\boldsymbol{\Sigma}^{-1}_{\boldsymbol{u}}\boldsymbol{\mu}_{\boldsymbol{u}} + \boldsymbol{y}^T_{l}\boldsymbol{\Sigma}^{-1}_{l}\boldsymbol{y}_{l} \Biggr] \Biggr\}
     \end{gathered}
\end{equation}

\noindent From Eq. \eqref{Eq262} the multivariate square with respect to the inducing variables $ \boldsymbol{u} $ can be completed by adding and subtracting the term $ \boldsymbol{\mu}^T_{\boldsymbol{u}}\boldsymbol{\Sigma}^{-1}_{\boldsymbol{u}}\boldsymbol{\mu}_{\boldsymbol{u}} $ such that

\begin{equation}\label{Eq263}
\begin{gathered}
     p(\boldsymbol{u})\exp \Bigl\{\Omega_{l}\bigl(\boldsymbol{u},q(\boldsymbol{\tau}),q(\boldsymbol{\gamma}) ; \{\boldsymbol{\zeta}, \boldsymbol{\sigma} \}\bigr) \Bigr\} = 
     % c_3 \exp \Biggl\{-\frac{1}{2} \Biggl[\boldsymbol{u}^T\boldsymbol{\Sigma}^{-1}_{\boldsymbol{u}}\boldsymbol{u} - 2\boldsymbol{u}^T\boldsymbol{\Sigma}^{-1}_{\boldsymbol{u}}\boldsymbol{\mu}_{\boldsymbol{u}} + \boldsymbol{\mu}^T_{\boldsymbol{u}}\boldsymbol{\Sigma}^{-1}_{\boldsymbol{u}}\boldsymbol{\mu}_{\boldsymbol{u}} - \boldsymbol{\mu}^T_{\boldsymbol{u}}\boldsymbol{\Sigma}^{-1}_{\boldsymbol{u}}\boldsymbol{\mu}_{\boldsymbol{u}} + \boldsymbol{y}^T_{l}\boldsymbol{\Sigma}^{-1}_{l}\boldsymbol{y}_{l} \Biggr] \Biggr\} = \\[1.5ex]
     % c_3 \exp \Biggl\{-\frac{1}{2} \Biggl[\boldsymbol{y}^T_{l}\boldsymbol{\Sigma}^{-1}_{l}\boldsymbol{y}_{l} - \boldsymbol{\mu}^T_{\boldsymbol{u}}\boldsymbol{\Sigma}^{-1}_{\boldsymbol{u}}\boldsymbol{\mu}_{\boldsymbol{u}} \Biggr] \Biggr\} \exp \Biggl\{-\frac{1}{2} \Biggl[ (\boldsymbol{u} - \boldsymbol{\mu}_{\boldsymbol{u}})^T\boldsymbol{\Sigma}^{-1}_{\boldsymbol{u}}(\boldsymbol{u} - \boldsymbol{\mu}_{\boldsymbol{u}}) \Biggr] \Biggr\} = \\[1.5ex] 
     c_4 \exp \Biggl\{-\frac{1}{2} \Biggl[ (\boldsymbol{u} - \boldsymbol{\mu}_{\boldsymbol{u}})^T\boldsymbol{\Sigma}^{-1}_{\boldsymbol{u}}(\boldsymbol{u} - \boldsymbol{\mu}_{\boldsymbol{u}}) \Biggr] \Biggr\}
     \end{gathered}
\end{equation}

\noindent The quantity $ c_4 $ in Eq. \eqref{Eq263} is defined as follows

\begin{equation}\label{Eq264}
\begin{gathered}
     c_4 = c_3 \exp \Biggl\{-\frac{1}{2} \Biggl[\boldsymbol{y}^T_{l}\boldsymbol{\Sigma}^{-1}_{l}\boldsymbol{y}_{l} - \boldsymbol{\mu}^T_{\boldsymbol{u}}\boldsymbol{\Sigma}^{-1}_{\boldsymbol{u}}\boldsymbol{\mu}_{\boldsymbol{u}} \Biggr] \Biggr\} 
     \end{gathered}
\end{equation}

Recall from Eq. \eqref{Eq259} that the posterior density normalisation constant $ \mathcal{Z}_{l} \bigl(q(\boldsymbol{\tau}),q(\boldsymbol{\gamma}) ; \{\boldsymbol{\zeta}, \boldsymbol{\sigma} \} \bigr) $ is required to compute the optimal variational posterior density $ q_l(\boldsymbol{u}; \{ q(\boldsymbol{\tau}),q(\boldsymbol{\gamma})\} ; \{\boldsymbol{\zeta}, \boldsymbol{\sigma} \}) $ (see Eq. (68)). From Eq. \eqref{Eq259}, using the results from Eq. \eqref{Eq263}, the posterior density normalisation constant can be computed as follows

\begin{equation}\label{Eq265}
    \mathcal{Z}_{l} \bigl(q(\boldsymbol{\tau}),q(\boldsymbol{\gamma}) ; \{\boldsymbol{\zeta}, \boldsymbol{\sigma} \} \bigr) = c_4 \mathlarger{ \int\limits_{\boldsymbol{u}} } \exp \Bigl\{-\frac{1}{2} \Bigl[ (\boldsymbol{u} - \boldsymbol{\mu}_{\boldsymbol{u}})^T\boldsymbol{\Sigma}^{-1}_{\boldsymbol{u}}(\boldsymbol{u} - \boldsymbol{\mu}_{\boldsymbol{u}}) \Bigr] \Bigr\} d\boldsymbol{u}
\end{equation}

\noindent The integral in Eq. \eqref{Eq265} is recognised as the normalisation constant for a multivariate Gaussian density and evaluates to 

\begin{equation}\label{Eq266}
    \mathcal{Z}_{l} \bigl(q(\boldsymbol{\tau}),q(\boldsymbol{\gamma}) ; \{\boldsymbol{\zeta}, \boldsymbol{\sigma} \} \bigr) = c_4 (2\pi)^{\frac{M}{2}} \vert \boldsymbol{\Sigma}_{\boldsymbol{u}} \vert^{\frac{1}{2}}
\end{equation}

Using the results from Eqs. \eqref{Eq263} and \eqref{Eq266} in conjunction with Eq. (68), and noting that $ \ln p(\boldsymbol{y} \vert \boldsymbol{f}) $ has been substituted with $ \ln p_l(\boldsymbol{y} \vert \boldsymbol{f} ; \{\boldsymbol{\zeta}, \boldsymbol{\sigma} \})$, the optimal variational inducing variable posterior density $ q_l(\boldsymbol{u}; \{ q(\boldsymbol{\tau}),q(\boldsymbol{\gamma})\} ; \{\boldsymbol{\zeta}, \boldsymbol{\sigma} \}) $ can be computed as follows

\begin{equation}\label{Eq267}
\begin{gathered}
    q_l(\boldsymbol{u}; \{ q(\boldsymbol{\tau}),q(\boldsymbol{\gamma})\} ; \{\boldsymbol{\zeta}, \boldsymbol{\sigma} \}) = \cfrac{\bcancel{c_4} \exp \Bigl\{-\frac{1}{2} \Bigl[ (\boldsymbol{u} - \boldsymbol{\mu}_{\boldsymbol{u}})^T\boldsymbol{\Sigma}^{-1}_{\boldsymbol{u}}(\boldsymbol{u} - \boldsymbol{\mu}_{\boldsymbol{u}}) \Bigr] \Bigr\}}{\bcancel{c_4} (2\pi)^{\frac{M}{2}} \vert \boldsymbol{\Sigma}_{\boldsymbol{u}} \vert^{\frac{1}{2}}}
    \end{gathered}
\end{equation} 

Observe from the results outlined in Eq. \eqref{Eq267} that the optimal approximate inducing variable posterior density  $ q_l(\boldsymbol{u}; \{ q(\boldsymbol{\tau}),q(\boldsymbol{\gamma})\} ; \{\boldsymbol{\zeta}, \boldsymbol{\sigma} \}) $ is recognised as a multivariate Gaussian density function that is parameterised by Eqs. (72) to (74) in Sect. 5.4.

\vspace{0.5cm}

\subsection{Analytically Computing The Optimal (Collapsed) Secondary Variational Lower Bound}\label{SI_1_2}

To analytically compute the optimal variational lower bound for the case where the practitioner has access to a censored observational data set, recall Eq. (71) from Sect. 5.4 which is repeated below for convenience

\begin{equation}\label{Eq271}
    \begin{gathered}           
    \mathcal{F}_l^\ast \bigl( \boldsymbol{\theta};\{ q(\boldsymbol{\tau}),q(\eta_\tau),q(\boldsymbol{\gamma}) \} ; \{\boldsymbol{\zeta}, \boldsymbol{\sigma} \}\}\bigr) \text{ } = \text{ } \ln \int\limits_{\boldsymbol{u}} p(\boldsymbol{u})\exp \Bigl\{\Omega_{l}\bigl(\boldsymbol{u},q(\boldsymbol{\tau}),q(\boldsymbol{\gamma}) ; \{\boldsymbol{\zeta}, \boldsymbol{\sigma} \}\bigr) \Bigr\} d\boldsymbol{u} \text{ } + \\[1.5ex] \text{ } \mathbb{E}_{q(\eta_\tau)}\Bigl[-\mathcal{KL} \bigl[q(\boldsymbol{\tau}) \vert \vert p(\boldsymbol{\tau} \vert \eta_\tau) \bigr] \Bigr] \text{ } - \text{ } \mathcal{KL} \bigl[q(\boldsymbol{\gamma}) \vert \vert p(\boldsymbol{\gamma}) \bigr] \text{ } - \text{ } \mathcal{KL} \bigl [q(\eta_\tau) \vert \vert p(\eta_\tau) \bigr]
    \end{gathered}       
\end{equation} 

From Eq. \eqref{Eq271}, notice that the integral associated with the inducing variables $ \boldsymbol{u} $ is also the variational posterior density normalisation constant (see Eq. \eqref{Eq259}) that normalises the density in Eq. (72). Consequently, Eq. \eqref{Eq271} can be rewritten to obtain 

\begin{equation}\label{Eq272}
    \begin{gathered}           
    \mathcal{F}_l^\ast \bigl( \boldsymbol{\theta};\{ q(\boldsymbol{\tau}),q(\eta_\tau),q(\boldsymbol{\gamma}) \} ; \{\boldsymbol{\zeta}, \boldsymbol{\sigma} \}\}\bigr) \text{ } = \text{ } \ln \mathcal{Z}_{l} \bigl(q(\boldsymbol{\tau}),q(\boldsymbol{\gamma}) ; \{\boldsymbol{\zeta}, \boldsymbol{\sigma} \} \bigr) \text{ } + \\[1.5ex] \text{ } \mathbb{E}_{q(\eta_\tau)}\Bigl[-\mathcal{KL} \bigl[q(\boldsymbol{\tau}) \vert \vert p(\boldsymbol{\tau} \vert \eta_\tau) \bigr] \Bigr] \text{ } - \text{ } \mathcal{KL} \bigl[q(\boldsymbol{\gamma}) \vert \vert p(\boldsymbol{\gamma}) \bigr] \text{ } - \text{ } \mathcal{KL} \bigl [q(\eta_\tau) \vert \vert p(\eta_\tau) \bigr]
    \end{gathered}       
\end{equation} 

\noindent The natural logarithmic term $ \ln(\cdot) $ in Eq. \eqref{Eq272} can be further decomposed, using the results from Eqs. \eqref{Eq261}, \eqref{Eq264}, and \eqref{Eq266}, to obtain 

\begin{equation}\label{Eq273}
    \begin{gathered}           
    \ln \mathcal{Z}_{l} \bigl(q(\boldsymbol{\tau}),q(\boldsymbol{\gamma}) ; \{\boldsymbol{\zeta}, \boldsymbol{\sigma} \} \bigr) =  c_2 - \frac{1}{2}\ln \vert \boldsymbol{K}_{MM} \vert  - \frac{1}{2}\Biggl[\boldsymbol{y}^T_{l}\boldsymbol{\Sigma}^{-1}_{l}\boldsymbol{y}_{l} - \boldsymbol{\mu}^T_{\boldsymbol{u}}\boldsymbol{\Sigma}^{-1}_{\boldsymbol{u}}\boldsymbol{\mu}_{\boldsymbol{u}} \Biggr] + \frac{1}{2}\ln \vert \boldsymbol{\Sigma}_{\boldsymbol{u}} \vert 
    \end{gathered}       
\end{equation} 

\noindent From Eq. (73), Eq. \eqref{Eq273} can be further simplified such that

\begin{equation}\label{Eq274}
    \begin{gathered}           
    \ln \mathcal{Z}_{l} \bigl(q(\boldsymbol{\tau}),q(\boldsymbol{\gamma}) ; \{\boldsymbol{\zeta}, \boldsymbol{\sigma} \} \bigr) = \frac{1}{2} \ln \vert \boldsymbol{K}_{MM} \vert - \frac{1}{2} \ln \vert \boldsymbol{Q} \vert + c_2 - \frac{1}{2}\Biggl[\boldsymbol{y}^T_{l}\boldsymbol{\Sigma}^{-1}_{l}\boldsymbol{y}_{l} - \boldsymbol{\mu}^T_{\boldsymbol{u}}\boldsymbol{\Sigma}^{-1}_{\boldsymbol{u}}\boldsymbol{\mu}_{\boldsymbol{u}} \Biggr]
    \end{gathered}       
\end{equation} 

\noindent The term in square brackets in Eq. \eqref{Eq274}, using the results in Eqs. \eqref{Eq262}, (73), and (74), can be rewritten to obtain

\begin{equation}\label{Eq275}
    \begin{gathered}           
    \Biggl[\boldsymbol{y}^T_{l}\boldsymbol{\Sigma}^{-1}_{l}\boldsymbol{y}_{l} - \boldsymbol{\mu}^T_{\boldsymbol{u}}\boldsymbol{\Sigma}^{-1}_{\boldsymbol{u}}\boldsymbol{\mu}_{\boldsymbol{u}} \Biggr] = 
    \Biggl[\boldsymbol{y}^T_{l}\boldsymbol{\Sigma}^{-1}_{l}\boldsymbol{y}_{l} - \boldsymbol{y}^T_{l}\boldsymbol{\Sigma}^{-1}_{l}\overline{\boldsymbol{\Psi}}_1\boldsymbol{Q}^{-1}\underbrace{\boldsymbol{K}_{MM} \boldsymbol{K}^{-1}_{MM}}_{\boldsymbol{I}_{MM}}\overline{\boldsymbol{\Psi}}^T_1\boldsymbol{\Sigma}^{-1}_{l}\boldsymbol{y}_{l} \Biggr] = \\[1.5ex]
    \Biggl[\boldsymbol{y}^T_{l}\underbrace{\bigl(\boldsymbol{\Sigma}^{-1}_{l} - \boldsymbol{\Sigma}^{-1}_{l}\overline{\boldsymbol{\Psi}}_1\boldsymbol{Q}^{-1}\overline{\boldsymbol{\Psi}}^T_1\boldsymbol{\Sigma}^{-1}_{l} \bigr)}_{\boldsymbol{A}}\boldsymbol{y}_{l}\Biggr] %= \\[1.5ex]
    %\boldsymbol{y}^T_{l}\boldsymbol{A}\boldsymbol{y}_{l}
    \end{gathered}       
\end{equation} 

\noindent Using the results from Eq. \eqref{Eq275}, the logarithmic term in Eq. \eqref{Eq274} simplifies to

\begin{equation}\label{Eq276}
    \begin{gathered}           
    \ln \mathcal{Z}_{l} \bigl(q(\boldsymbol{\tau}),q(\boldsymbol{\gamma}) ; \{\boldsymbol{\zeta}, \boldsymbol{\sigma} \} \bigr) =  \frac{1}{2} \ln \vert \boldsymbol{K}_{MM} \vert - \frac{1}{2} \ln \vert \boldsymbol{Q} \vert  - \frac{1}{2}\boldsymbol{y}^T_{l}\boldsymbol{A}\boldsymbol{y}_{l} + c_2
    \end{gathered}       
\end{equation} 

% \noindent From Eq. \eqref{Eq275}, the matrix $ \boldsymbol{A} $ is defined as follows

% \begin{equation}\label{Eq277}
%     \begin{gathered}           
%     \boldsymbol{A} = \boldsymbol{\Sigma}^{-1}_{l} - \boldsymbol{\Sigma}^{-1}_{l}\overline{\boldsymbol{\Psi}}_1\boldsymbol{Q}^{-1}\overline{\boldsymbol{\Psi}}^T_1\boldsymbol{\Sigma}^{-1}_{l}
%     \end{gathered}       
% \end{equation}

\noindent The last step that is required to simplify the logarithmic term in Eq. \eqref{Eq276} would be to substitute the expression associated with the quantity $ c_2 $. Using the results from Eqs. \eqref{Eq247} and \eqref{Eq256}, the logarithmic term in Eq. \eqref{Eq276} can be rewritten to obtain

\begin{equation}\label{Eq278}
    \begin{gathered}           
    \ln \mathcal{Z}_{l} \bigl(q(\boldsymbol{\tau}),q(\boldsymbol{\gamma}) ; \{\boldsymbol{\zeta}, \boldsymbol{\sigma} \} \bigr) = \\[1.5ex] \frac{1}{2} \ln \vert \boldsymbol{K}_{MM} \vert - \frac{1}{2} \ln \vert \boldsymbol{Q} \vert  - \frac{1}{2}\boldsymbol{y}^T_{l}\boldsymbol{A}\boldsymbol{y}_{l} + \Biggl[\frac{1}{2}\boldsymbol{b}^T\boldsymbol{\Sigma}^{-1}_c\boldsymbol{b} + \boldsymbol{c}^T\boldsymbol{\Sigma}^{-1}_c\boldsymbol{1}^{\ast} - \boldsymbol{b}^T\boldsymbol{\Sigma}^{-1}_c\boldsymbol{d} \Biggr] - \\[1.5ex] \text{ } \frac{1}{2} \Biggl[N_{1,[>q_1]} \ln(2\pi\sigma^2_{1}) + N_{2,[>q_2]} \ln(2\pi\sigma^2_{2}) \Biggr] - \frac{{\overline{\psi}_0}}{2} + \frac{1}{2}\text{tr$ \Bigl\{ \boldsymbol{K}^{-1}_{MM}\overline{\boldsymbol{\Psi}}_2 \Bigr\} $}  
    \end{gathered}       
\end{equation} 

From Eq. \eqref{Eq271}, the collapsed secondary variational lower bound, using the results from Eq. \eqref{Eq278}, is then given by Eq. (75) in Sect. 5.4. Note that all $ \mathcal{KL} $-divergence quantities associated with Eqs. \eqref{Eq271} and (75) are available in a closed-form solution for the choice of prior and approximate variational posterior densities used in this work.

\vspace{0.5cm}

\section{Derivation - BGP-LVM Predictions For River/Stream Networks}\label{Section8}

River/stream network-based BGP-LVM latent function predictions, collectively denoted with the latent function prediction vector $ \boldsymbol{f}^{\ast} = [(\boldsymbol{f}^{\ast}_{1})^T,(\boldsymbol{f}^{\ast}_{2})^T, \cdots, (\boldsymbol{f}^{\ast}_{K_f})^T]^T $, at unobserved temporal input locations and known sampled spatial locations are in line with the prediction framework outlined in \cite{Titsias2010}, \cite{Damianou2011}, \cite{Titsias2013}, and \cite{Damianou2016}. Starting from the joint density, the latent function predictive density can be derived as follows

\begin{equation}\label{Eq281}
    \begin{gathered}           
    p(\boldsymbol{f}^{\ast} \vert \boldsymbol{y}) =  \int\limits_{\boldsymbol{\tau}} \int\limits_{\boldsymbol{\gamma}} \int\limits_{\boldsymbol{u}} \int\limits_{\eta_{\tau}} \int\limits_{\boldsymbol{f}} p(\boldsymbol{f}^{\ast},\boldsymbol{f}, \boldsymbol{u}, \boldsymbol{\tau}, \eta_{\tau}, \boldsymbol{\gamma} \vert \boldsymbol{y}) d\boldsymbol{f} d\eta_{\tau} d\boldsymbol{u} d\boldsymbol{\gamma}d \boldsymbol{\tau} = \\[1.5ex]
    \int\limits_{\boldsymbol{\tau}} \int\limits_{\boldsymbol{\gamma}} \int\limits_{\boldsymbol{u}} \int\limits_{\eta_{\tau}} \int\limits_{\boldsymbol{f}} p(\boldsymbol{f}^{\ast} \vert \boldsymbol{f}, \boldsymbol{u}, \boldsymbol{\tau}, \eta_{\tau}, \boldsymbol{\gamma}, \boldsymbol{y})p(\boldsymbol{f}, \boldsymbol{u}, \boldsymbol{\tau}, \eta_{\tau}, \boldsymbol{\gamma} \vert \boldsymbol{y}) d\boldsymbol{f} d\eta_{\tau} d\boldsymbol{u} d\boldsymbol{\gamma} d\boldsymbol{\tau}
    \end{gathered}       
\end{equation}

Given that $ \boldsymbol{f}^{\ast} $ is conditionally independent of $ \boldsymbol{f},\eta_{\tau} $, and $ \boldsymbol{y} $ given $ \boldsymbol{u}, \boldsymbol{\gamma} $, and $ \boldsymbol{\tau} $, Eq. \eqref{Eq281} can be simplified to obtain

\begin{equation}\label{Eq282}
    \begin{gathered}           
    p(\boldsymbol{f}^{\ast} \vert \boldsymbol{y}) = 
    \int\limits_{\boldsymbol{\tau}} \int\limits_{\boldsymbol{\gamma}} \int\limits_{\boldsymbol{u}} \int\limits_{\eta_{\tau}} \int\limits_{\boldsymbol{f}} p(\boldsymbol{f}^{\ast} \vert \boldsymbol{u}, \boldsymbol{\tau}, \boldsymbol{\gamma})p(\boldsymbol{f}, \boldsymbol{u}, \boldsymbol{\tau}, \eta_{\tau}, \boldsymbol{\gamma} \vert \boldsymbol{y}) d\boldsymbol{f} d\eta_{\tau} d\boldsymbol{u} d\boldsymbol{\gamma} d\boldsymbol{\tau}
    \end{gathered}       
\end{equation}

Recall that the true underlying posterior density $ p(\boldsymbol{f}, \boldsymbol{u}, \boldsymbol{\tau}, \eta_{\tau}, \boldsymbol{\gamma} \vert \boldsymbol{y}) $ is approximated by the variational posterior density $ q(\boldsymbol{f}, \boldsymbol{u}, \boldsymbol{\tau}, \eta_{\tau}, \boldsymbol{\gamma}) $, in other words, 

\begin{equation}\label{Eq283}
    \begin{gathered}           
    p(\boldsymbol{f}, \boldsymbol{u}, \boldsymbol{\tau}, \eta_{\tau}, \boldsymbol{\gamma} \vert \boldsymbol{y}) \approx q(\boldsymbol{f}, \boldsymbol{u}, \boldsymbol{\tau}, \eta_{\tau}, \boldsymbol{\gamma}) 
    \end{gathered}       
\end{equation}

\noindent Consequently, from Eq. \eqref{Eq283}, Eq. \eqref{Eq282} can be approximated as follows

\begin{equation}\label{Eq284}
    \begin{gathered}           
    p(\boldsymbol{f}^{\ast} \vert \boldsymbol{y}) \approx  q(\boldsymbol{f}^{\ast}) = \int\limits_{\boldsymbol{\tau}} \int\limits_{\boldsymbol{\gamma}} \int\limits_{\boldsymbol{u}} \int\limits_{\eta_{\tau}} \int\limits_{\boldsymbol{f}} p(\boldsymbol{f}^{\ast} \vert \boldsymbol{u}, \boldsymbol{\tau}, \boldsymbol{\gamma})q(\boldsymbol{f}, \boldsymbol{u}, \boldsymbol{\tau}, \eta_{\tau}, \boldsymbol{\gamma})  d\boldsymbol{f} d\eta_{\tau} d\boldsymbol{u} d\boldsymbol{\gamma} d\boldsymbol{\tau}
    \end{gathered}       
\end{equation}

\noindent Using Eq. (63), Eq. \eqref{Eq284} can be further expanded to obtain

\begin{equation}\label{Eq285}
    \begin{gathered}           
    q(\boldsymbol{f}^{\ast}) =\int\limits_{\boldsymbol{\tau}} \int\limits_{\boldsymbol{\gamma}} \int\limits_{\boldsymbol{u}} \int\limits_{\eta_{\tau}} \int\limits_{\boldsymbol{f}} p(\boldsymbol{f}^{\ast} \vert \boldsymbol{u}, \boldsymbol{\tau}, \boldsymbol{\gamma})p(\boldsymbol{f} \vert \boldsymbol{u}, \boldsymbol{\tau}, \boldsymbol{\gamma})q(\boldsymbol{u})q(\boldsymbol{\tau})q(\eta_{\tau})q(\boldsymbol{\gamma})  d\boldsymbol{f} d\eta_{\tau} d\boldsymbol{u} d\boldsymbol{\gamma} d\boldsymbol{\tau}
    \end{gathered}       
\end{equation}

From Eq. \eqref{Eq285}, the following integral-based expression for the approximate latent function predictive posterior density can be obtained

\begin{equation}\label{Eq287}
    \begin{gathered}           
    q(\boldsymbol{f}^{\ast}) = 
    % \int\limits_{\boldsymbol{\tau}} \int\limits_{\boldsymbol{\gamma}} \int\limits_{\boldsymbol{u}} p(\boldsymbol{f}^{\ast} \vert \boldsymbol{u}, \boldsymbol{\tau}, \boldsymbol{\gamma})q(\boldsymbol{u})q(\boldsymbol{\tau})q(\boldsymbol{\gamma})d\boldsymbol{u} d\boldsymbol{\gamma} d\boldsymbol{\tau} \\[1.5ex]
    % \text{ } \text{ } \text{ } \text{ } \text{ } \text{ } \text{ } \text{ } \text{ } \text{ } \text{ } \text{ } \text{ } \text{ } \text{ }
    \int\limits_{\boldsymbol{\tau}} \int\limits_{\boldsymbol{\gamma}} q(\boldsymbol{\tau})q(\boldsymbol{\gamma}) \underbrace{\Biggl[ \text{ } \int\limits_{\boldsymbol{u}} p(\boldsymbol{f}^{\ast} \vert \boldsymbol{u}, \boldsymbol{\tau}, \boldsymbol{\gamma})q(\boldsymbol{u})d\boldsymbol{u} \Biggr]}_{q(\boldsymbol{f}^{\ast} \vert \boldsymbol{\tau}, \boldsymbol{\gamma})}d\boldsymbol{\gamma} d\boldsymbol{\tau} 
    \end{gathered}       
\end{equation}

\noindent The integral in square brackets, as associated with the density $ q(\boldsymbol{f}^{\ast} \vert \boldsymbol{\tau}, \boldsymbol{\gamma}) $ in Eq. \eqref{Eq287}, can be solved for analytically. From Eqs. (47) to (48), observe that the conditional density $ p(\boldsymbol{f}^{\ast} \vert \boldsymbol{u}, \boldsymbol{\tau}, \boldsymbol{\gamma}) $ also takes the form of a multivariate Gaussian density such that

\begin{equation}\label{Eq288}
    \begin{gathered}           
    p(\boldsymbol{f}^{\ast} \vert \boldsymbol{u}, \boldsymbol{\tau}, \boldsymbol{\gamma}) = \mathcal{N}(\boldsymbol{f}^{\ast} \vert \boldsymbol{K}_{N^{\ast}M}\boldsymbol{K}^{-1}_{MM}\boldsymbol{u},\boldsymbol{K}_{N^{\ast}N^{\ast}} - \boldsymbol{K}_{N^{\ast}M}\boldsymbol{K}^{-1}_{MM}\boldsymbol{K}_{MN^{\ast}})
    \end{gathered}       
\end{equation}

\noindent The matrices $ \boldsymbol{K}_{N^{\ast}N^{\ast}},\boldsymbol{K}_{N^{\ast}M}, $ and $ \boldsymbol{K}_{MN^{\ast}} $ are computed analogously to $ \boldsymbol{K}_{NN},\boldsymbol{K}_{NM}, $ and $ \boldsymbol{K}_{MN} $, however, now at temporal instances where the practitioner wishes to make predictions about the sampled spatial locations, for example, $ s_1, s_2, $ and $ s_3 $ in Figs. \ref{fig:Figure7} and 4. From Eqs. (72) to (74), \eqref{Eq287}, and \eqref{Eq288}, note that the density $ q(\boldsymbol{f}^{\ast} \vert \boldsymbol{\tau}, \boldsymbol{\gamma}) $, for the case of censored observational data, can be computed as follows

\begin{equation}\label{Eq289}
    \begin{gathered}           
     q(\boldsymbol{f}^{\ast} \vert \boldsymbol{\tau}, \boldsymbol{\gamma}) = \int\limits_{\boldsymbol{u}} p(\boldsymbol{f}^{\ast} \vert \boldsymbol{u}, \boldsymbol{\tau}, \boldsymbol{\gamma})q_l(\boldsymbol{u}; \{ q(\boldsymbol{\tau}),q(\boldsymbol{\gamma})\} ; \{\boldsymbol{\zeta}, \boldsymbol{\sigma} \})d\boldsymbol{u} = \\[1.5ex] 
     \int\limits_{\boldsymbol{u}} \mathcal{N}(\boldsymbol{f}^{\ast} \vert \boldsymbol{K}_{N^{\ast}M}\boldsymbol{K}^{-1}_{MM}\boldsymbol{u},\boldsymbol{K}_{N^{\ast}N^{\ast}} - \boldsymbol{K}_{N^{\ast}M}\boldsymbol{K}^{-1}_{MM}\boldsymbol{K}_{MN^{\ast}}) \mathcal{N}(\boldsymbol{u} \vert \boldsymbol{\mu}_{\boldsymbol{u}},\boldsymbol{\Sigma}_{\boldsymbol{u}})d\boldsymbol{u} 
    \end{gathered}       
\end{equation}

\noindent Note that, from a notational perspective, the conditional density $ q(\boldsymbol{f}^{\ast} \vert \boldsymbol{\tau}, \boldsymbol{\gamma}) $ in Eq. \eqref{Eq289} should be denoted by the symbol $ q(\boldsymbol{f}^{\ast} \vert \boldsymbol{\tau}, \boldsymbol{\gamma}; \{ q(\boldsymbol{\tau}),q(\boldsymbol{\gamma})\}; \{\boldsymbol{\zeta}, \boldsymbol{\sigma} \}) $ based on the optimal variational posterior density notation associated with $ q_l(\boldsymbol{u}; \{ q(\boldsymbol{\tau}),q(\boldsymbol{\gamma})\} ; \{\boldsymbol{\zeta}, \boldsymbol{\sigma} \}) $. However, for notational convenience, the authors denote the conditional density as $ q(\boldsymbol{f}^{\ast} \vert \boldsymbol{\tau}, \boldsymbol{\gamma}) $. 
%Furthermore, for the case of fully observed or missing data (see Sect. \ref{Section5.1}), the same integral expression in Eq. \eqref{Eq289} can be used, however, $ q_l(\boldsymbol{u}; \{ q(\boldsymbol{\tau}),q(\boldsymbol{\gamma})\} ; \{\boldsymbol{\zeta}, \boldsymbol{\sigma} \}) $ can be substituted with $ q(\boldsymbol{u}; \{ q(\boldsymbol{\tau}),q(\boldsymbol{\gamma}) \}) $ (see Eq. \eqref{Eq79}). 
From Eq. \eqref{Eq289}, it can be shown that the conditional density $ q(\boldsymbol{f}^{\ast} \vert \boldsymbol{\tau}, \boldsymbol{\gamma}) $ corresponds to a multivariate Gaussian density parameterised by

\begin{equation}\label{Eq290}
    \begin{gathered}           
     q(\boldsymbol{f}^{\ast} \vert \boldsymbol{\tau}, \boldsymbol{\gamma}) = \mathcal{N}\bigl(\boldsymbol{f}^{\ast} \vert \boldsymbol{\mu}_{{f}^{\ast}}(\boldsymbol{\tau}, \boldsymbol{\gamma}),\boldsymbol{\Sigma}_{{f}^{\ast}}(\boldsymbol{\tau}, \boldsymbol{\gamma})\bigr) \\[1.5ex] \boldsymbol{\mu}_{{f}^{\ast}}(\boldsymbol{\tau}, \boldsymbol{\gamma}) = \boldsymbol{K}_{N^{\ast}M}\boldsymbol{Q}^{-1}\overline{\boldsymbol{\Psi}}^T_1\boldsymbol{\Sigma}^{-1}_{l}\boldsymbol{y}_{l}; \text{ } \boldsymbol{\Sigma}_{{f}^{\ast}}(\boldsymbol{\tau}, \boldsymbol{\gamma}) = \boldsymbol{K}_{N^{\ast}N^{\ast}} - \boldsymbol{K}_{N^{\ast}M}\boldsymbol{K}^{-1}_{MM}\boldsymbol{K}_{MN^{\ast}} + \boldsymbol{K}_{N^{\ast}M}\boldsymbol{Q}^{-1}\boldsymbol{K}_{MN^{\ast}}
    \end{gathered}       
\end{equation}

\noindent Observe from Eq. \eqref{Eq290} that the mean vector $ \boldsymbol{\mu}_{{f}^{\ast}}(\boldsymbol{\tau}, \boldsymbol{\gamma}) $ and covariance matrix $ \boldsymbol{\Sigma}_{{f}^{\ast}}(\boldsymbol{\tau}, \boldsymbol{\gamma}) $ have an explicit dependence on the uncertain inputs associated with $ \boldsymbol{\tau} $ and $ \boldsymbol{\gamma} $. The functional dependency structure is introduced through the matrices $ \boldsymbol{K}_{N^{\ast}N^{\ast}},\boldsymbol{K}_{N^{\ast}M}, $ and $ \boldsymbol{K}_{MN^{\ast}} $ that explicitly depends on $ \boldsymbol{\tau} $ and $ \boldsymbol{\gamma} $. With Eq. \eqref{Eq290}, Eq. \eqref{Eq287} can be rewritten to obtain 

\begin{equation}\label{Eq291}
    \begin{gathered}           
    q(\boldsymbol{f}^{\ast})  = \int\limits_{\boldsymbol{\tau}} \int\limits_{\boldsymbol{\gamma}} q(\boldsymbol{\tau})q(\boldsymbol{\gamma}) \mathcal{N}\bigl(\boldsymbol{f}^{\ast} \vert \boldsymbol{\mu}_{{f}^{\ast}}(\boldsymbol{\tau}, \boldsymbol{\gamma}),\boldsymbol{\Sigma}_{{f}^{\ast}}(\boldsymbol{\tau}, \boldsymbol{\gamma}) \bigr) d\boldsymbol{\gamma} d\boldsymbol{\tau} 
    \end{gathered}       
\end{equation}

\noindent Note that, in general, the integral associated with Eq. \eqref{Eq291} is analytically intractable since the mean vector $ \boldsymbol{\mu}_{{f}^{\ast}}(\boldsymbol{\tau}, \boldsymbol{\gamma}) $ and covariance matrix $ \boldsymbol{\Sigma}_{{f}^{\ast}}(\boldsymbol{\tau}, \boldsymbol{\gamma}) $ are nonlinear functions of $ \boldsymbol{\tau} $ and $ \boldsymbol{\gamma} $. To make progress in finding a potential mechanism for approximating the integral in Eq. \eqref{Eq291}, consider the running example associated with Figs. \ref{fig:Figure7} and 4. Associated with the running example are the following approximate variational posterior densities that have been rewritten, purely for mathematical convenience, using properties of the Gaussian density function

\begin{equation}\label{Eq292}
    \begin{gathered}           
    q(\boldsymbol{\tau}) = q(\tau_1)q(\tau_2)q(\tau_3) =  %\mathcal{N} \bigl(\tau_1 \vert \mu_{\tau_{1}},\sigma^2_{\tau_{1}} \bigr) \mathcal{N} \bigl(\tau_2 \vert \mu_{\tau_{2}},\sigma^2_{\tau_{2}} \bigr) \mathcal{N} \bigl(\tau_3 \vert \mu_{\tau_{3}},\sigma^2_{\tau_{3}} \bigr) = \\[1.5ex]
    \mathcal{N}\mathlarger{\Biggl(} \underbrace{\begin{bmatrix} \tau_1 \\ \tau_2 \\ \tau_3  \end{bmatrix}}_{\boldsymbol{\tau}} \Biggl\vert \underbrace{\begin{bmatrix} \mu_{\tau_{1}} \\ \mu_{\tau_{2}} \\ \mu_{\tau_{3}}  \end{bmatrix}}_{\boldsymbol{\mu}_{\tau}}, \underbrace{\begin{bmatrix} \sigma^2_{\tau_{1}} & 0 & 0 \\ 0 & \sigma^2_{\tau_{2}} & 0 \\ 0 & 0 & \sigma^2_{\tau_{3}}  \end{bmatrix}}_{\boldsymbol{\Sigma}_{\tau}} \mathlarger{\Biggr)} = 
    \mathcal{N}( \boldsymbol{\tau} \vert \boldsymbol{\mu}_{\tau}, \boldsymbol{\Sigma}_{\tau} )
    \end{gathered}       
\end{equation}

\begin{equation}\label{Eq293}
    \begin{gathered}           
    q(\boldsymbol{\gamma}) = q(\gamma_2)q(\gamma_3) = %\mathcal{N} \bigl(\gamma_2 \vert \mu_{\gamma_{2}},\sigma^2_{\gamma_{2}} \bigr) \mathcal{N} \bigl(\gamma_3 \vert \mu_{\gamma_{3}},\sigma^2_{\gamma_{3}} \bigr) = \\[1.5ex]
    \mathcal{N}\Biggl( \underbrace{\begin{bmatrix} \gamma_2 \\ \gamma_3  \end{bmatrix}}_{\boldsymbol{\gamma}} \Bigl\vert \underbrace{\begin{bmatrix}  \mu_{\gamma_{2}} \\ \mu_{\gamma_{3}}  \end{bmatrix}}_{\boldsymbol{\mu}_{\gamma}}, \underbrace{\begin{bmatrix}  \sigma^2_{\gamma_{2}} & 0 \\ 0 & \sigma^2_{\gamma_{3}}  \end{bmatrix}}_{\boldsymbol{\Sigma}_{\gamma}} \Biggr) = 
    \mathcal{N}( \boldsymbol{\gamma} \vert \boldsymbol{\mu}_{\gamma}, \boldsymbol{\Sigma}_{\gamma} )
    \end{gathered}       
\end{equation}

\noindent The densities in Eqs. \eqref{Eq292} and \eqref{Eq293}, again using properties of the multivariate Gaussian density function, can be compactly written as

\begin{equation}\label{Eq294}
    \begin{gathered}           
    q(\boldsymbol{x}) = q(\boldsymbol{\tau})q(\boldsymbol{\gamma}) =  \mathcal{N}( \boldsymbol{\tau} \vert \boldsymbol{\mu}_{\tau}, \boldsymbol{\Sigma}_{\tau} )\mathcal{N}( \boldsymbol{\gamma} \vert \boldsymbol{\mu}_{\gamma}, \boldsymbol{\Sigma}_{\gamma} )  =  \mathcal{N}\Biggl( \underbrace{\begin{bmatrix} \boldsymbol{\tau} \\ \boldsymbol{\gamma}  \end{bmatrix}}_{\boldsymbol{x}} \Bigl\vert \underbrace{\begin{bmatrix}  \boldsymbol{\mu}_{\tau} \\ \boldsymbol{\mu}_{\gamma}  \end{bmatrix}}_{\boldsymbol{\mu}_{x}}, \underbrace{\begin{bmatrix}  \boldsymbol{\Sigma}_{\tau} & \boldsymbol{0} \\ \boldsymbol{0} & \boldsymbol{\Sigma}_{\gamma}  \end{bmatrix}}_{\boldsymbol{\Sigma}_{x}} \Biggr) = 
    \mathcal{N}(\boldsymbol{x} \vert \boldsymbol{\mu}_{x}, \boldsymbol{\Sigma}_{x})
    \end{gathered}       
\end{equation}

\noindent With the definitions from Eq. \eqref{Eq292} to \eqref{Eq294}, the integral-based expression for the approximate latent function predictive density (see Eq. \eqref{Eq291}) can be rewritten as 

\begin{equation}\label{Eq295}
    \begin{gathered}           
    q(\boldsymbol{f}^{\ast})  = \int\limits_{\boldsymbol{x}} q(\boldsymbol{x}) \mathcal{N}\bigl(\boldsymbol{f}^{\ast} \vert \boldsymbol{\mu}_{{f}^{\ast}}(\boldsymbol{x}),\boldsymbol{\Sigma}_{{f}^{\ast}}(\boldsymbol{x}) \bigr) d\boldsymbol{x} = 
    \int\limits_{\boldsymbol{x}} \mathcal{N}(\boldsymbol{x} \vert \boldsymbol{\mu}_{x}, \boldsymbol{\Sigma}_{x}) \mathcal{N}\bigl(\boldsymbol{f}^{\ast} \vert \boldsymbol{\mu}_{{f}^{\ast}}(\boldsymbol{x}),\boldsymbol{\Sigma}_{{f}^{\ast}}(\boldsymbol{x}) \bigr) d\boldsymbol{x} 
    \end{gathered}       
\end{equation}

Note that despite rewriting the integral-based expression for the approximate predictive posterior density $ q(\boldsymbol{f}^{\ast}) $ in terms of the vector $ \boldsymbol{x} $, the resulting integral in Eq. \eqref{Eq295} is still analytically intractable. One potential mechanism for approximating the predictive density in Eq. \eqref{Eq295} would be to apply Monte Carlo integration which requires drawing independent samples from the Gaussian density $ q(\boldsymbol{x}) $ given by Eq. \eqref{Eq294}. A second approach to approximating the predictive density in Eq. \eqref{Eq295} would be to use a deterministic approximation which is the avenue that the authors pursue in this work. From Eq. \eqref{Eq290}, consider the $ a^{\text{th}} $ conditional latent function (marginal) predictive density, which explicitly depends on $ \boldsymbol{x} $, as associated with a sampled spatial location $ s \in \sample_a^{(s)} $ at arbitrary temporal prediction input location $ t^{\ast} $,  given by

\begin{equation}\label{Eq296}
    \begin{gathered}           
     q({f}^{\ast}_{a}(s,t^{\ast}) \vert \boldsymbol{x}) = \mathcal{N}\bigl({f}^{\ast}_{a}(s,t^{\ast}) \vert {\mu}_{{f}^{\ast}_{a}}(s,t^{\ast};\{\boldsymbol{x}\}),{\sigma}^2_{{f}^{\ast}_{a}}(s,t^{\ast};\{\boldsymbol{x}\})\bigr) \\[1.5ex] {\mu}_{{f}^{\ast}_{a}}(s,t^{\ast};\{\boldsymbol{x}\}) = \boldsymbol{k}^{s,t^{\ast}}_{N^{\ast}M}\boldsymbol{Q}^{-1}\overline{\boldsymbol{\Psi}}^T_1\boldsymbol{\Sigma}^{-1}_{l}\boldsymbol{y}_{l}
     \\[1.5ex] {\sigma}^2_{{f}^{\ast}_{a}}(s,t^{\ast};\{\boldsymbol{x}\}) = {k}^{s,t^{\ast}}_{N^{\ast}N^{\ast}} - \boldsymbol{k}^{s,t^{\ast}}_{N^{\ast}M}\boldsymbol{K}^{-1}_{MM}\boldsymbol{k}^{s,t^{\ast}}_{MN^{\ast}} + \boldsymbol{k}^{s,t^{\ast}}_{N^{\ast}M}\boldsymbol{Q}^{-1}\boldsymbol{k}^{s,t^{\ast}}_{MN^{\ast}}
    \end{gathered}       
\end{equation}

In Eq. \eqref{Eq296}, the scalar $ {k}^{s,t^{\ast}}_{N^{\ast}N^{\ast}} $ and $ 1 \times M $ row vector $ \boldsymbol{k}^{s,t^{\ast}}_{N^{\ast}M} $, with $ \boldsymbol{k}^{s,t^{\ast}}_{MN^{\ast}} = \bigl(\boldsymbol{k}^{s,t^{\ast}}_{N^{\ast}M} \bigr)^T $, denote the variance and cross-covariance, respectively, as associated with sampled spatial location $ s $ at temporal prediction input location $ t^{\ast} $, and can be computed using the covariance/cross-covariance function construction procedure from Sect. 5.2. The $ a^{\text{th}} $ latent function marginal predictive density $ q({f}^{\ast}_{a}(s,t^{\ast})) $, using the results from Eq. \eqref{Eq296}, can then be computed as follows

\begin{equation}\label{Eq297}
    \begin{gathered}           
    q({f}^{\ast}_{a}(s,t^{\ast})) = 
    \int\limits_{\boldsymbol{x}} \mathcal{N}(\boldsymbol{x} \vert \boldsymbol{\mu}_{x}, \boldsymbol{\Sigma}_{x}) \mathcal{N}\bigl({f}^{\ast}_{a}(s,t^{\ast}) \vert {\mu}_{{f}^{\ast}_{a}}(s,t^{\ast};\{\boldsymbol{x}\}),{\sigma}^2_{{f}^{\ast}_{a}}(s,t^{\ast};\{\boldsymbol{x}\})\bigr) d\boldsymbol{x} 
    \end{gathered}       
\end{equation}

Following the ideas outlined in \cite{Girard2003}, \cite{Quinonero2003}, \cite{Titsias2010}, \cite{Damianou2011}, \cite{Titsias2013}, and \cite{Damianou2016}, the mean and variance of the predictive density $ q({f}^{\ast}_{a}(s,t^{\ast})) $ in Eq. \eqref{Eq297}, for a select few covariance/cross-covariance functions, can be computed analytically using $ \{1\}$ the law of iterated expectations and $ \{2\} $ the law of total variance. With the computed mean and variance, the marginal predictive density in Eq. \eqref{Eq297} can be approximated with a Gaussian density such that 

\begin{equation}\label{Eq298}
    \begin{gathered}           
    q({f}^{\ast}_{a}(s,t^{\ast})) \approx \mathcal{N}({f}^{\ast}_{a}(s,t^{\ast}) \vert {\mu}_{{f}^{\ast}_{a}}(s,t^{\ast}),{\sigma}^2_{{f}^{\ast}_{a}}(s,t^{\ast})) \\[1.5ex] {\mu}_{{f}^{\ast}_{a}}(s,t^{\ast}) = \mathbb{E}_{q(\boldsymbol{x})}\bigl[{\mu}_{{f}^{\ast}_{a}}(s,t^{\ast};\{\boldsymbol{x}\}) \bigr]
    \\[1.5ex]
    {\sigma}^2_{{f}^{\ast}_{a}}(s,t^{\ast}) = \mathbb{E}_{q(\boldsymbol{x}))} \bigl[{\sigma}^2_{{f}^{\ast}_{a}}(s,t^{\ast};\{\boldsymbol{x}\})\bigr) \bigr] + \mathbb{E}_{q(\boldsymbol{x})}\bigl[\bigl({\mu}_{{f}^{\ast}_{a}}(s,t^{\ast};\{\boldsymbol{x}\})\bigr)^2 \bigr] - \bigl( {\mu}_{{f}^{\ast}_{a}}(s,t^{\ast})\bigr)^2
    \end{gathered}       
\end{equation}

\noindent The authors would like to explicitly point out that assuming that the marginal predictive density in Eq. \eqref{Eq297} can be approximated with a Gaussian density function is a limitation of the current framework and its predictive capacity. A lot of time and effort has been given to developing a consistent, closed-form, deterministic approximation as a means of performing latent function inference, however, this consistent approach is violated in the model latent function prediction stage when assuming that the marginal predictive density in Eq. \eqref{Eq297} can be approximated with a Gaussian density function. The authors anticipate that the Gaussian density function approximation may give inaccurate marginal latent function prediction results if the density in Eq. \eqref{Eq297} is skewed, however, in this work, and the work of the authors listed previously, the Gaussian density function approximation has been implemented with great success.

The mean $ {\mu}_{{f}^{\ast}_{a}}(s,t^{\ast}) $ in Eq. \eqref{Eq298}, using the result from Eq. \eqref{Eq296}, can be computed as follows

\begin{equation}\label{Eq299}
    \begin{gathered}           
    {\mu}_{{f}^{\ast}_{a}}(s,t^{\ast}) =  \underbrace{\mathbb{E}_{q(\boldsymbol{x})} \Bigl[\boldsymbol{k}^{s,t^{\ast}}_{N^{\ast}M} \Bigr]}_{\overline{\boldsymbol{\Psi}}^{s,t^{\ast}}_{1}} \underbrace{\boldsymbol{Q}^{-1}\overline{\boldsymbol{\Psi}}^T_1\boldsymbol{\Sigma}^{-1}_{l}\boldsymbol{y}_{l}}_{\boldsymbol{\beta}_l } \text{ } \text{ }
    \therefore {\mu}_{{f}^{\ast}_{a}}(s,t^{\ast}) = \overline{\boldsymbol{\Psi}}^{s,t^{\ast}}_{1}\boldsymbol{\beta}_l 
    \end{gathered}       
\end{equation}

\noindent In a similar manner, the variance $ {\sigma}^2_{{f}^{\ast}_{a}}(s,t^{\ast}) $ can be computed as follows

\begin{equation}\label{Eq300}
    \begin{gathered}           
    {\sigma}^2_{{f}^{\ast}_{a}}(s,t^{\ast}) = \\[1.5ex] \text{ tr} \Bigl\{\Bigl[\boldsymbol{Q}^{-1} - \boldsymbol{K}^{-1}_{MM} + \boldsymbol{\beta}_l \boldsymbol{\beta}^T_l \Bigr] \underbrace{\mathbb{E}_{q(\boldsymbol{x})} \Bigl[\boldsymbol{k}^{s,t^{\ast}}_{MN^{\ast}}\boldsymbol{k}^{s,t^{\ast}}_{N^{\ast}M} \Bigr]}_{\overline{\boldsymbol{\Psi}}^{s,t^{\ast}}_{2}} \Bigr\} \text{ } +  \underbrace{\mathbb{E}_{q(\boldsymbol{x})} \Bigl[{k}^{s,t^{\ast}}_{N^{\ast}N^{\ast}}  \Bigr]}_{\overline{\psi}^{s,t^{\ast}}_{0}} - \Bigl({\mu}_{{f}^{\ast}_{a}}(s,t^{\ast}) \Bigr)^2 \\[1.5ex]
    \therefore {\sigma}^2_{{f}^{\ast}_{a}}(s,t^{\ast}) =  \text{ tr} \Bigl\{\Bigl[\boldsymbol{Q}^{-1} - \boldsymbol{K}^{-1}_{MM} + \boldsymbol{\beta}_l \boldsymbol{\beta}^T_l \Bigr] \overline{\boldsymbol{\Psi}}^{s,t^{\ast}}_{2}\Bigr\} + \overline{\psi}^{s,t^{\ast}}_{0} - \Bigl({\mu}_{{f}^{\ast}_{a}}(s,t^{\ast}) \Bigr)^2
    \end{gathered}       
\end{equation}

\vspace{0.25CM}

\section{Model Specification and Case Study Minutiae}\label{Section9.1}

\subsection{Selecting The Moving-Average Functions}\label{SI_3_1}

For each case study, either single or multiple data sets are generated by sampling underlying latent functions from the conditional GP prior (see Eq. (10)), with the uncertain input $ \boldsymbol{\tau} $ and $ \boldsymbol{\gamma} $ set to known values, followed by artificially corrupting the underlying latent function samples by adding zero-mean Gaussian distributed noise with known standard deviation parameters. The covariance matrix $ \boldsymbol{K}_{NN} $ for the conditional GP prior has been constructed using the spatio-temporal covariance results from Eqs. (25) and (26) with $ K_f = 2 $ underlying latent functions. For the spatial and temporal components of the underlying latent functions, the authors selected the exponential and exponentiated quadratic moving-average functions, respectively, parameterised as follows

\begin{equation}\label{Eq302}
    \begin{gathered}    
    g_a(x \vert \boldsymbol{\theta}_a^{(s)}) = \frac{\nu_{a,s}}{l_{a,s}^2}\exp\Bigl\{ -\frac{x}{2l_{a,s}^2}\Bigr\} \text{ } \text{ } \forall \text{ } a = 1,2; \text{ } \text{ } G_a(t_q \vert \boldsymbol{\theta}_{a}^{(t)}) = \frac{\nu_{a,t}}{l_{a,t}}\exp\Bigl\{ -\frac{t_q^2}{2l_{a,t}^2}\Bigr\} \text{ } \text{ } \forall \text{ } a = 1,2
    \end{gathered}       
\end{equation}

\noindent For the inducing variables $ \boldsymbol{u} $, $ K_u = 2 $ variational inducing functions are constructed by also making use of the exponential and exponentiated quadratic moving-average functions parameterised, respectively, as 

\begin{equation}\label{Eq310}
    \begin{gathered}    
    g'_a(x \vert \boldsymbol{\theta}_a^{'(s)}) = \frac{\nu'_{a,s}}{l_{a,s}^{'2}}\exp\Bigl\{ -\frac{x}{2l_{a,s}^{'2}}\Bigr\} \text{ } \text{ } \forall \text{ } a = 1,2; \text{ } \text{ } G'_a(t_q \vert \boldsymbol{\theta}_{a}^{'(t)}) = \frac{\nu'_{a,t}}{l_{a,t}^{'}}\exp\Bigl\{ -\frac{t_q^2}{2l_{a,t}^{'2}}\Bigr\} \text{ } \text{ } \forall \text{ } a = 1,2
    \end{gathered}       
\end{equation}

\noindent For the selected spatial and temporal component moving-average functions, the summary statistics outlined in Eq. \ref{Eq254} are available in a closed-form solution. Furthermore, the parameters associated with Eqs. \eqref{Eq302} and \eqref{Eq310}, for example, $ \nu_{a,s}, l_{a,s}, \nu'_{a,t}, l'_{a,t} $, and so forth, are freely optimisable quantities that can be estimated by maximising the variational lower bound given by Eq. (75) using gradient-based optimisation. As a concrete example, using Eqs. \eqref{Eq302} and \eqref{Eq310}, consider computing the temporal covariance component

\begin{equation}\label{Eq304}
\begin{split}
    C_{f_{1},f_{{2}}}^{(t)}(t_p, t_q \vert \boldsymbol{\theta}_{1}^{(t)},\boldsymbol{\theta}_{2}^{(t)}) = \int\limits_{-\infty}^{\infty} G_{1}(t_p - z \vert \boldsymbol{\theta}_{1}^{(t)})G_{{2}}(t_q - z\vert \boldsymbol{\theta}_{2}^{(t)}) dz = \frac{\sqrt{2\pi}\nu_{1,t}\nu_{2,t}}{\sqrt{l_{1,t}^2 + l_{2,t}^2}} \exp \Biggl\{ -\frac{(t_p - t_q)^2}{2(l_{1,t}^2 + l_{2,t}^2)} \Biggr\}
    \end{split}
\end{equation}

\noindent As another example, from Eqs. \eqref{Eq302} and (42), for sampled spatial locations $ s_1 $ and $ s_2 $, it can be shown that the unweighted spatial covariance component corresponds to

\begin{equation}\label{Eq306}
\begin{gathered}
    C_{f_1,f_2}^{(s)} \Biggl(\mathlarger{\sum}_{j \in \{1,2\}} \tau_j^2 \mathlarger{\mathlarger{\vert}} \boldsymbol{\theta}_1^{(s)},\boldsymbol{\theta}_2^{(s)} \Biggr) = \int\limits_{0}^{\infty} g_1(x \text{ } + \mathlarger{\sum}_{j \in \{1,2\}} \tau_j^2  \vert \boldsymbol{\theta}_1^{(s)})g_{{2}}(x \vert \boldsymbol{\theta}_{2}^{(s)})  dx  = 
    \frac{2\nu_{1,s}\nu_{2,s}}{l_{1,s}^2 + l_{2,s}^2} \exp \Biggl\{ -\frac{(\tau_1^2 + \tau_2^2)}{2l_{1,s}^2} \Biggr\}
    \end{gathered}
\end{equation}

\noindent For each case study considered, to generate latent function samples and noise-corrupted observational data from the conditional GP prior, the moving-average function parametric values and deterministic input values for $ \boldsymbol{\tau}, \boldsymbol{\gamma}, \sigma_{1} $, and $ \sigma_{2} $, as outlined in Table \ref{Table3} and Table \ref{Table1}, respectively, have been used. Table \ref{Table1} also outlines the measured/estimated values for the model inputs used throughout all the case studies, as mapped through the alternative latent variable parameterisations (see Sect. 5.1), that serve as the uncertain inputs for $ \boldsymbol{\tau} $ and $ \boldsymbol{\gamma} $.

% Table 3
\bgroup
\def\arraystretch{1.3} %  1 is the default, change whatever you need
\begin{table}[ht!]
    \centering
    \begin{tabular}{|c|c|c|}
    \hline
    \multicolumn{2}{|c|}{\textbf{Moving-Average Function Parameter}} & \textbf{Parameter Numeric Value} \\ \hline
      \parbox[t]{2mm}{\multirow{4}{*}{\rotatebox[origin=c]{90}{\textbf{Spatial}}}} & $ \nu_{1,s} $ & 15.6250 \\ \cline{2-3}
        & $ \nu_{2,s} $ & 18.7500 \\ \cline{2-3}
        & $ l_{1,s} $ & 15.0000 \\ \cline{2-3}
        & $ l_{2,s} $ & 20.0000 \\ \hline
        \parbox[t]{2mm}{\multirow{4}{*}{\rotatebox[origin=c]{90}{\textbf{Temporal}}}} & $ \nu_{1,t} $ & 0.4950 \\ \cline{2-3}
        & $ \nu_{2,t} $ & 1.3200 \\ \cline{2-3}
        & $ l_{1,t} $ & 0.5000 \\ \cline{2-3}
        & $ l_{2,t} $ & 1.7000\\ \hline
    \end{tabular}
    \caption{Spatial and temporal moving-average function parametric values that were used to generate the underlying latent function samples for the simulation-based case studies considered in this work.}
    \label{Table3}
\end{table}

% Table 1
\bgroup
\def\arraystretch{1.3} %  1 is the default, change whatever you need
\begin{table}[ht!]
    \centering    
    \begin{tblr}{|c|c|c|c|}
    \hline 
        \textbf{Deterministic Input} & \textbf{Value} & \textbf{Measured/Estimated Input} & \textbf{Value} \\ \hline
         $ \tau_1 $ & 3.8730 & $ d_{ \tau_1 } $ & 3.7093 \\ \hline
         $ \tau_2 $ & 2.2361 & $ d_{ \tau_2 } $ & 2.0828 \\ \hline
         $ \tau_3 $ & 3.1623 & $ d_{ \tau_3 } $ & 3.2979 \\ \hline
         $ \gamma_2 $ & 0.9808 & $ d_{ \gamma_2 } $ & 0.7899 \\ \hline
         $ \gamma_3 $ & 0.1199 & $ d_{ \gamma_3 } $ & 0.3035 \\ \hline
         $ \sigma_{1} $ & 0.3500 & - & - \\ \hline
         $ \sigma_{2} $ & 0.2500 & - & -  \\ \hline
    \end{tblr}
    \caption{Ground truth model parameter and uncertain measurement/estimated input values that are used throughout the simulation-based case studies considered in this work.}
    \label{Table1}
\end{table}

\noindent For the $ \tau_j $ measured value, denoted as $ d_{\tau_j} $, a coefficient of variation corresponding to 0.10 was used in conjunction with the deterministic input value, which serves as the mean, to compute a standard deviation parameter. A sample for the measured input was then drawn from a univariate Gaussian density centred on the deterministic value with the computed standard deviation parameter. The same process was repeated for the $ \gamma_2 $ measured/estimated value, denoted as $ d_{\gamma_2} $, however, a coefficient of variation corresponding to 0.30 was used. The measured/estimated value for $ d_{\gamma_3} $ was obtained by solving the equation $ \Phi^2(d_{\gamma_2}) + \Phi^2(d_{\gamma_3}) = 1 $ to ensure that the measured/estimated $ d_{\gamma_a} $ values satisfy the variance stationarity constraint. The prior standard deviation parameter $ \sigma_{\gamma} $, as associated with $ p(\boldsymbol{\gamma}) $ (see Eqs. (44)), was set to a value of 0.25 throughout all the simulation-based case studies. The mean and standard deviation associated with the hyperprior $ p(\eta_{\tau}) $ were set to -1 and 0.75, respectively, as this parameter configuration, after mapping through the alternative latent variable parameterisation, places probability mass over the region $ 0 < \sigma_{\tau}^2 < 2 $. In other words, a priori, the authors believe that the hydrological distance variance parameter is in the range $ 0 < \sigma_{\tau}^2 < 2 $.

\subsection{Deriving The Variational Constraints}\label{SI_3_2}

\noindent Recall from Sect. 3.1 that the original tails-up SSN model of \cite{VerHoef2006} requires that $ w_2 + w_3 = 1 $ to maintain stationarity of the variances. Under the alternative latent variable parameterisation introduced in this work, the constraint was reformulated as $ \Phi^2(\gamma_2) + \Phi^2(\gamma_3) = 1  $. However, recall that $ \gamma_2 $ and $ \gamma_3 $ are modelled as uncertain inputs in this work and are associated with the variational posterior densities $ q(\gamma_2) $ and $ q(\gamma_3) $. Consequently, after propagating the uncertainty associated with $ q(\gamma_2) $ and $ q(\gamma_3) $, it is possible to derive a variational inference-based analogue for the constraint $ \Phi^2(\gamma_2) + \Phi^2(\gamma_3) = 1  $. More specifically, after \{1\} constructing the required covariance matrix $ \boldsymbol{K}_{NN} $ following the upstream construction procedure, and \{2\} computing the expectation with respect to $ q(\boldsymbol{\gamma}) $, as associated with the statistic $ \overline{\boldsymbol{\Psi}}_0 = \mathbb{E}_{q(\boldsymbol{\tau})} \bigl[\mathbb{E}_{q(\boldsymbol{\gamma})}\bigl[\boldsymbol{K}^{l}_{NN} \bigr] \bigr] $, the variational analogue for the constraint $ \Phi^2(\gamma_2) + \Phi^2(\gamma_3) = 1 $, as a consequence of modelling the input uncertainty in $ \boldsymbol{\gamma} $, can be shown to correspond to

\begin{equation}\label{Eq307}
\begin{gathered}
    \biggl[\Phi(\mu_{\gamma_2} \vert 0,1 + \sigma^2_{\gamma_2} ) - 2T(a_{\gamma_2},b_{\gamma_2})\biggr] + \biggl[\Phi(\mu_{\gamma_3} \vert 0,1 + \sigma^2_{\gamma_3} ) - 2T(a_{\gamma_3},b_{\gamma_3})\biggr] = 1
\end{gathered}
\end{equation}

\noindent The quantities $ a_{\gamma_2}, a_{\gamma_3},b_{\gamma_2} $, and $ b_{\gamma_3} $ are defined as follows

\begin{equation}\label{Eq308}
\begin{gathered}
    a_{\gamma_2} = \frac{\mu_{\gamma_2}}{\sqrt{1 + \sigma^2_{\gamma_2}}}; \text{ } a_{\gamma_3} = \frac{\mu_{\gamma_3}}{\sqrt{1 + \sigma^2_{\gamma_3}}}; \text{ } b_{\gamma_2} = \frac{1}{\sqrt{1 + 2\sigma^2_{\gamma_2}}}; \text{ } b_{\gamma_3} = \frac{1}{\sqrt{1 + 2\sigma^2_{\gamma_3}}}
\end{gathered}
\end{equation}

% \begin{equation}\label{Eq309}
% \begin{gathered}
%     b_{\gamma_2} = \frac{1}{\sqrt{1 + 2\sigma^2_{\gamma_2}}}; \text{ } b_{\gamma_3} = \frac{1}{\sqrt{1 + 2\sigma^2_{\gamma_3}}}
% \end{gathered}
% \end{equation}

\noindent The function $ T(\cdot) $ in Eq. \eqref{Eq307} denotes the Owen's T-function (see \citealp{Owen1980}) and is defined by the following integral-based expression

\begin{equation}\label{Eq312}
\begin{gathered}
    T(a_{\gamma_j},b_{\gamma_j}) = \int\limits_{0}^{b_{\gamma_j}} \frac{\mathcal{N}(a_{\gamma_j} \vert 0,1 )\mathcal{N}(a_{\gamma_j}x \vert 0,1 )}{1 + x^2}dx
\end{gathered}
\end{equation}

\noindent As a consequence of the uncertainty propagation associated with $ q(\boldsymbol{\gamma}) $, observe that the variational inference-based constraint in Eq. \eqref{Eq307} must be satisfied during the gradient-based optimisation routine to ensure stationarity of the variances. In this work, the authors rely on constraint-based optimisation, such as fmincon in MATLAB (see Sect. 6), to enforce the required model constraints. Furthermore, also note that the variational inspired constraint in Eq. \eqref{Eq307} explicitly depends on the freely optimisable variational parameters associated with $ q(\gamma_2) $ and $ q(\gamma_3) $, respectively. 

Another set of implicit and less obvious constraints that arise as a result of applying the BGP-LVM framework to river/stream networks are associated with the variational hydrological distances $ \boldsymbol{h}^{\prime} = [h_1^{\prime},h_2^{\prime},h_3^{\prime}]^T $. From the definition of the spatial inducing input locations in Sect. 5.2, notice that the variational hydrological distance $ h_j^{\prime} $, as associated with a spatial inducing input location, will be smaller (in numeric value) than the hydrological distance $ \tau^2_j $, as associated with its corresponding sampled spatial location. For example, from Fig. 4, observe that $ h_1^{\prime} < \tau_1^2 $ where $ \tau_1^2 = h_1 $. Similarly, from Fig. 4, observe that $ h_2^{\prime} < \tau_2^2 $ and $ h_3^{\prime} < \tau_3^2 $. What is important to note is that $ h_1^{\prime},h_2^{\prime} $, and $ h_3^{\prime} $ govern the placement of the spatial inducing input locations $ s_1^{\prime},s_2^{\prime} $, and $ s_3^{\prime} $, respectively. In turn, the placement of $ s_1^{\prime},s_2^{\prime} $, and $ s_3^{\prime} $ are confined relative to the sampled spatial locations $ s_1,s_2 $, and $ s_3 $ which depend on $ \tau_1, \tau_2 $, and $ \tau_3 $, respectively. 

From a purely upstream construction procedure perspective, the fact that $ h_1^{\prime} < \tau_1^2, h_2^{\prime} < \tau_2^2 $, and $ h_3^{\prime} < \tau_3^2 $ must be satisfied seems like a trivial observation. However, $ \tau_1, \tau_2 $, and $ \tau_3 $ are uncertain inputs to the model and are associated with the variational posterior densities $ q(\tau_1), q(\tau_2) $, and $ q(\tau_3) $, respectively. Consequently, from a gradient-based optimisation perspective, what can happen is that certain combinations of the variational lower bound model parameters can violate the upstream construction procedure. 
To remedy the situation, variational inference-based constraints inspired by $ h_1^{\prime} < \tau_1^2, h_2^{\prime} < \tau_2^2 $, and $ h_3^{\prime} < \tau_3^2 $ can be derived and imposed. As an example, from Eqs. (54), (55), \eqref{Eq302}, and \eqref{Eq310}, consider computing %the spatial component integral associated with $ \mathbb{C}\text{ov}[u_{2}(s_2^{\prime}),f_{1}(s_2)] = C^{\ast}_{2,1} \Bigl(d(s_2^{\prime},s_2;\boldsymbol{\tau},\boldsymbol{h}^{\prime}) \vert \boldsymbol{\theta}^{\ast}_{2,1} \Bigr) = C^{\ast}_{2,1} \Bigl(\tau_2^2 - h_2^{\prime}\vert \boldsymbol{\theta}^{\ast}_{2,1} \Bigr) $ which evaluates to

\vspace{-0.25cm}

\begin{equation}\label{Eq313}
\begin{gathered}
   \mathbb{C}\text{ov}[u_{2}(s_2^{\prime}),f_{1}(s_2)] = \int\limits_{0}^{\infty} g^{\prime}_2(x + [\tau_2^2 - h_2^{\prime}] \vert \boldsymbol{\theta}_{2}^{'(s)})g_1(x \vert \boldsymbol{\theta}_{1}^{(s)}) dx = \frac{2\nu_{1,s}\nu'_{2,s}}{l_{1,s}^2 + l_{2,s}^{'2}} \exp \Biggl\{ -\frac{(\tau_2^2 - h_2^{\prime})}{2l_{2,s}^{'2}} \Biggr\} 
\end{gathered}
\end{equation}

Recall from the definition of the spatial inducing input locations outlined in Sect. 5.2 that spatial inducing input location $ s_2^{\prime} $ must maintain a small distance $ \epsilon_{s_2^{\prime},s_2} $ away from the sampled spatial location $ s_2 $. From the upstream construction procedure perspective, this can be achieved by requiring that the variational hydrological distance $ h_2^{\prime} $ be smaller than $ \tau_2^2 $, in other words, $ h_2^{\prime} < \tau_2^2 $, since $ h_2^{\prime} $ and $ \tau_2^2 $ govern the placement of $ s_2^{\prime} $ and $ s_2 $ on the stream segment. However, the variational hydrological distance $ h_2^{\prime} $ is a freely optimisable parameter that can be tuned to place spatial inducing input location $ s_2^{\prime} $ anywhere on the stream segment/branch as long as the junction point $ u_1 $ is not traversed and $ h_2^{\prime} < \tau_2^2 $ is satisfied. Mathematically, the aforementioned properties can be imposed by requiring that

\vspace{-0.20cm}

\begin{equation}\label{Eq314}
\begin{gathered}
   \tau_2^2 - h_2^{\prime} \geq \epsilon_{s_2^{\prime},s_2} 
\end{gathered}
\end{equation}

Similarly, for the remaining spatial inducing input locations and physically sampled spatial locations, the following constraints are imposed to preserve the upstream construction methodology 

\begin{equation}\label{Eq315}
\begin{gathered}
   \tau_1^2 - h_1^{\prime} \geq \epsilon_{s_1^{\prime},s_1}; \text{ }
   \tau_3^2 - h_3^{\prime} \geq \epsilon_{s_3^{\prime},s_3} 
\end{gathered}
\end{equation}

Throughout all the simulation-based case studies considered in this paper, the minimum distance requirements $ \epsilon_{s_1^{\prime},s_1}, \epsilon_{s_2^{\prime},s_2} $, and $ \epsilon_{s_3^{\prime},s_3} $ are set to a numeric value of $ 10^{-6} $. Note that deriving variational inference-based analogues for the constraints in Eq. \eqref{Eq314} and \eqref{Eq315} is not a trivial task. How might one go about deriving these analogue constraints? Interestingly enough, the analogue constraints are already present in the BGP-LVM model for river/stream networks. More specifically, from the work of \cite{Titsias2008,Titsias2009}, it is a well-known fact that the term $ -\text{tr}(\boldsymbol{K}_{NN} - \boldsymbol{K}_{NM}\boldsymbol{K}_{MM}^{-1}\boldsymbol{K}_{MN}) $ serves as a regularisation/penalty term that influences the placement of the inducing variables. In the absence of any uncertain inputs, note that the trace term $ \text{tr}(\boldsymbol{K}_{NN} - \boldsymbol{K}_{NM}\boldsymbol{K}_{MM}^{-1}\boldsymbol{K}_{MN})  $ represents the total variance of the conditional GP prior $ p(\boldsymbol{f} \vert \boldsymbol{u}) $. When the trace term is zero, in other words, $ \boldsymbol{K}_{NN} = \boldsymbol{K}_{NM}\boldsymbol{K}_{MM}^{-1}\boldsymbol{K}_{MN} $, the approximate posterior density $ p(\boldsymbol{f} \vert \boldsymbol{u})q(\boldsymbol{u}) $, in the absence of the uncertain inputs, exactly matches the true underlying posterior density. The emphasis here is that the trace term regulates the quality of the inducing variable approximation and the placement of the inducing input locations. With the introduction of the uncertain inputs $ \boldsymbol{\tau} $ and $ \boldsymbol{\gamma} $, the same theoretical arguments hold, however, now the uncertain inputs are variationally integrated over with respect to the approximate variational posterior densities $ q(\boldsymbol{\tau}) $ and $ q(\boldsymbol{\gamma}) $. Consequently, from Eqs. \eqref{Eq254} or (75), the trace term that regulates the quality of the inducing variable approximation and the selection of the spatio-temporal inducing input locations, correspond to

\begin{equation}\label{Eq316}
\begin{gathered}
   \text{tr}\Bigl\{ \overline{\boldsymbol{\Psi}}_0 \boldsymbol{\Sigma}^{-1}_{l} -  \boldsymbol{K}^{-1}_{MM}\overline{\boldsymbol{\Psi}}_2 \Bigr\}
\end{gathered}
\end{equation}

Recall that the goal is to find variational inference-inspired analogues for the constraints associated with Eq. \eqref{Eq314} and \eqref{Eq315}. The only matrix in Eq. \eqref{Eq316} that considers the interaction between $ \tau_j $ and $ h^{\prime}_j $ is the statistic matrix $ \overline{\boldsymbol{\Psi}}_2 $. Consequently, the matrix $ \overline{\boldsymbol{\Psi}}_2 $ can be used to derive and impose the required constraints. However, since the interaction between $ \tau_j $ and $ h^{\prime}_j $ happens through the separable spatial covariance component contribution, it is sufficient to only consider the spatial component contribution to $ \overline{\boldsymbol{\Psi}}_2 $. For example, when constructing the matrix $ \overline{\boldsymbol{\Psi}}_2 $, the element-wise product interaction between $ \mathbb{C}\text{ov}[u_{2}(s_2^{\prime}),f_{1}(s_2)] $ with itself, as associated with the physically sampled spatial location $ s_2 $ and its spatial inducing input location $ s_2^{\prime} $, arises such that

\begin{equation}\label{Eq317}
\begin{gathered}
   \mathbb{C}\text{ov}[u_{2}(s_2^{\prime}),f_{1}(s_2)] \times \mathbb{C}\text{ov}[u_{2}(s_2^{\prime}),f_{1}(s_2)] = \frac{4\nu_{1,s}^2\nu_{2,s}^{'2}}{l_{1,s}^2 + l_{2,s}^{'2}} \exp \Biggl\{ -\frac{(\tau_2^2 - h_2^{\prime})}{2 l_{2,s}^{'2}} \Biggr\} 
\end{gathered}
\end{equation}

Note that to arrive at Eq. \eqref{Eq317}, the result from Eq. \eqref{Eq313} has been used. To compute the matrix $ \overline{\boldsymbol{\Psi}}_2 $, it is necessary to compute expected values with respect to the approximate variational posterior density $ q(\boldsymbol{\gamma}) $ followed by computing expected values with respect to $ q(\boldsymbol{\tau}) $.
In doing so, the following expected value result is obtained for Eq. \eqref{Eq317}

\vspace{-0.25cm}

\begin{equation}\label{Eq321}
\begin{gathered}
   \mathbb{E}_{q(\boldsymbol{\tau})} \Bigl[ \mathbb{E}_{q(\boldsymbol{\gamma})} \bigl[\mathbb{C}\text{ov}[u_{2}(s_2^{\prime}),f_{1}(s_2)] \times \mathbb{C}\text{ov}[u_{2}(s_2^{\prime}),f_{1}(s_2)]\bigr] \Bigr] = \\[1.5ex] \frac{4l_{2,s}^{'2}\nu_{1,s}^2\nu_{2,s}^{'2}}{(l_{1,s}^2 + l_{2,s}^{'2})^2\sqrt{2\sigma^2_{\tau_{2}} + l_{2,s}^{'2}}} \exp  \Biggl\{-\frac{l_{2,s}^{'2}\mu^2_{\tau_{2}} - h_2^{\prime}(2\sigma^2_{\tau_{2}} + l_{2,s}^{'2})}{l_{2,s}^{'2}(2\sigma^2_{\tau_{2}} + l_{2,s}^{'2})} \Biggr\} 
\end{gathered}
\end{equation}

\noindent From Eq. \eqref{Eq314} it has been established that $ \tau_2^2 - h_2^{\prime} \geq \epsilon_{s_2^{\prime},s_2} $. Furthermore, note that the difference $ (\tau_2^2 - h_2^{\prime}) $ forms part of the $ \exp(\cdot) $ term in Eqs. \eqref{Eq317}. Consequently, to derive a variational inference-inspired analogue for the constraint in Eq. \eqref{Eq314}, the expression in the $ \exp(\cdot) $ term after taking the expected values can be compared with the $ \exp(\cdot) $ term before taking the expected value. In other words, from Eq. \eqref{Eq317} and \eqref{Eq321} 

\begin{equation}\label{Eq322}
\begin{gathered}
   -\frac{(\tau_2^2 - h_2^{\prime})}{l_{2,s}^{'2}} \text{ } {\overset{\mathbb{E}_{q(\tau_2)}[\cdot]}{\mathlarger{\mathlarger{\leadsto}}}} \text{ } -\frac{l_{2,s}^{'2}\mu^2_{\tau_{2}} - h_2^{\prime}(2\sigma^2_{\tau_{2}} + l_{2,s}^{'2})}{l_{2,s}^{'2}(2\sigma^2_{\tau_{2}} + l_{2,s}^{'2})} \text{ } \therefore
   (\tau_2^2 - h_2^{\prime}) \text{ } {\overset{\mathbb{E}_{q(\tau_2)}[\cdot]}{\mathlarger{\mathlarger{\leadsto}}}} \text{ } \frac{l_{2,s}^{'2}\mu^2_{\tau_{2}} }{(2\sigma^2_{\tau_{2}} + l_{2,s}^{'2})} - h_2^{\prime}
\end{gathered}
\end{equation}

\noindent In Eq. \eqref{Eq322}, the symbol $ \leadsto $ should be interpreted as 'leads to the result' after computing the required expecteds values. Since it is required that $ \tau_2^2 - h_2^{\prime} \geq \epsilon_{s_2^{\prime},s_2} $, from Eq. \eqref{Eq322} it then follows that 

\begin{equation}\label{Eq323}
\begin{gathered}
   \frac{l_{2,s}^{'2}\mu^2_{\tau_{2}} }{(2\sigma^2_{\tau_{2}} + l_{2,s}^{'2})} - h_2^{\prime} \geq \epsilon_{s_2^{\prime},s_2}
\end{gathered}
\end{equation}

Using similar arguments as outlined above, for example, by identifying the element-wise product interactions $ \mathbb{C}\text{ov}[u_{1}(s_1^{\prime}),f_{1}(s_1)] \times \mathbb{C}\text{ov}[u_{1}(s_1^{\prime}),f_{1}(s_1)] $, $ \mathbb{C}\text{ov}[u_{1}(s_3^{\prime}),f_{1}(s_3)] \times \mathbb{C}\text{ov}[u_{2}(s_3^{\prime}),f_{1}(s_3)] $, and so forth, as associated with a physically sampled spatial location and its corresponding spatial inducing input location in matrix $ \overline{\boldsymbol{\Psi}}_2 $, the following additional variational inference-inspired constraints can be derived from the matrix $ \overline{\boldsymbol{\Psi}}_2 $

\begin{equation}\label{Eq324}
\begin{gathered}
   \frac{l_{2,s}^2\mu^2_{\tau_{1}} }{(2\sigma^2_{\tau_{1}} + l_{2,s}^2)} - h_1^{\prime} \geq \epsilon_{s_1^{\prime},s_1} \text{ } ; \text{ } \frac{l_{2,s}^{'2}\mu^2_{\tau_{3}} }{(2\sigma^2_{\tau_{3}} + l_{2,s}^{'2})} - h_3^{\prime} \geq \epsilon_{s_3^{\prime},s_3} \\[1.5ex]
   \frac{l_{1,s}^2\mu^2_{\tau_{1}} }{(2\sigma^2_{\tau_{1}} + l_{1,s}^2)} - h_1^{\prime} \geq \epsilon_{s_1^{\prime},s_1} \text{ } ; \text{ } \frac{l_{1,s}^{'2}\mu^2_{\tau_{2}} }{(2\sigma^2_{\tau_{2}} + l_{1,s}^{'2})} - h_2^{\prime} \geq \epsilon_{s_2^{\prime},s_2} \text{ } ; \text{ } \frac{l_{1,s}^{'2}\mu^2_{\tau_{3}} }{(2\sigma^2_{\tau_{3}} + l_{1,s}^{'2})} - h_3^{\prime} \geq \epsilon_{s_3^{\prime},s_3} \\[1.5ex]
   \frac{l_{1,s}^{'2}l_{2,s}^{'2}\mu^2_{\tau_{2}} }{(l_{1,s}^{'2} + l_{2,s}^{'2})\sigma^2_{\tau_{2}} + l_{1,s}^{'2}l_{2,s}^{'2}} - h_2^{\prime} \geq \epsilon_{s_2^{\prime},s_2} \text{ } ; \text{ } \frac{l_{1,s}^{'2}l_{2,s}^{'2}\mu^2_{\tau_{3}} }{(l_{1,s}^{'2} + l_{2,s}^{'2})\sigma^2_{\tau_{3}} + l_{1,s}^{'2}l_{2,s}^{'2}} - h_3^{\prime} \geq \epsilon_{s_3^{\prime},s_3}
\end{gathered}
\end{equation}

\noindent The constraints in Eq. \eqref{Eq323} and \eqref{Eq324} can then be used in conjunction with the variational lower bound (see Eq. (75)) to preserve the upstream construction procedure associated with the BGP-LVM for river/stream networks during gradient-based optimisation. Recall that to induce a computationally efficient and analytically tractable variational lower bound, the concept of the inducing variable $ \boldsymbol{u} $ was introduced and used. Based on the upstream construction procedure associated with $ \boldsymbol{u} $ (see Sect. 5.2), the following additional constraint arises that is required to maintain stationarity of the variances

\vspace{-0.5 cm}

\begin{equation}\label{Eq325}
\begin{gathered}
    \Phi^2(\alpha_2) + \Phi^2(\alpha_3) = 1
\end{gathered}
\end{equation}

Furthermore, due to the ad hoc nature of the heteroskedastic-based procedure outlined in \cite{Basson2023}, it can be difficult to learn reasonable estimates for the additional heteroskedastic variance parameters $ \sigma_{q_ad_a}^2 $ and $ \sigma_{d_a}^2 $, respectively, per $ a^\text{th} $ latent function. To circumvent this problem, this paper imposes the following artificial constraints on the heteroskedastic variance parameters 

\vspace{-0.25cm}

\begin{equation}\label{Eq133}
    \begin{gathered}
        \sigma_{q_ad_a}^2 \leq \sigma_a^2 + \rho_a \text{ }\forall \text{ } a = 1,\cdots,K_f; \text{ } \text{ }
        \sigma_{d_a}^2 \leq \sigma_a^2 + \rho_a \text{ } \forall \text{ } a = 1,\cdots,K_f
    \end{gathered}
\end{equation}

In Eq. \eqref{Eq133} the symbol $ \rho_a $ denotes the offset parameter which was set to $ \rho_a = 10^{-3} \text{ } \forall \text{ } a = 1,\cdots,K_f $ in this paper. Given this setting, the heteroskedastic variance parameters $ \sigma_{q_ad_a}^2 $ and $ \sigma_{d_a}^2 $ are allowed to take on a maximum value approximately equal to the measurement noise variance parameter $ \sigma_a^2 $. Consequently, the maximum variance parameter allowed for each Normal cdf factor is approximately $ 2\sigma_a^2 $. Empirically, the authors find that the artificially imposed constraints in Eq. \eqref{Eq133} produce sensible parameter estimates for the heteroskedastic variance parameters during gradient-based optimisation. Note that these constraints, like the heteroskedastic strategy of \cite{Basson2023}, are also ad hoc. 

The constraints in Eqs. \eqref{Eq133}, \eqref{Eq307}, \eqref{Eq323}, \eqref{Eq324}, and \eqref{Eq325}, in conjunction with the variational lower bound (see Eq. (75)), have been implemented for all the simulation-based case studies explored in this work. When computing the statistics associated with Eq. \eqref{Eq254} and the covariance matrix $ \boldsymbol{K}_{MM} $, using the selected moving-average functions from Eqs. \eqref{Eq302} and \eqref{Eq310}, the practitioner will find that certain products of parameter combinations repeat as a consequence of the assumed separable spatio-temporal covariance form. The repeating parameter combinations can be reparameterised to reduce the overall size of the parameter space that is searched over during the gradient-based optimisation routine. Table \ref{Table2} outlines the repeating products of parameter combinations as well as the proposed re-parameterisation that is used in this work. 

% Table 2
\bgroup
\def\arraystretch{1.3} %  1 is the default, change whatever you need
\begin{table}[h!]
    \centering
    \begin{tblr}{|c|c|}
    \hline
         \textbf{Repeating Parameter Combination} & \textbf{Re-parameterisation} \\ \hline
        $  \nu_{1,t}\nu_{1,s} $ & $ \xi_{1} $ \\ \hline
        $ \nu_{2,t}\nu_{2,s} $ & $ \xi_{2} $ \\ \hline
        $ \nu'_{1,t}\nu'_{1,s} $ & $ \xi_1' $ \\ \hline
        $ \nu'_{2,t}\nu'_{2,s} $ & $ \xi'_2 $ \\ \hline 
    \end{tblr}
    \caption{Separable spatio-temporal covariance re-parameterisations that are used during the gradient-based optimisation routine for the simulation-based case studies considered in this work.}
    \label{Table2}
\end{table}

\vspace{1.0cm}

\subsection{Inverting The Alternative Latent Variable Parameterisations}\label{SI_3_3}

\noindent Since the alternative latent variable parameterisations associated with $ \eta_{\tau}, \boldsymbol{\tau} $, and $ \boldsymbol{\gamma} $ are invertible functions, the change of variables rule for continuous probability density functions can be used to derive the corresponding approximate variational posterior densities for the original set of latent variables $ \sigma^2_{\tau}, \boldsymbol{h} $, and $ \boldsymbol{w} $ (\citealp{Bishop2009}). Using the change of variables rule for continuous probability density functions, it can be shown that the density functions associated with $ \sigma^2_{\tau}, h_j $, and $ w_k $, respectively, correspond to

\vspace{-0.30cm}

\begin{equation}\label{Eq329}
\begin{gathered}
    q(\sigma^2_{\tau}) = \frac{1}{\sigma^2_{\tau}} \text{ } \mathcal{N}\bigl(\ln{\sigma^2_{\tau}} \vert \mu_{\eta_{\tau}},\sigma^2_{\eta_{\tau}}\bigr) = \mathcal{LN}\bigl(\sigma^2_{\tau} \vert \text{ } \mu_{\eta_{\tau}},\sigma^2_{\eta_{\tau}}\bigr)
\end{gathered}
\end{equation}

\vspace{-0.30cm}

\begin{equation}\label{Eq330}
\begin{gathered}
    q(h_j) = \frac{1}{2\sqrt{h_j}} \Biggl[\mathcal{N}\Bigl(-\sqrt{h_j} \text{ } \mathlarger{\mathlarger{\vert}} \text{ } \mu_{\tau_j},\sigma^2_{\tau_j} \Bigr) + \mathcal{N}\Bigl(\sqrt{h_j} \text{ } \mathlarger{\mathlarger{\vert}} \text{ } \mu_{\tau_j},\sigma^2_{\tau_j} \Bigr) \Biggr] \text{ }; \text{ } h_j > 0
\end{gathered}
\end{equation}

\vspace{-0.30cm}

\begin{equation}\label{Eq331}
\begin{gathered}
    q(w_k) = \frac{1}{2\sqrt{w_k}\text{ }\mathcal{N}\bigl(\Phi^{-1}(\sqrt{w_k}) \vert 0,1^2 \bigr)} \text{ } \mathcal{N}\bigl(\Phi^{-1}(\sqrt{w_k}) \text{ } \vert \text{ } \mu_{\gamma_k},\sigma^2_{\gamma_k} \bigr)  \text{ }; \text{ } w_k > 0
\end{gathered}
\end{equation}

\noindent In Eq. \eqref{Eq329}, the symbol $ \mathcal{LN}(\cdot) $ denotes the Log-Normal density function with parameters $ \mu_{\eta_{\tau}} $ and $ \sigma^2_{\eta_{\tau}} $ whereas in Eq. \eqref{Eq331} the symbol $ \Phi^{-1}(\cdot) $ denotes the standard Gaussian inverse cumulative distribution function. 

\vspace{-0.5cm}

\subsection{Assessing The Predictive Performance}\label{SI_3_4}

\noindent To assess the predictive performance of the BGP-LVM for river/stream networks against competing benchmarks, the root mean squared error (RMSE, see Eq. \eqref{Eq326}), the mean absolute error (MAE, see Eq. \eqref{Eq327}), and the mean negative log-loss (MNLL, see Eq. \eqref{Eq328}) are reported and compared (see \citealp{Rasmussen2006}, \citealp{Lazaro2010}, \citealp{Groot2012}, \citealp{Zhao2016}, and \citealp{Basson2023}). 

\begin{equation}\label{Eq326}
\begin{gathered}
    \text{RMSE$\bigl( \boldsymbol{f^{\ast}} $ , $ {\boldsymbol{\mu}}_{{f}^{\ast}} \bigr)$} = \sqrt{\frac{1}{K_f} \mathlarger{\sum}\limits_{a = 1}^{K_f} \Biggl[\frac{1}{N^{\ast}_a} \mathlarger{\sum}\limits_{n = 1}^{N^{\ast}_a} \Bigl(\bigl[\boldsymbol{f^{\ast}}_{a}\bigr]_n - \bigl[{\boldsymbol{\mu}}_{{f}^{\ast}_{a}}]_n \Bigr)^2 \Biggr]}
\end{gathered}
\end{equation}

\vspace{0.25cm}

\begin{equation}\label{Eq327}
\begin{gathered}
    \text{MAE$ \bigl( \boldsymbol{f^{\ast}} $ , $ {\boldsymbol{\mu}}_{{f}^{\ast}} \bigr)$} = \frac{1}{K_f} \mathlarger{\sum}\limits_{a = 1}^{K_f} \Biggl[\frac{1}{N^{\ast}_a} \mathlarger{\sum}\limits_{n = 1}^{N^{\ast}_a} \Bigl\vert\bigl[\boldsymbol{f^{\ast}}_{a}\bigr]_n - \bigl[{\boldsymbol{\mu}}_{{f}^{\ast}_{a}}]_n \Bigr\vert \Biggr]
\end{gathered}
\end{equation}

\vspace{0.25cm}

\begin{equation}\label{Eq328}
\begin{gathered}
    \text{MNLL$ \bigl( \boldsymbol{f^{\ast}} $ , $ {\boldsymbol{\mu}}_{{f}^{\ast}} $ , $ {\boldsymbol{\sigma}}_{{f}^{\ast}} \bigr)$} = \\[1.5ex] \frac{1}{K_f} \mathlarger{\sum}\limits_{a = 1}^{K_f} \mathlarger{\Biggl[}\frac{1}{N^{\ast}_a} \mathlarger{\sum}\limits_{n = 1}^{N^{\ast}_a} \mathlarger{\Biggl[} \frac{1}{2} \ln{ \Bigl(2\pi\bigl(\bigl[{\boldsymbol{\sigma}}_{{f}^{\ast}_{a}}]_n\bigr)^2 \Bigr)} + \frac{\Bigl(\bigl[\boldsymbol{f^{\ast}}_{a}\bigr]_n - \bigl[{\boldsymbol{\mu}}_{{f}^{\ast}_{a}}]_n \Bigr)^2}{2\bigl(\bigl[{\boldsymbol{\sigma}}_{{f}^{\ast}_{a}}]_n \bigr)^2} \mathlarger{\Biggr]} \mathlarger{\Biggr]}
\end{gathered}
\end{equation}

\vspace{0.25cm}

\noindent For all three criteria, smaller values imply better model predictive performance. In Eqs. \eqref{Eq326} to \eqref{Eq328}, the symbol $ N^{\ast}_a $ denotes the total number of predicted latent function values per a$^{\text{th}}$ underlying latent function. The symbol $ \vert \cdot \vert $ in Eq. \eqref{Eq327} requires computing the absolute value function. Furthermore, since all the case studies considered in this work are simulation-based, the authors have access to the values for the ground truth latent functions, collectively denoted as $ \boldsymbol{f}^{\ast} = [(\boldsymbol{f}^{\ast}_{1})^T,(\boldsymbol{f}^{\ast}_{2})^T]^T $ with $ K_f = 2 $, at the spatio-temporal prediction input locations of interest. The column vectors $ {\boldsymbol{\mu}}_{{f}^{\ast}} $ and $ {\boldsymbol{\sigma}}_{{f}^{\ast}} $ follow the same structural definition as $ \boldsymbol{f}^{\ast} $, in other words, $ {\boldsymbol{\mu}}_{{f}^{\ast}} = [\boldsymbol{\mu}^T_{f^{\ast}_{1}},\boldsymbol{\mu}^T_{f^{\ast}_{2}}]^T $ and $ {\boldsymbol{\sigma}}_{{f}^{\ast}} = [\boldsymbol{\sigma}^T_{f^{\ast}_{1}},\boldsymbol{\sigma}^T_{f^{\ast}_{2}}]^T $. The symbol $ {\boldsymbol{\mu}}_{{f}^{\ast}} $ denotes the mean of the latent function predictive posterior density whereas $ {\boldsymbol{\sigma}}_{{f}^{\ast}} $ denotes the associated marginal predictive standard deviation. Since all the latent function predictive posterior densities considered in this work are of a Gaussian functional form, the mean $ {\boldsymbol{\mu}}_{{f}^{\ast}} $ is also the predictive density maximum a posteriori (MAP) estimate. For the BGP-LVM framework considered in this work, the elements of the column vectors $ {\boldsymbol{\mu}}_{{f}^{\ast}} $ and $ {\boldsymbol{\sigma}}_{{f}^{\ast}} $ can be computed from Eq. \eqref{Eq299} and the square root of Eq. \eqref{Eq300}, respectively. Note that the ordering associated with $ \boldsymbol{f}^{\ast}, {\boldsymbol{\mu}}_{{f}^{\ast}}, $ and $ {\boldsymbol{\sigma}}_{{f}^{\ast}} $ is not of particular importance. However, the selected ordering should be consistent. For example, if the n$^{\text{th}}$ element associated with $ \boldsymbol{f}^{\ast} $ pertains to making a prediction about the a$^{\text{th}}$ underlying latent function at spatial location $ s $ and temporal input $ t^{\ast} $, then the n$^{\text{th}}$ element of $ {\boldsymbol{\mu}}_{{f}^{\ast}} $ and $ {\boldsymbol{\sigma}}_{{f}^{\ast}} $ must also pertain to the a$^{\text{th}}$ underlying latent function at spatial location $ s $ and temporal input $ t^{\ast} $.

\section{Specifying The Number Of Temporal Inducing Input Locations}\label{SI_5}

Recall that the multi-output BGP-LVM, and by association the independent counterpart, requires the practitioner to specify the number of temporal inducing input locations $ M_t $ (see Sect. 5.2). To determine the number of temporal inducing input locations, the authors generated an additional data set and trained both the multi-output and independent BGP-LVM frameworks for various values of the number of temporal inducing input locations $ M_t $. For the randomly generated data set, 7 temporal inducing input locations were selected as a starting point, followed by running the gradient-based optimiser to find parameter point estimates for the unknown model parameters for each BGP-LVM framework. From an implementation perspective, the authors minimised the negative variational lower bound using fmincon, in conjunction with the MultiStart algorithm, in MATLAB.

The MultiStart algorithm allows the user to explore multiple model parameter starting points and, for the case study under consideration, the authors randomly selected 500 starting points. The MultiStart algorithm returns multiple model parameter point estimates, each associated with an objective function local minimum, ranked according to the objective function value. The authors recorded the lowest objective function value for each of the BGP-LVM frameworks as the associated values correspond to the best local minimum found by the optimiser. This procedure was repeated for each of the BGP-LVM frameworks with the number of temporal inducing input locations incrementally increased until $ M_t = 25 $ was reached. Figure \ref{fig:Figure14} visually depicts the objective function values for the multi-output (top panel) and independent (bottom panel) BGP-LVM frameworks against the number of temporal inducing input locations.

% Temporal Inducing Input Selection
%\begin{figure*}[h!tb]%[!t]]
%   %\hspace{-1.15in}
%   \hspace{-0.20in}
%   \sbox0{\includegraphics{Fig14.jpeg}}%
%   \includegraphics[scale = 0.050]{Fig14.jpeg}
%   \caption{Plot of the negative variational lower bound for the multi-output (top panel) and independent (bottom panel) BGP-LVM frameworks as the number of temporal inducing input locations $ M_t $ are incrementally increased. Results are displayed for up to $ M_t = 22 $ temporal inducing input locations.} %from 7 to 25.}
%   \label{fig:Figure14}
%\end{figure*}

From Fig. \ref{fig:Figure14}, observe that, as the number of temporal inducing input locations $ M_t $ increases, for both the BGP-LVM frameworks the objective function value decreases until it stabilises between 15 to 17 temporal inducing input locations. This indicates that the variational lower bound for both BGP-LVM frameworks has reached a point where it is sufficiently tight (see \cite{Titsias2008,Titsias2009} and \cite{Basson2023} for more details). 
For the case study under consideration, the authors opted for a more conservative estimate for $ M_t $ and set the number of temporal inducing input locations to $ M_t = 20 $. Note that throughout all the repeated simulation-based experiments performed for Case Study 1, $ M_t $ was fixed to the conservative value of $ M_t = 20 $. 

% Temporal Inducing Input Selection
\begin{figure*}[h!tb]%[!t]]
   %\hspace{-1.15in}
   \hspace{-0.20in}
   \sbox0{\includegraphics{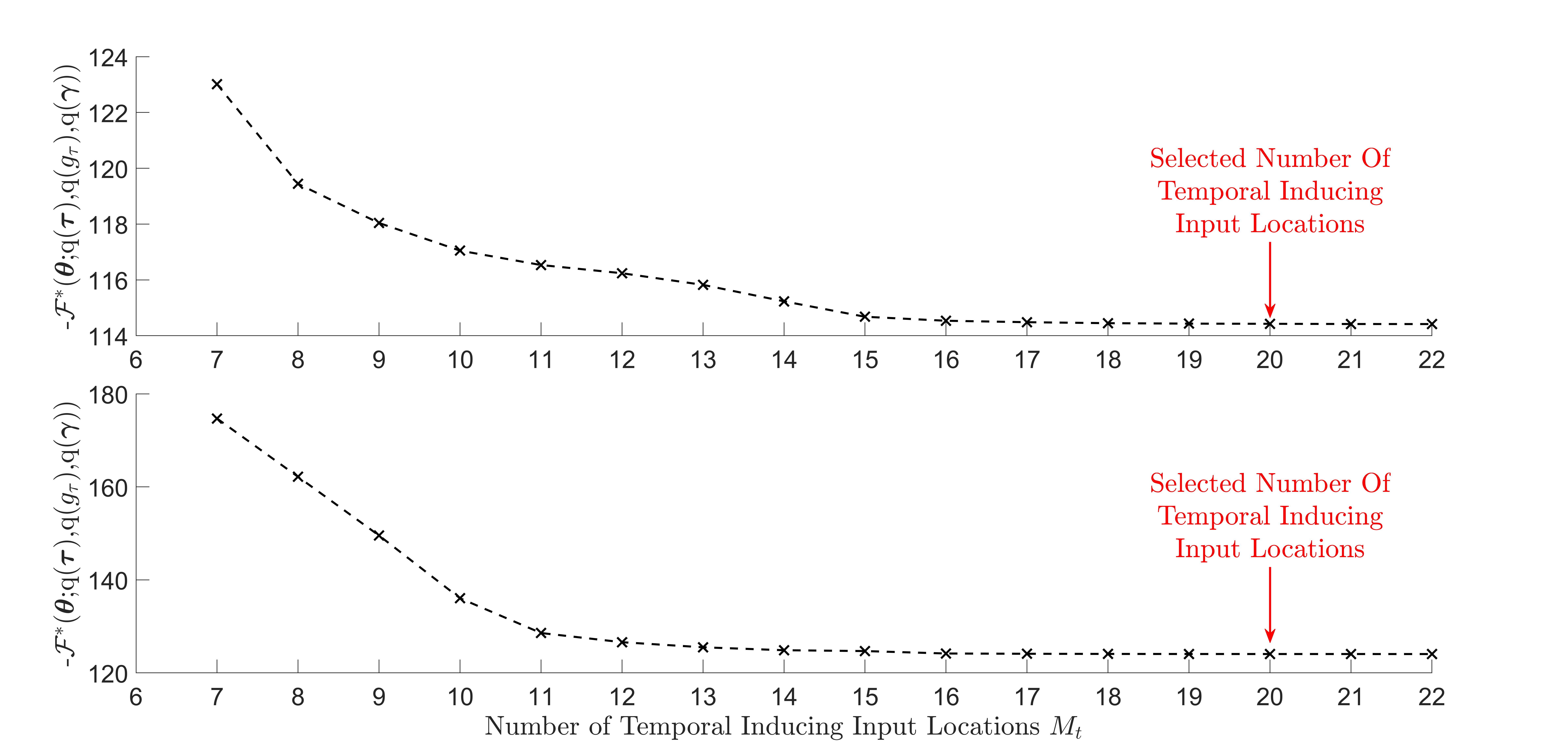}}%
   \includegraphics[scale = 0.050]{Fig14.jpeg}
   \caption{Plot of the negative variational lower bound for the multi-output (top panel) and independent (bottom panel) BGP-LVM frameworks as the number of temporal inducing input locations $ M_t $ are incrementally increased. Results are displayed for up to $ M_t = 22 $ temporal inducing input locations.} %from 7 to 25.}
   \label{fig:Figure14}
\end{figure*}

\newpage

\section{Additional Supplementary Figures}\label{SI_4}

% Figure 7
\begin{figure*}[!ht]%[!t]]
    \centering
    \includegraphics[scale = 1.2]{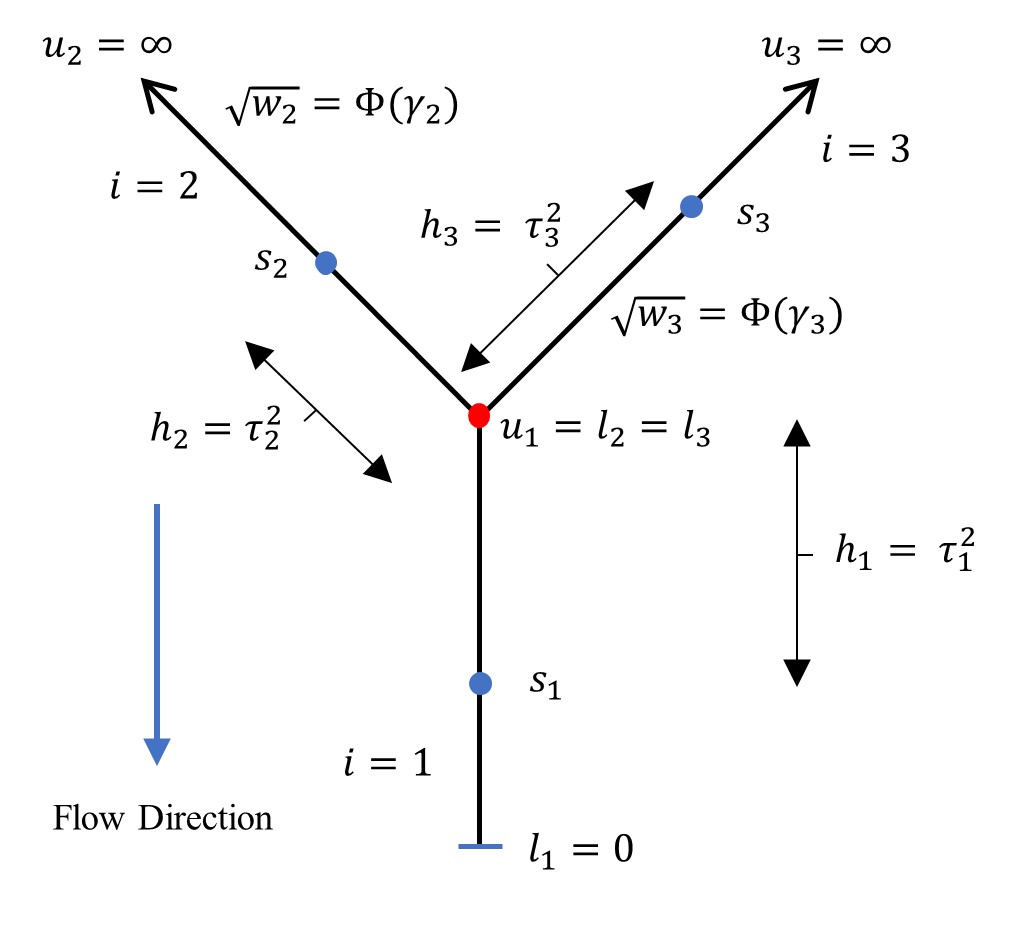}
    \caption{The hypothetical stream network depicted in Fig. 1 parameterised in terms of the alternative set of latent variables. Here $ \Phi(\gamma_2) $ and $ \Phi(\gamma_3) $ govern the stationarity of the variances whereas $ \tau_1 $, $  \tau_2 $, and $ \tau_3 $ characterise the stream distance between the sampled spatial locations $ s_1 $, $ s_2 $, and $ s_3 $, respectively, relative to the junction point $ u_1 $. Figure \ref{fig:Figure7} has been reproduced and adjusted from the work of \cite{VerHoef2006}.}
    \label{fig:Figure7}
\end{figure*}

\end{document}